\documentclass[twocolumn,tighten,times]{aastex631_arXiv}

\hypersetup{
           breaklinks=true,   
           colorlinks=true,   
           pdfusetitle=true,  
           linkcolor=red,
           citecolor=blue,
           filecolor=cyan,
           urlcolor=purple
           }

\graphicspath{{./}{./figures/}} 

\usepackage{amsmath}
\usepackage{advdate}
\usepackage{float}

\usepackage{verbatim}

\newcommand{\gvec}[1]{\boldsymbol{#1}}               

\newcommand{\Mearth}{\mbox{$M_{\oplus}$}}            
\newcommand{\Mp}{\mbox{$M_{p}$}}                     
\newcommand{\Rp}{\mbox{$R_{p}$}}                     
\newcommand{\Ms}{\mbox{$M_{\star}$}}                 
\newcommand{\Msun}{\mbox{$M_{\odot}$}}               
\newcommand{\dMp}{\dot{M}_{p}}                       
\newcommand{\Rhill}{\mbox{$R_\mathrm{H}$}}           
\newcommand{\Rbondi}{\mbox{$R_\mathrm{B}$}}          
\newcommand{\sigu}{\mbox{$\mathrm{g\,cm}^{-2}$}}     
\newcommand{\rhou}{\mbox{$\mathrm{g\,cm}^{-3}$}}     
\newcommand{\K}{\mbox{$\mathrm{K}$}}                 
\newcommand{\AU}{\mbox{au}}                          
\newcommand{\Myr}{\mbox{Myr}}                        
\newcommand{\cisec}[1]{Section~\ref{#1}}             
\newcommand{\cifig}[1]{Figure~\ref{#1}}              
\newcommand{\cieq}[1]{Equation~(\ref{#1})}           
\newcommand{\citab}[1]{Table~\ref{#1}}               

\newcommand{\T}{\rule{0pt}{2.6ex}}                   
\newcommand{\B}{\rule[-1.2ex]{0pt}{0pt}}             
\received{January 12, 2024}
\revised{April 3, 2024}
\accepted{April 6, 2024}
\published{May 24, 2024}

\shorttitle{Planet Formation by Accretion of Small Solids}
\shortauthors{D'Angelo \& Bodenheimer}

\begin{document}

\title{\Large{Planet Formation by Gas-Assisted Accretion of Small Solids}}

\author[0000-0002-2064-0801]{Gennaro D'Angelo}
\affiliation{Theoretical Division, Los Alamos National Laboratory, Los Alamos, NM 87545, USA; \href{mailto:gennaro@lanl.gov}{gennaro@lanl.gov}}

\author[0000-0001-6093-3097]{Peter Bodenheimer}
\affiliation{UCO/Lick Observatory, Department of Astronomy and Astrophysics, University of California, Santa Cruz, CA 95064, USA; \href{mailto:peter@ucolick.org}{peter@ucolick.org}}
\begin{abstract}
\noindent%
We compute the accretion efficiency of small solids, with radii 
$1\,\mathrm{cm}\le R_{s}\le 10\,\mathrm{m}$, on planets embedded
in gaseous disks. Planets have masses $3\le\Mp\le 20$ Earth masses
($\Mearth$) and orbit within $10\,\AU$ of a solar-mass star.
Disk thermodynamics is modeled via three-dimensional radiation-hydrodynamic
calculations that typically resolve the planetary envelopes.
Both icy and rocky solids are considered, explicitly modeling their
thermodynamic evolution. 
The maximum efficiencies of $1\le R_{s}\le 100\,\mathrm{cm}$ particles
are generally $\lesssim 10$\%, whereas $10\,\mathrm{m}$ 
solids tend to accrete efficiently or be segregated beyond the planet's orbit. 
A simplified approach is applied to compute the accretion efficiency 
of small cores, with masses $\Mp\le 1\,\Mearth$ and without envelopes, 
for which efficiencies are approximately proportional to $M_{p}^{2/3}$.
The mass flux of solids, estimated from unperturbed drag-induced
drift velocities, provides typical accretion rates
$d\Mp/dt\lesssim 10^{-5}\,\mathrm{\Mearth\,yr^{-1}}$.
In representative disk models with an initial gas-to-dust mass ratio
of $70$--$100$ and total mass of $0.05$--$0.06\,\Msun$, solids' accretion
falls below $10^{-6}\,\mathrm{\Mearth\,yr^{-1}}$ after $1$--$1.5\,\Myr$.
The derived accretion rates, as functions of time and planet mass, are 
applied to formation calculations that compute dust opacity self-consistently
with the delivery of solids to the envelope.
Assuming dust-to-solid coagulation times of $\approx 0.3\,\Myr$ and disk
lifetimes of $\approx 3.5\,\Myr$, heavy-element inventories in the range
$3$--$7\,\Mearth$ require that $\approx 90$--$150\,\Mearth$ of solids
cross the planet's orbit.
The formation calculations encompass a variety of outcomes, from
planets a few times \Mearth, predominantly composed of heavy elements,
to giant planets. 
The peak luminosities during the epoch of solids' accretion range from
$\approx 10^{-7}$ to $\approx 10^{-6}\,L_{\odot}$.
\end{abstract}
\keywords{
\href{http://astrothesaurus.org/uat/1579}{Stellar accretion disks (1579)}; 
\href{http://astrothesaurus.org/uat/2009}{Radiative magnetohydrodynamics (2009)};
\href{http://astrothesaurus.org/uat/101}{Astrophysical fluid dynamics (101)}; 
\href{http://astrothesaurus.org/uat/1965}{Computational methods (1965)}; 
\href{http://astrothesaurus.org/uat/1241}{Planet formation (1241)}; 
\href{http://astrothesaurus.org/uat/2204}{Planetary-disk interactions (2204)}; 
\href{http://astrothesaurus.org/uat/1259}{Planetesimals (1259)}; 
\href{http://astrothesaurus.org/uat/1300}{Protoplanetary disks (1300)}; 
\href{http://astrothesaurus.org/uat/511}{Extrasolar rocky planets (511)}; 
\href{http://astrothesaurus.org/uat/2172}{Extrasolar gaseous planets (2172)}; 
\href{http://astrothesaurus.org/uat/509}{Extrasolar gaseous giant planets (509)};
\href{http://astrothesaurus.org/uat/1244}{Planetary atmospheres (1244)}
}

\section{Introduction}
\label{sec:Intro}

\defcitealias{gennaro2015}{DP15}
\defcitealias{gennaro2013}{DB13}

Classic theories of planet formation assume that initial growth
proceeds by collisions with planetesimals, bodies ranging in
size from a fraction to hundreds of kilometers \citep[e.g.,][]{lissauer1993a}.
Relict populations of these bodies are found in the asteroid and Kuiper
belts \citep[e.g.,][]{morbidelli2020,BdM2022}.
The path to formation of planetesimals around stars remains mysterious,
but conventional wisdom generally assumes that they are assembled
from aggregates of much smaller particles, which are
collected through some hydrodynamical process (in the circumstellar
gas) and eventually clump together under their own gravity 
\citep[see, e.g.,][and references therein]{weiss2022}.
These smaller particles, whose presence can be inferred from observations
of protoplanetary disks \citep[e.g.,][]{natta2007,andrews2020}, would
form through coagulation of primordial dust grains entrained in the gas.

This scenario for planetesimal formation has led to the suggestion
that swarms of ``small particles'' orbiting a star in a protoplanetary
disk may directly supply solid material to emerging planetary embryos
and planets, determining their heavy-element inventories
\citep{ormel2010,lambrechts2012}.
Over the past decade, such delivery mode of solids has been referred
to as ``pebble accretion''. However, despite the terminology, these
solids are unrelated to clastic rocks and are not required to fulfill
the geological definition \citep[][]{gog1972}.
In fact, they are not characterized by size or composition, but rather
by drag interaction with circumstellar gas, corresponding to Stokes
numbers (see below) in the range between $\sim 10^{-3}$ and $\approx 1$
\citep{drazkowska2022}.
These numbers quantify the coupling timescale, in units of the orbital
time (i.e., the inverse of the Keplerian frequency), between particle
and gas dynamics.
The Stokes number is proportional to the radius and material density 
of a particle, and inversely proportional to the gas density.
Therefore, according to current terminology, astrophysical pebbles orbiting
in a solar-nebula type of disk, between $\approx 1$ and $\approx 10\,\AU$,
can range from submillimeter silicate grains 
\citep[i.e., astrophysical dust,][]{dalessio2001} to meter-sized ice blocks.
At larger distances, as the gas density lowers, smaller particles would
be considered as astrophysical pebbles (and the opposite would occur
closer to the star).
An overview of the typical outcomes from pebble accretion calculations
can be found in, e.g., \citet[][]{johansen2017}, \citet[][]{drazkowska2022},
and references therein.

Herein, we build first-principles numerical models of gas and solids 
thermodynamics. Since Stokes numbers cannot be constrained in
the calculations (because they depend on primitive variables), we consider
specific particle sizes (from $1\,\mathrm{cm}$ to $10\,\mathrm{m}$)
and compositions (SiO$_{2}$ and H$_{2}$O), and we will refer to these
particles simply as ``small solids''.
Overlap with the domain of astrophysical pebbles varies according
to local gas conditions.

The largest difference between planetesimal and small-solid accretion
scenarios rests on their dynamics as they evolve in circumstellar gas.
On the one hand, drag forces only provide a correction to the Keplerian
orbits of the large bodies \citep[][]{whipple1973}. The drag-induced
drifting timescale of $1\,\mathrm{km}$ planetesimals is $\sim 10^{6}$
orbital periods (in the $1$--$10\,\AU$ region), hence radial mobility
due to aerodynamic drag can be largely ignored during planet formation.
The accretion rate on a planet is determined by the local surface density
of solids and by a characteristic cross-section for collisions, which can 
be enhanced by gravitational focusing and atmospheric drag 
\citep[see, e.g.,][]{pollack1996,inaba2003a}.
On the other hand, drag forces strongly affect the motion of small solids
\citep[][]{weidenschilling1977b}. Radial drift occurs on much shorter
timescales than those of planetesimals. 
Hence, accretion is a non-local process.
The global transport of solids through the disk determines the availability
of the supply and the gas thermodynamics at and below the orbital length-scale
determines the probability of accretion.
The former cannot be evaluated without a global model of the circumstellar
gas. The latter requires modeling tidal interactions between the planet
and the disk \citep[e.g.,][]{morbidelli2012} and gas dynamics in 
proximity of the planet \citep[e.g.,][]{picogna2018,popovas2018}.
Therefore, any outcome of a planet formation calculation based 
on small-solid accretion is bound to depend on both local and remote
gas thermodynamics.
The applicability of generic formulations of accretion rates, found in 
the literature, is obviously limited by underlying assumptions
\citep[see, e.g.,][]{lambrechts2012,drazkowska2022}.

Planetesimal accretion proceeds as long as there is material in 
a region, the ``feeding zone'', spanning a few Hill radii on either
side of the planet's orbit. Once this zone is depleted, accretion is
much reduced and limited to the resupply rate into the region
\citep[e.g.,][]{pollack1996,gennaro2014}.
Small-solid accretion can proceed as long as there is transport of
material across the planet's orbit, which can be hindered by the overall
depletion of solids, dissipating gas, and disk-planet tidal interactions.

In terms of interior compositions, the heavy-element inventory provided
by the two accretion scenarios may differ, since solids are accreted
locally in one case and (effectively) remotely in the other. However,
if planetesimals form out of small solids, which have drifted considerable
distances prior to clumping together, planetesimals too can have (largely 
or to some extent) non-local compositions. Hence, diversity in compositions
may not discriminate between the two scenarios.
Differences in heavy-element stratification may also be difficult 
to ascertain.
Clearly, small and large solids may coexist \citep[e.g.,][]{kessler2023},
although mass partitioning would depend on the mechanism that converts 
one population into the other
\citep[see, e.g.,][and references therein]{drazkowska2022,weiss2022}.

Herein, we present direct calculations of small-solid accretion, based on 
three-dimensional radiation-hydrodynamics calculations of disks with
embedded planets that resolve the planetary envelope. Models also account for
the thermal evolution of the particles. We focus on planet masses between
$3$ and $20\,\Mearth$. These calculations provide the accretion efficiency
(or probability) of the particles on the planets at different orbital 
locations ($1$ to $10\,\AU$) and disk conditions. 
The accretion efficiency is the fraction of the local accretion
rate of solids intercepted by the planet.
The results are complemented with those obtained from a simpler
approach, adopted to compute the accretion efficiency of bare cores 
(i.e., without a substantial gaseous envelope)
less massive than one Earth mass.
The global transport of small solids, computed from drag-induced drift
velocities in unperturbed disks, is combined with the accretion efficiencies
to provide the accretion rates of the planets. These rates are then applied
to actual formation and structure calculations, which compute dust opacity
in the planet's envelope self-consistently with the delivery of solids.
These latter calculations are intended as demonstrations of the framework
provided herein and do not target any specific planet.

In the following, the orbital dynamics of small solids in unperturbed and
perturbed disks is described in Section~\ref{sec:DIG}. The radiation-hydrodynamics
models and accretion efficiencies of envelope-bearing planets are presented in 
Section~\ref{sec:RHD}; the accretion efficiencies of small, bare-core planets
are discussed in Section~\ref{sec:SPC}.
The formation calculations are presented in Section~\ref{sec:FSM}.
Finally, our conclusions are summarized in Section~\ref{sec:DC}.


\section{Drag-Induced Orbital Decay of Solids}
\label{sec:DIG}

\subsection{Dynamics in Unperturbed Disks}
\label{sec:DUK}

Small solids orbiting a star in a smooth unperturbed disk may experience 
strong drag forces exerted by the gas, and thus undergo rapid orbital decay.
Assuming that the solids' mass is locally much smaller than the gas mass,
the time required for the velocity of a solid particle, $\gvec{u}_{s}$, 
to converge toward the gas velocity, $\gvec{u}_{g}$, is
\begin{equation}
  \tau_{s}=\frac{|\gvec{u}_{g}-\gvec{u}_{s}|}{|\gvec{a}_{D}|},
  \label{eq:tau_s_def}
\end{equation}
where $\gvec{a}_{D}$ is the drag acceleration. The time $\tau_{s}$ 
is referred to as the ``stopping time'' of the particle
\citep{whipple1973,weidenschilling1977b}. 
Indicating with $\rho_{g}$ and $\rho_{s}$ respectively the gas and 
solid density, $R_{s}$ the particle radius, and $a_{s}$ its orbital 
distance, \cieq{eq:tau_s_def} becomes 
\citep{whipple1973,weidenschilling1977b}
\begin{equation}
  \tau_{s}=
        \frac{8}{3}
  \left(\frac{1}{C_{D}}\right)
  \left(\frac{\rho_{s}}{\rho_{g}}\right)
  \left(\frac{R_{s}}{|\gvec{u}_{g}-\gvec{u}_{s}|}\right).
  \label{eq:tau_s}
\end{equation}
The drag coefficient, $C_{D}$, is a function of the 
thermodynamic properties of both gas and solids.
In the free molecular flow regime (i.e., for very small particles),
$C_{D} |\gvec{u}_{g}-\gvec{u}_{s}|$ tends to a constant, hence
$\tau_{s}\propto R_{s}$.

The rotation rate of the gas, $\Omega_{g}$, is generally different 
from the Keplerian rate, $\Omega_{\mathrm{K}}$, and depends on the radial 
gradient of the gas pressure, $P_{g}$, typically a function of gas 
density and temperature, $T_{g}$. 
By requiring conservation of the gas linear momentum in the disk's radial
direction, $r$, it can be approximated as 
\begin{equation}
 \Omega_{g}^{2}\approx\frac{1}{r}\left(\frac{\partial \Phi}{\partial r}\right)%
                                    +\frac{1}{r\rho_{g}}\left(\frac{\partial P_{g}}{\partial r}\right),
 \label{eq:OPHI}
\end{equation}
$\Phi$ being the gravitational potential in the disk. 
Neglecting the self-gravity of gas and solids, 
$\partial\Phi/\partial r=r\Omega^{2}_{\mathrm{K}}$ and \cieq{eq:OPHI} 
applied to the midplane reduces to
\begin{equation}
 \Omega_{g}\approx\Omega_{\mathrm{K}}\sqrt{1-\eta\left(\frac{H_{g}}{a_{s}}\right)^{2}}, 
 \label{eq:Omega}
\end{equation}
where $H_{g}$ is the disk's pressure scale-height at the midplane.
The quantity $\eta$ depends on the normalized gradients 
$\partial\ln{\rho_{g}/\partial\ln{r}}$ and
$\partial\ln{T_{g}/\partial\ln{r}}$ 
\citep[see, e.g.,][]{takeuchi2002,tanaka2002}, and is of order unity. 
Hereafter, $\eta$ is included in the ratio $H_{g}/a_{s}$.
For typical protoplanetary disks,
$(\Omega_{\mathrm{K}}-\Omega_{g})/\Omega_{\mathrm{K}}\approx%
(H_{g}/a_{s})^{2}/2$ is on the order of a percent or less
\citep[e.g.,][]{dalessio1998}.

The midplane (non-transient) velocity of a particle subject to drag
can be approximated to 
\citep[e.g.,][and references therein]{chiang2010}
\begin{eqnarray}
  \dot{a}_{s} &\approx&-\frac{a_{s}\Omega^{2}_{\mathrm{K}}(H_{g}/a_{s})^{2}\tau_{s}}%
                                {1+(\tau_{s}\Omega_{\mathrm{K}})^{2}} \label{eq:ur_s} \\
  \dot{\phi}_{s} &\approx&\Omega_{\mathrm{K}}-\left(\frac{1}{2}\right)%
                                \frac{\Omega_{\mathrm{K}}(H_{g}/a_{s})^{2}}%
                                {1+(\tau_{s}\Omega_{\mathrm{K}})^{2}},\label{eq:uphi_s}
\end{eqnarray}
$\phi_{s}$ being the azimuthal angle of the particle. Equations~(\ref{eq:ur_s})
and (\ref{eq:uphi_s}) assume a negligible radial component of $\gvec{u}_{g}$.
Quantity $\tau_{s}\Omega_{\mathrm{K}}$ is referred to as the Stokes number
of the particle.
For $\tau_{s}\Omega_{\mathrm{K}}\gg 1$, the orbital decay timescale is 
$\tau_{D}=a_{s}/|\dot{a}_{s}|\approx (a_{s}/H_{g})^{2}\tau_{s}$ ($\gg \tau_{s}$).
For $\tau_{s}\Omega_{\mathrm{K}}= 1$, the decay timescale attains its minimum, 
$\tau_{D}\approx 2(a_{s}/H_{g})^{2}/\Omega_{\mathrm{K}}$.
For $\tau_{s}\Omega_{\mathrm{K}}\ll 1$, the limit of a fully coupled solid,
$|\dot{a}_{s}|/a_{s}\approx \Omega^{2}_{\mathrm{K}}(H_{g}/a_{s})^{2}\tau_{s}$
and the radial displacement converges to that of the gas.

\begin{figure}[t]
\centering%
\resizebox{\linewidth}{!}{\includegraphics[clip]{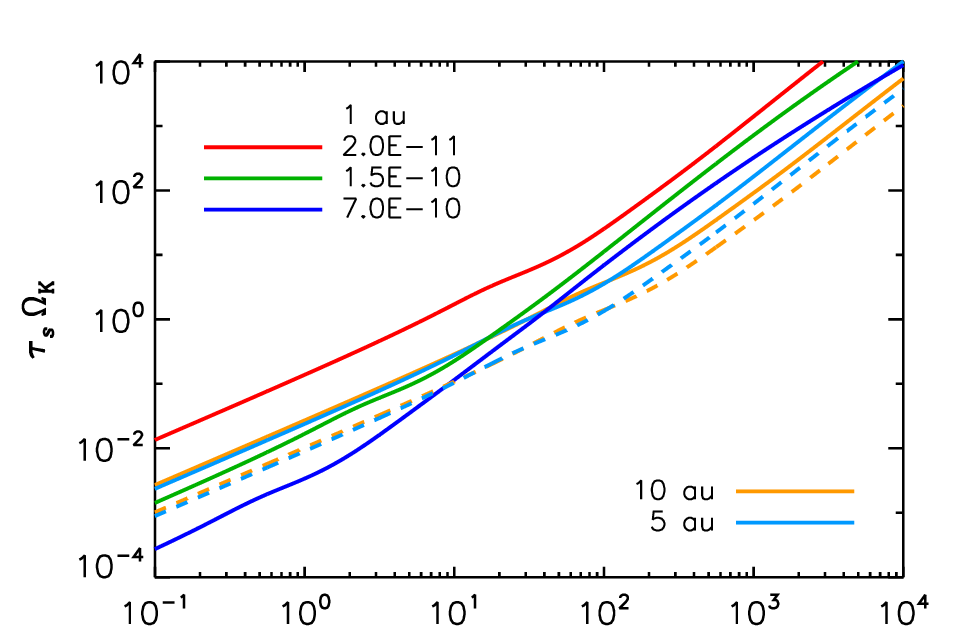}}
\resizebox{\linewidth}{!}{\includegraphics[clip]{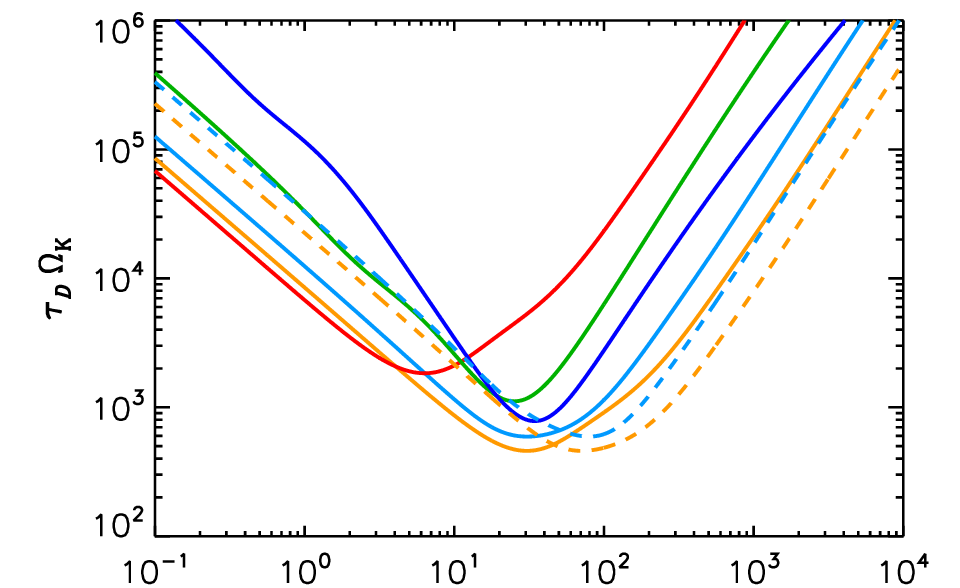}}
\resizebox{\linewidth}{!}{\includegraphics[clip]{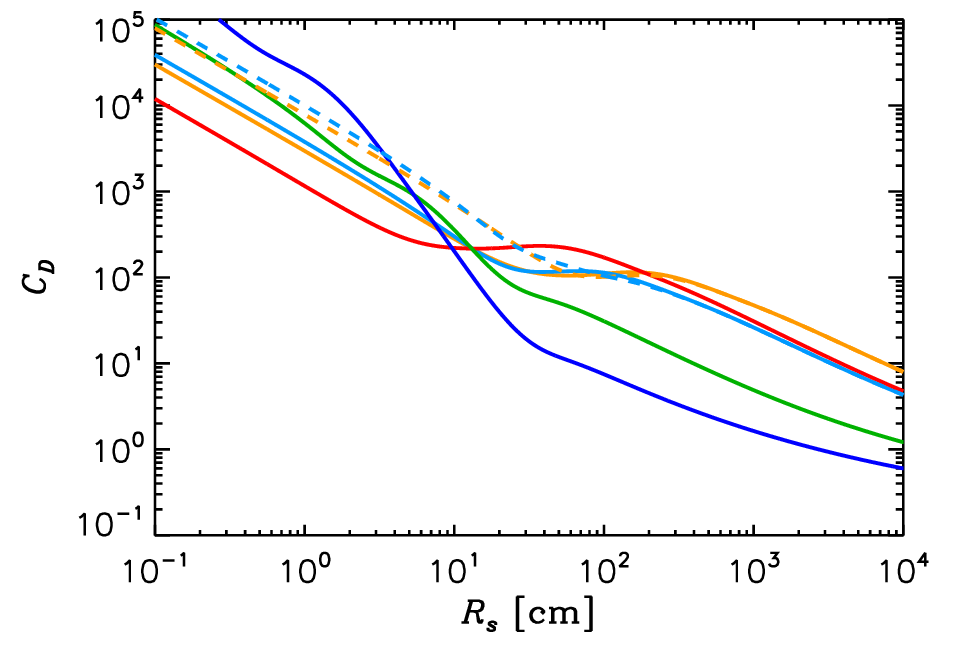}}
\caption{%
             Stopping time $\tau_{s}$ (top) and orbital decay time
             $\tau_{D}$ (middle) of rocky (solid lines) and icy
             (dashed lines) particles of radius $R_{s}$, orbiting 
             in a smooth and unperturbed protoplanetary disk. 
             The time $\tau_{s}$ is computed by inverting \cieq{eq:tau_s} 
             and $\tau_{D}=a_{s}/|\dot{a}_{s}|$ (see \cieq{eq:ur_s}).
             The drag coefficient $C_{D}$ \citepalias{gennaro2015} 
             is also shown (bottom).
             The different lines refer to different disk locations and/or gas
             densities (as indicated in $\rhou$) and temperatures
             (see also \citab{table:dat}).
             }
\label{fig:taud}
\end{figure}
Using the coefficient $C_{D}$ reported in
\citet[][hereafter, \citetalias{gennaro2015}]{gennaro2015}, 
which depends on $|\gvec{u}_{g}-\gvec{u}_{s}|$ and hence
on $\tau_{s}$, \cieq{eq:tau_s} can be inverted to provide
a function $\tau_{s}=\tau_{s}(R_{s})$, plotted in \cifig{fig:taud} 
(top panel).
Through \cieq{eq:ur_s}, one can then recover the timescale
for drag-induced orbital decay, $\tau_{D}$ (middle panel).
For reference, $C_{D}$ is plotted in the bottom panel.
The results in \cifig{fig:taud} are based on disk thermodynamic
conditions relevant to the radiation-hydrodynamics calculations
described in \cisec{sec:RHD}, also detailed in \citab{table:dat}.
Under such conditions, $\tau_{s}\Omega_{\mathrm{K}}\approx 1$ 
when $R_{s}$ is a few times $10$ to $\approx 100\,\mathrm{cm}$. 
Smaller particles may satisfy the condition 
$\tau_{s}\Omega_{\mathrm{K}}\approx 1$ in a low density disk.
The timescale $\tau_{D}$ has a minimum of a few hundred orbital 
periods or less, and diverges as $R_{s}\rightarrow 0$
($\tau_{s}\Omega_{\mathrm{K}}\rightarrow 0$) 
because the drift velocity
$|\dot{a}_{s}|\propto \Omega^{2}_{\mathrm{K}}\tau_{s} a_{s}$ 
approaches zero. In this limit, however, $\dot{a}_{s}$ converges 
to the radial velocity of the gas, $u^{r}_{g}$, which cannot be 
neglected. In such cases, a corrected version of \cieq{eq:ur_s}
is \citep[see, e.g.,][]{takeuchi2002}
\begin{equation}
  \dot{a}_{s} \approx\frac{u^{r}_{g}-a_{s}\Omega^{2}_{\mathrm{K}}(H_{g}/a_{s})^{2}\tau_{s}}%
                                {1+(\tau_{s}\Omega_{\mathrm{K}})^{2}}.
  \label{eq:ur_s_corr}
\end{equation}

In the classic theory of accretion disks \citep[e.g.,][]{lynden-bell1974,pringle1981},
\begin{equation}
  u^{r}_{g} =-\frac{3}{2}\frac{\nu_{g}}{r}\left[1+
                    2\frac{\partial\ln{(\nu_{g}\Sigma_{g})}}{\partial \ln{r}}\right],
  \label{eq:ugr}
\end{equation}
in which $\nu_{g}$ and $\Sigma_{g}$ are respectively the kinematic 
viscosity and surface density of the gas. 
Writing the kinematic viscosity as
$\nu_{g}=\alpha_{g}H^{2}_{g}\Omega_{\mathrm{K}}$ \citep{S&S1973}, 
if $\partial\ln{(\nu_{g}\Sigma_{g})}/{\partial \ln{r}}\ll 1$ then the bracket 
in \cieq{eq:ugr} is $\approx 1$, and the correction term 
in \cieq{eq:ur_s_corr} is negligible as long as
$\tau_{s}\Omega_{\mathrm{K}}\gg \alpha_{g}$.

\begin{figure}
\centering%
\resizebox{\linewidth}{!}{\includegraphics[clip]{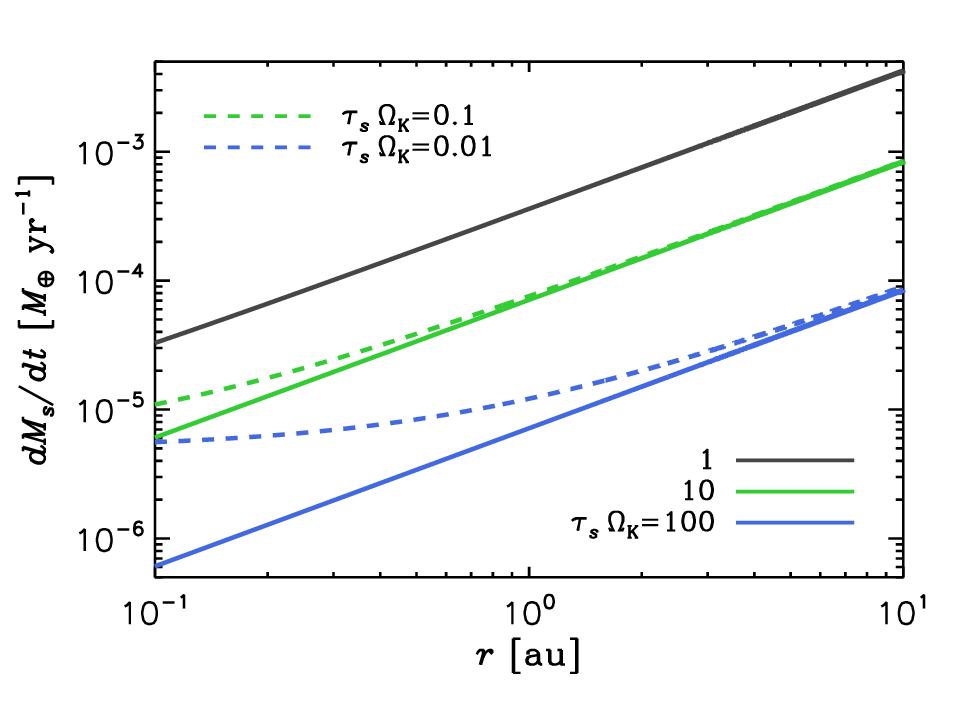}}
\caption{%
             Accretion rate of solids through a disk according to 
             \cieq{eq:dotM_s}, with $u^{r}_{g}\approx -(3/2)\nu_{g}/r$ 
             (see text), for particles of different stopping times, 
             $\tau_{s}$, as indicated. 
             The disk's gas scale-height is $H_{g}\propto r^{9/7}$
             \citep{chiang2010}. 
             A reference surface density of solids of $\Sigma_{s}=1\,\sigu$ 
             is applied throughout ($dM_{s}/dt\propto \Sigma_{s}$).
             }
\label{fig:dotM_s}
\end{figure}
Indicating with $\Sigma_{s}$ the solids' surface density, the mass accretion
rate of solids through an unperturbed disk is 
$dM_{s}/dt=-2\pi \Sigma_{s} a_{s} \dot{a}_{s}$.
Applying \cieq{eq:ur_s_corr}, the accretion rate becomes
\begin{equation}
  \frac{dM_{s}}{dt} 
  \approx 2\pi\!\left[\frac{(H_{g}/a_{s})^{2}\Omega_{\mathrm{K}}\tau_{s}
                           -u^{r}_{g}/(a_{s}\Omega_{\mathrm{K}})
                          }%
                          {
                      1+(\tau_{s}\Omega_{\mathrm{K}})^{2}}\right]\!%
                      a^{2}_{s}\Sigma_{s}\Omega_{\mathrm{K}
                          }.
  \label{eq:dotM_s}
\end{equation}
Neglecting the term in $u^{r}_{g}$, maximum accretion 
occurs for $\tau_{s}\Omega_{\mathrm{K}}\approx 1$, when 
$dM_{s}/dt\approx\pi a^{2}_{s}\Sigma_{s}\Omega_{\mathrm{K}}(H_{g}/a_{s})^{2}$.
At this rate, a mass of solids equal to $\pi a^{2}_{s}\Sigma_{s}$ 
would be transported through the radius $r=a_{s}$ in a timescale
$(a_{s}/H_{g})^{2}/\Omega_{\mathrm{K}}$, 
or $\sim 10^{3}$ years at $r=10\,\AU$. 
In the decoupled regime, $\tau_{s}\Omega_{\mathrm{K}}\gg 1$,
$dM_{s}/dt\approx%
2\pi a^{2}_{s}\Sigma_{s}\Omega_{\mathrm{K}}(H_{g}/a_{s})^{2}/(\tau_{s}\Omega_{\mathrm{K}})$
whereas in the opposite limit, $\tau_{s}\Omega_{\mathrm{K}}\ll 1$, 
$dM_{s}/dt\approx%
 2\pi a^{2}_{s}\Sigma_{s}\Omega_{\mathrm{K}}(H_{g}/a_{s})^{2}\tau_{s}\Omega_{\mathrm{K}}$.
In the well-coupled regime, if 
$\tau_{s}\Omega_{\mathrm{K}}\lesssim \alpha_{g}$,
the term in $u^{r}_{g}$ cannot be neglected and \cieq{eq:dotM_s} implies 
possible stalling or outward transport of solids in expanding disk regions
\citep[where $u^{r}_{g}>0$,][]{lynden-bell1974} or in transition regions 
with steep density gradients (see \cieq{eq:ugr}).

\cieq{eq:dotM_s} is plotted in \cifig{fig:dotM_s}, for a radially constant 
value of $\Sigma_{s}=1\,\sigu$ and $u^{r}_{g}\approx -(3/2) \nu_{g}/r$, in
which $\nu_{g}$ is constant and corresponds to $\alpha_{g}\approx 0.005$ 
at $1\,\AU$ ($H_{g}\propto r^{9/7}$). 
The effect of $u^{r}_{g}$ can be seen when 
$\tau_{s}\Omega_{\mathrm{K}} \lesssim \alpha_{g}$
($\alpha_{g}$ is roughly proportional to $1/r$).
The curves in the figure should only be interpreted as 
the value at $r$ for the imposed $\Sigma_{s}$. 
In fact, if $\partial \dot{M}_{s}/\partial r\neq 0$, mass conservation
requires that
\begin{equation}
  \frac{\partial\Sigma_{s}}{\partial t}=%
  \frac{1}{2\pi r}\frac{\partial\dot{M}_{s}}{\partial r}+
  \Lambda(t,r),
  \label{eq:sigs}
\end{equation}
where $\Lambda$ includes source/sink terms to account for, e.g., 
the ongoing coagulation of solids from dust or fragmentation and erosion.

\begin{figure}
\centering%
\resizebox{\linewidth}{!}{\includegraphics[clip]{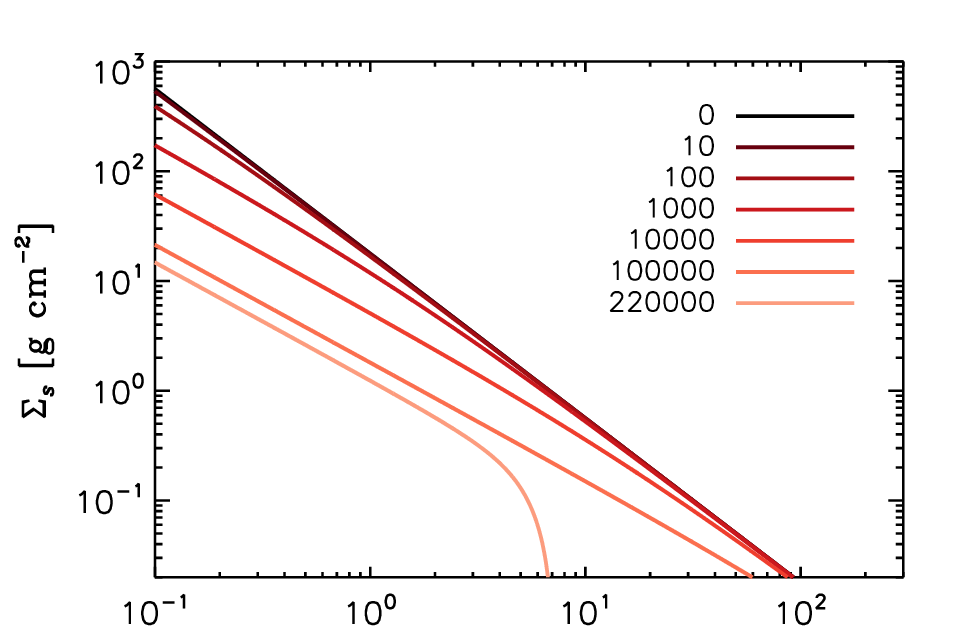}}
\resizebox{\linewidth}{!}{\includegraphics[clip]{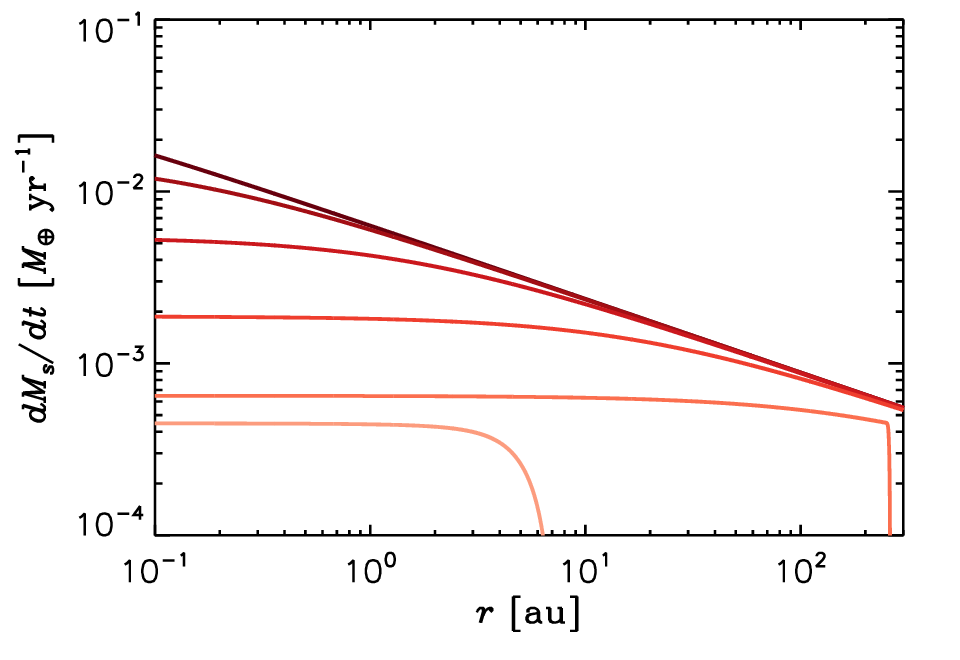}}
\caption{%
             Top: Solution of \cieq{eq:sigs} by imposing
             $\tau_{s}\Omega_{\mathrm{K}}= 1$ in \cieq{eq:dotM_s} and 
             the initial gaseous nebula model of \citet{chiang2010} 
             (see the text for details). The times in the legend are 
             in years. 
             The disk's gas scale-height $H_{g}$ is constant in time. 
             Bottom: Accretion rate of solids corresponding to 
             the evolution of $\Sigma_{s}$ in the top panel.
             }
\label{fig:sigs_ts}
\end{figure}
A solution of \cieq{eq:sigs} is plotted in \cifig{fig:sigs_ts}, requiring
that $\tau_{s}\Omega_{\mathrm{K}}= 1$ in \cieq{eq:dotM_s}.
The initial conditions are based on the nebula model of \citet{chiang2010},
with a total gas mass of $0.055\,\Msun$ 
($\Sigma_{g}\propto r^{-3/2}$ and gas temperature $T_{g}\propto r^{-3/7}$) 
and a total solids' mass of $\approx 183\,\Mearth$ (for an initial gas-to-solids
mass ratio of $100$). 
The conversion of dust into the larger particles is assumed to have occurred 
at time $t=0$, with no further growth or fragmentation (i.e., $\Lambda=0$).
Over the first $\approx 10^{5}$ years, the disk loses $\approx 120\,\Mearth$ 
of solids to the star, with the remainder removed during the next 
$\approx 1.2\times 10^{5}$ years. 
For comparison, the condition $\tau_{s}\Omega_{\mathrm{K}}= 0.1$ removes
$\approx 90\,\Mearth$ in $\approx 3\times 10^{5}$ years and the entire
reservoir of solids within $\approx 10^{6}$ years, whereas
the condition $\tau_{s}\Omega_{\mathrm{K}}= 0.01$ causes a depletion of 
$\approx 50\,\Mearth$ in $\approx 10^{6}$ years.  
Clearly, these numbers are determined by the initial mass and distribution 
of solids.
Provided that $H_{g}$ remains roughly constant in time, 
the condition $\tau_{s}\Omega_{\mathrm{K}}=\mathrm{constant}$ 
allows one to ignore the disk's gas evolution. 
However, due to the changing thermodynamic properties of the gas, 
this condition does not characterize any single solid's size,
but it rather corresponds to a changing $R_{s}$, in space and time.

\begin{figure*}
\centering%
\resizebox{\linewidth}{!}{\includegraphics[clip]{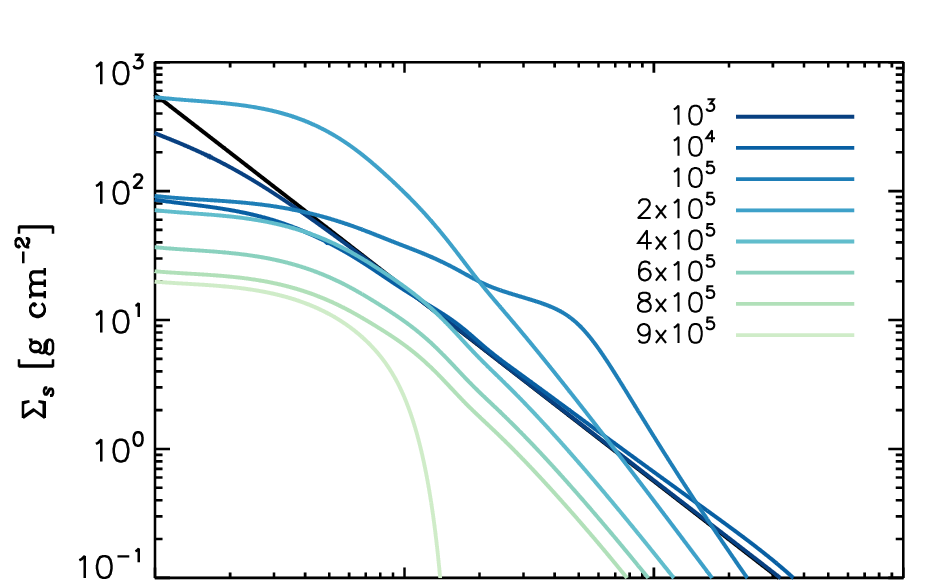}%
                                      \includegraphics[clip]{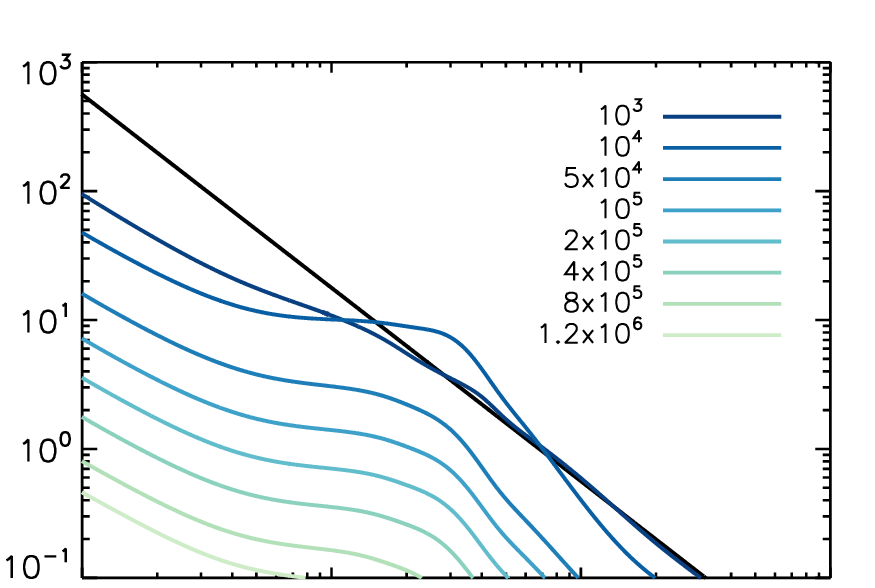}%
                                      \includegraphics[clip]{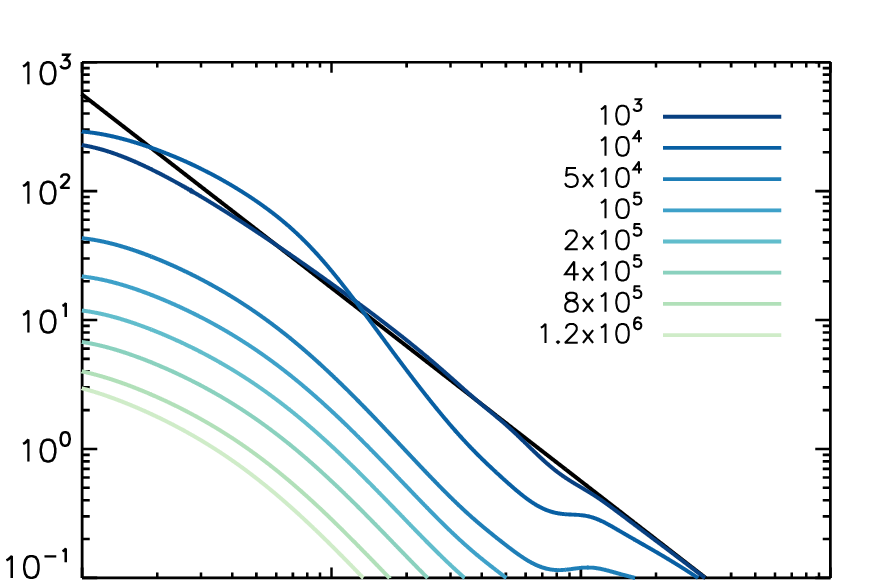}}
\resizebox{\linewidth}{!}{\includegraphics[clip]{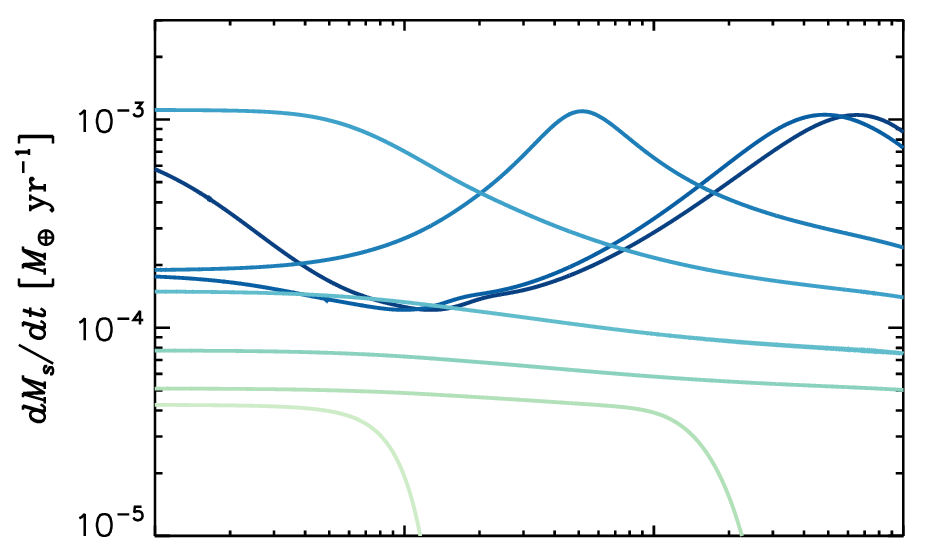}%
                                      \includegraphics[clip]{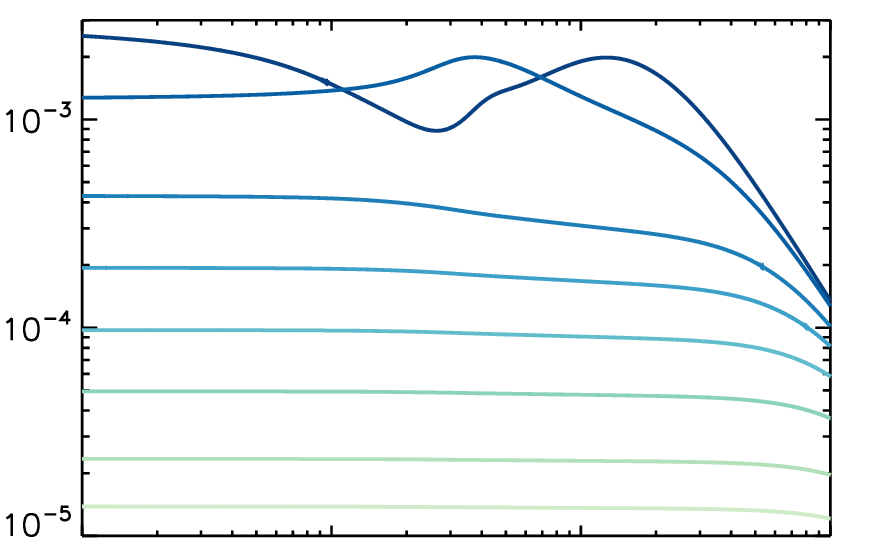}%
                                      \includegraphics[clip]{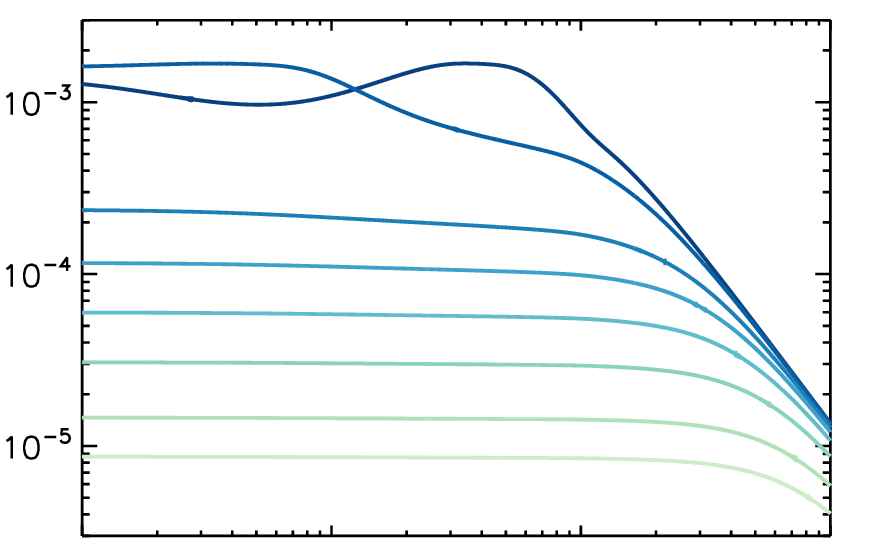}}
\resizebox{\linewidth}{!}{\includegraphics[clip]{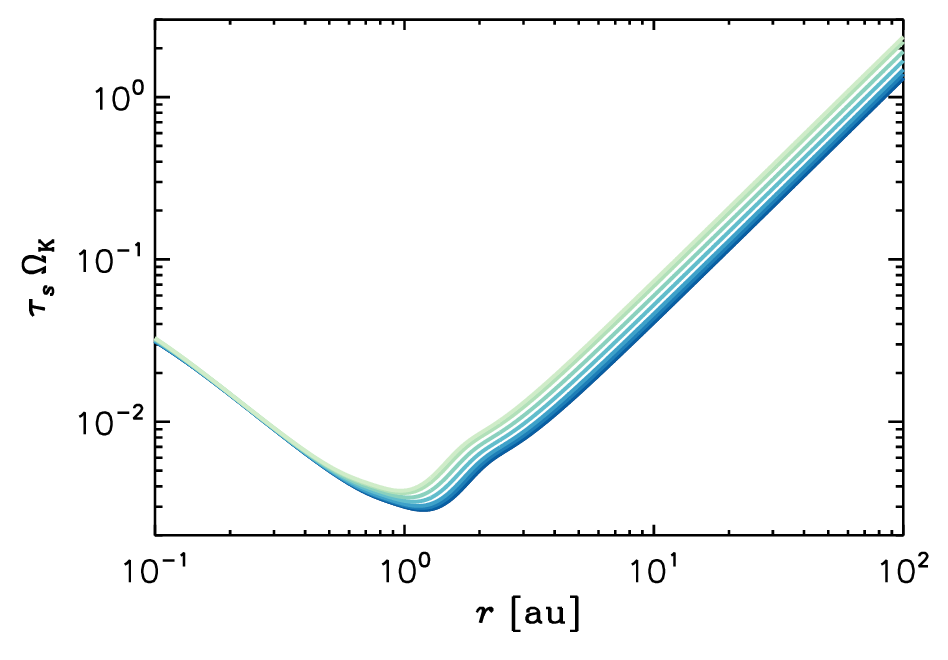}%
                                      \includegraphics[clip]{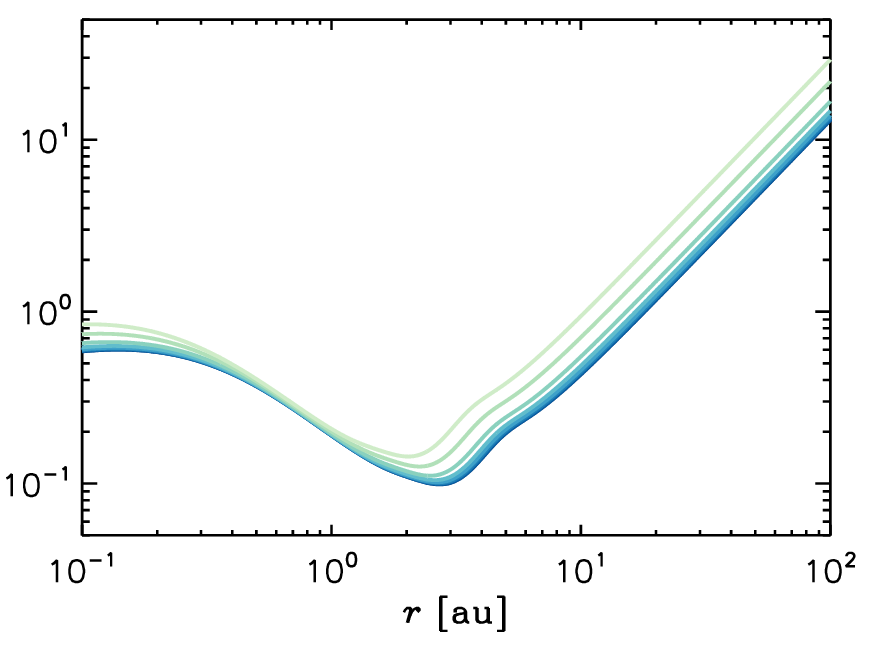}%
                                      \includegraphics[clip]{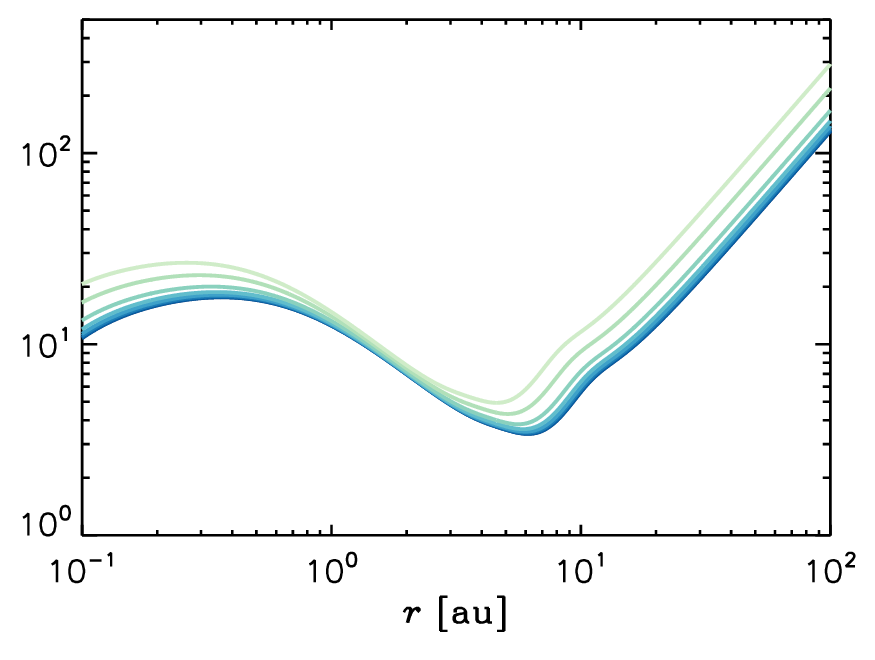}}
\caption{%
         Solutions of \cieq{eq:sigs} for rocky solids of radius
         $R_{s}$, equal to $1$ (left), $10$ (center), and
         $100\,\mathrm{cm}$ (right). 
         From top to bottom, the panels show $\Sigma_{s}$, $dM_{s}/dt$, and
         $\tau_{s}\Omega_{\mathrm{K}}$. 
         The times in the legends are in years. 
         The black curves in the top panels represent the initial
         distribution of solids. See the text for further details.
        }
\label{fig:sigs_rs}
\end{figure*}
The evolution of solids of a particular size can be obtained by solving 
\cieq{eq:sigs} for a fixed $R_{s}$. In this case, $\tau_{s}\Omega_{\mathrm{K}}$
is a function of both $t$ and $r$, and \cieq{eq:tau_s} must be inverted at 
any time and distance in order to self-consistently determine 
$\dot{a}_{s}$ and $\tau_{s}$.
The evolution of the disk's gas cannot be neglected in this case.
Thus, it is assumed that gas is depleted via accretion on the star 
and through photo-evaporation by hard radiation from the star. 
The gas kinematic viscosity is taken as 
$\nu_{g}\propto 1/\Sigma_{g}$ ($\alpha_{g}\approx 10^{-3}$ 
at $1\,\AU$), so that $d(\nu_{g}\Sigma_{g})/dr=0$ and
\cieq{eq:ugr} reduces to $u^{r}_{g}=-(3/2) \nu_{g}/r$. 
For simplicity, the time evolution of $\Sigma_{g}$ is obtained by
rescaling the initial condition (the same as in \cifig{fig:sigs_ts})
according to the current gas mass ($\partial\ln{\Sigma_{g}}/\partial\ln{r}$ 
is constant in time).
The disk's gas is entirely depleted in $\approx 3.5\,\Myr$
\citep[e.g.,][]{weiss2021}.

\cifig{fig:sigs_rs} illustrates the evolution for rocky solids of $1$ 
(left), $10$ (center), and $100\,\mathrm{cm}$ (right) radius. 
The initial inventory of solids, $\approx 183\,\Mearth$,
is entirely removed within $\approx 9\times 10^{5}$ years 
for the case shown in the left panels. 
The reason is that maximal accretion ($\tau_{s}\Omega_{\mathrm{K}}\approx 1$)
is initially sustained at distances of several tens of astronomical units, rapidly 
transferring large amounts of solids toward the star (see the top left 
and middle left panels). Consequently, two-thirds of the initial mass 
is removed in $\approx 3\times 10^{5}$ years. 
For particles with $R_{s}=10$ and $100\,\mathrm{cm}$, respectively,
$\approx 60$ and $\approx 110\,\Mearth$ are left at $1\,\Myr$, but 
$\lesssim 1\,\Mearth$ remains inside $20\,\AU$. 
At $t\approx 5\times 10^{5}$ years, $dM_{s}/dt$ inside $100\,\AU$ 
ranges from a few to several $10^{-5}\,\mathrm{\Mearth\,yr^{-1}}$, 
dropping below $10^{-5}\,\mathrm{\Mearth\,yr^{-1}}$ after $1\,\Myr$. 

An experiment conducted with $R_{s}=10\,\mathrm{m}$ rocks indicates 
that over $75$\% of the initial solids' mass survives after $3\,\Myr$
(because $\tau_{s}\Omega_{\mathrm{K}} \gg 1$), but only an amount of
$\approx 5\,\Mearth$ orbits within $10\,\AU$, where 
$dM_{s}/dt\lesssim 10^{-5}\,\mathrm{\Mearth\,yr^{-1}}$ 
after $10^{5}$ years. Beyond $20\,\AU$, $dM_{s}/dt$ never exceeds 
$\approx 10^{-5}\,\mathrm{\Mearth\,yr^{-1}}$ during the disk's 
evolution.

The bottom panels of \cifig{fig:sigs_rs} indicate that
for $0.1< r < 100\,\AU$, 
$\tau_{s}\Omega_{\mathrm{K}}$ can vary by up to three orders of magnitude 
for $R_{s}=1$ and $10\,\mathrm{cm}$ particles and by over an order of 
magnitude for $R_{s}=1\,\mathrm{m}$ (and $10\,\mathrm{m}$) bodies. 
The largest particles also display a significant increase 
of $\tau_{s}$, at any given $r$, as the disk's gas dissipates.
Clearly, the condition $\tau_{s}\Omega_{\mathrm{K}}=\mathrm{constant}$ 
is only indicative of a particle size for a given gas state. 
In these examples,
$\tau_{s}\Omega_{\mathrm{K}}= 1$ corresponds to $R_{s}\approx 1\,\mathrm{cm}$
particles between $r\approx 50$ and $\approx 80\,\AU$, and to 
$R_{s}\approx 10\,\mathrm{cm}$ particles between $r\approx 10$ 
and $\approx 20\,\AU$. 
Larger particles (in the $\approx 10$--$100\,\mathrm{cm}$ range) 
would satisfy this condition inside $10\,\AU$.

\begin{figure}
\centering%
\resizebox{\linewidth}{!}{\includegraphics[clip]{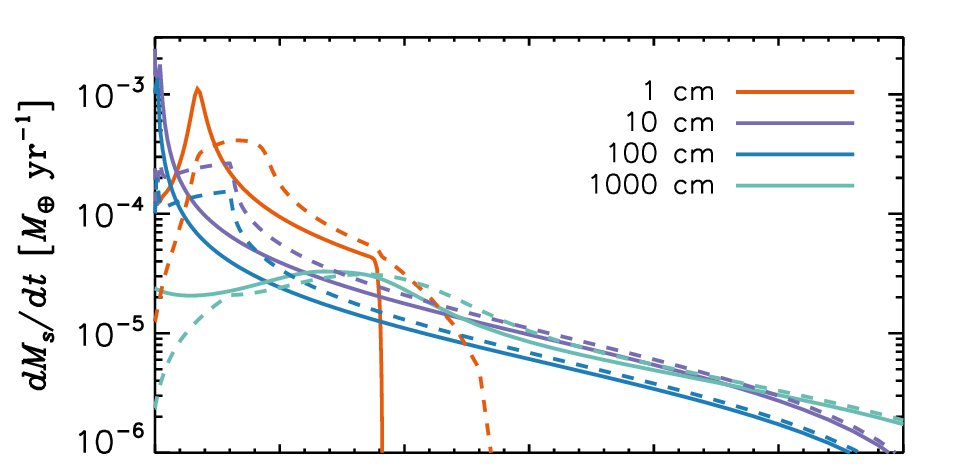}}
\resizebox{\linewidth}{!}{\includegraphics[clip]{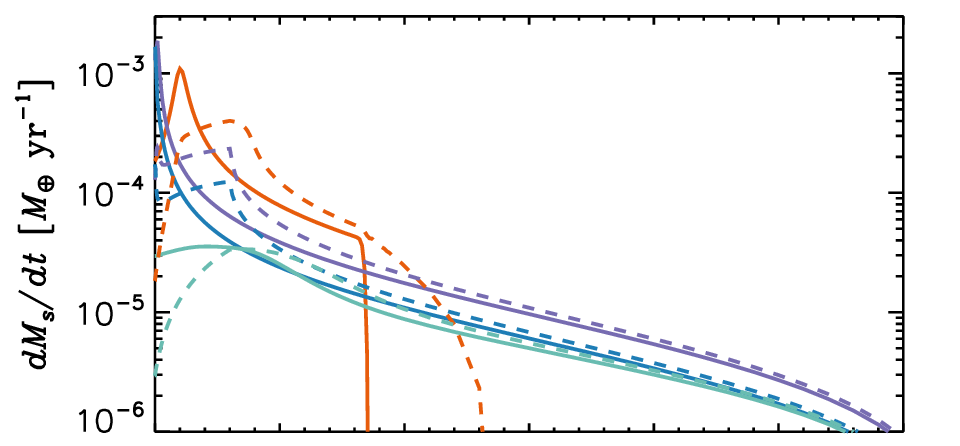}}
\resizebox{\linewidth}{!}{\includegraphics[clip]{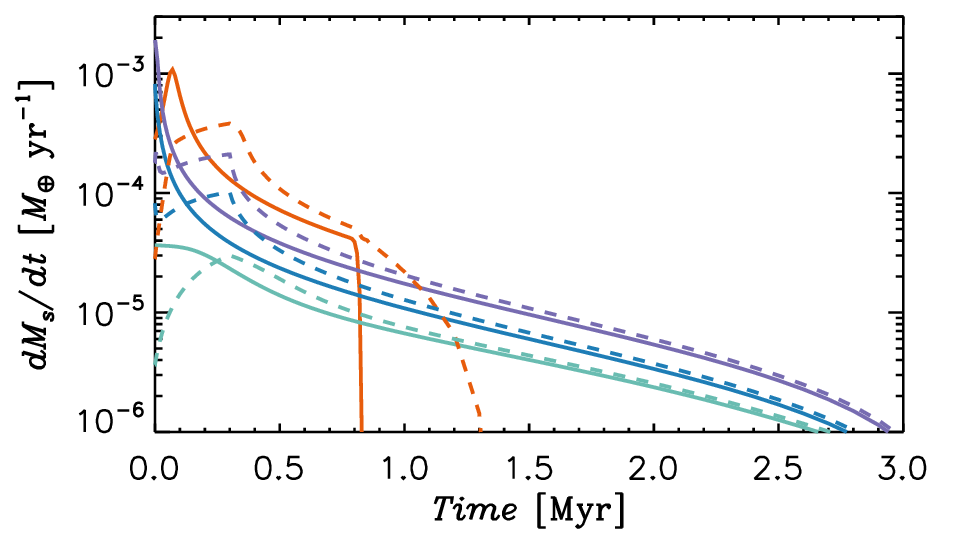}}
\caption{%
             Solutions of \cieq{eq:sigs} for solids of radius $R_{s}$, 
             as indicated, for the accretion rate through
             the disk, $dM_{s}/dt$. The panels illustrate the evolution 
             at distance $r=1$ (top), $5$ (middle), and $10\,\AU$
             (bottom).
             The solid lines represent results from the disk model in
             \cifig{fig:sigs_rs}.
             The dashed lines indicate results from similar models with 
             a source term, $\Lambda$ (see the text for details).
             }
\label{fig:dMsdt_rs}
\end{figure}
\cifig{fig:dMsdt_rs} shows $dM_{s}/dt$ versus time at $r=1$ (top), $5$ (middle), 
and $10\,\AU$ (bottom), for the calculations presented in \cifig{fig:sigs_rs} 
(solid lines).
For the first few $10^{5}$ years, $dM_{s}/dt$ can be
$\gtrsim 10^{-4}\,\mathrm{\Mearth\,yr^{-1}}$, but it becomes 
$\lesssim 10^{-5}\,\mathrm{\Mearth\,yr^{-1}}$ by $\approx 1\,\Myr$.
The same calculations were executed applying the source term
$\Lambda(t,r)=0.9\,\Sigma_{s}(0,r)/\tau_{c}$ in \cieq{eq:sigs}, with
a ``coagulation time'' $\tau_{c}=3\times 10^{5}\,\mathrm{yr}$
\citep{voelkel2020} and $\Lambda=0$ for $t>\tau_{c}$. 
At time $t=0$, $10$\% of the initial solids' mass is already in the form 
of particles of radius $R_{s}$. These latter results are illustrated in
\cifig{fig:dMsdt_rs} as dashed lines. Compared to the solid curves, 
peaks in $dM_{s}/dt$ are lower and broader, but differences become 
small after $\approx 0.5\,\Myr$.
The results presented in Figures~\ref{fig:sigs_ts}-\ref{fig:dMsdt_rs} 
are approximate but representative of the evolution of the solids' 
density and accretion rates in an unperturbed disk.

\subsection{Dynamics in Perturbed Disks}
\label{sec:DPK}

In the presence of a perturbing body, e.g., a planet, \cieq{eq:dotM_s} 
can be interpreted as the rate at which solids can be supplied 
to the proximity of its orbit, thus determining the maximum
rate of the solids' accretion on the planet.
Assuming that the thickness of the disk of solids, $H_{s}$, is 
infinitesimal, then the accretion rate on a small planet is essentially
two-dimensional and can be approximated by
\begin{equation}
 \frac{d\Mp}{dt}\approx\frac{1}{\pi}\left(\frac{R_{\mathrm{eff}}}{a_{p}}\right)\frac{dM_{s}}{dt},
 \label{eq:dotMp}
\end{equation}
in which $\Mp$ and $a_{p}$ are the planet mass and orbital radius,
respectively. 
This expression assumes that the mass flux of solids in the radial
direction can be approximated as azimuth-independent, so  
the fraction of the integrated flux intercepted by the planet is given 
by the ratio of the planet's ``effective'' size to the orbit circumference. 
This approximation is useful when working in one-dimensional, radial 
disks. 
Corrections due to azimuthal variations of the solids' transport can 
be accounted for in the definition of the planet's effective radius 
for accretion. In fact, the radius $R_{\mathrm{eff}}$ can be much 
larger than the planet's physical radius, $R_{p}$ 
(the two-dimensional assumption implies that $R_{\mathrm{eff}}> H_{s}$).

Applying a formalism developed for the accretion of planetesimals 
in the two-body problem framework \citep[see, e.g.,][]{lissauer1987},
the effective radius for the capture of solids can be written as
$R_{\mathrm{eff}}\approx R_{p} (u_{\mathrm{esc}}/u_{\mathrm{rel}})$,
if $u_{\mathrm{esc}}\gg u_{\mathrm{rel}}$ (otherwise
$R_{\mathrm{eff}}\approx R_{p}$), 
where $u_{\mathrm{esc}}=\sqrt{2G\Mp/R_{p}}$ is the escape velocity 
from the planet surface and $u_{\mathrm{rel}}$ is the relative velocity 
(in magnitude) between the planet and the solids.
This relation originates from the conservation of relative energy and angular
momentum of the approaching particle.
If $\tau_{s}\Omega_{\mathrm{K}}\gg \alpha_{g}$, the relative velocity
can be obtained directly from Equations~(\ref{eq:ur_s}) and (\ref{eq:uphi_s}),
\begin{equation}
u_{\mathrm{rel}}\approx\left(\frac{H_{g}}{a_{p}}\right)^{2}%
                          \frac{\sqrt{1/4+(\tau_{s}\Omega_{\mathrm{K}})^{2}}}{%
                          1+(\tau_{s}\Omega_{\mathrm{K}})^{2}}\Omega_{\mathrm{K}}a_{p},
 \label{eq:urel}
\end{equation}
assuming that the planet moves on a circular orbit.

\vspace*{2pt}
The efficiency, or probability, of the accretion of solids onto a planet is defined as
\begin{equation}
  \zeta=\frac{\dMp}{\dot{M}_{s}},
 \label{eq:zetadef}
\end{equation}
where $\dot{M}_{s}$ is the accretion rate of solids toward the planet,
in the proximity of its orbit (and $\dMp\le\dot{M}_{s}$ by assumption).
Inserting the effective radius given above in \cieq{eq:dotMp}, we find  
\begin{equation}
  \zeta\approx \frac{\sqrt{2}}{\pi}\!\left(\frac{a_{p}}{H_{g}}\right)^{2}\!\!%
                       \frac{1+(\tau_{s}\Omega_{\mathrm{K}})^{2}}%
                              {\sqrt{1/4+(\tau_{s}\Omega_{\mathrm{K}})^{2}}}%
                              \sqrt{\!\left(\frac{\Mp}{\Ms}\right)\!\!\left(\frac{R_{p}}{a_{p}}\right)},
 \label{eq:zeta}
\end{equation}
in which $\Ms$ is the stellar mass.
For $\tau_{s}\Omega_{\mathrm{K}}\gtrsim1$, \cieq{eq:zeta} scales with planet mass 
and stopping time as the accretion probability estimated by \citet{kary1993} 
in the ``non-gravitating'' planet limit, 
$\zeta\propto \tau_{s}\Omega_{\mathrm{K}}\Mp^{2/3}%
\propto \tau_{s}\Omega_{\mathrm{K}}(\Rhill/a_{p})^{2}$,
where $\Rhill$ is the Hill radius of the planet.
Maximum values of the large-scale transport of solids, i.e., 
$dM_{s}/dt$ in \cieq{eq:dotM_s}, 
occur when $\tau_{s}\Omega_{\mathrm{K}}\approx 1$, at which 
$\zeta\approx (4/\pi) (a_{p}/H_{g})^{2}\sqrt{(2/5)(\Mp/\Ms)(R_{p}/a_{p})}$.
For a Moon-sized planetary embryo orbiting at $5\,\AU$, the efficiency would be 
$\approx 10^{-4}$.
Therefore, the values of $dM_{s}/dt$ reported in \cifig{fig:dMsdt_rs} would result 
in $d\Mp/dt < 10^{-7}\,\Mearth\,\mathrm{yr^{-1}}$, and in a mass-doubling 
timescale exceeding $10^{5}$ years.

\vspace*{2pt}
Alternatively, the effective radius for the capture of solids can be
estimated by equating the gravitational and relative kinetic energies 
\citep{lambrechts2012}, a condition required for escape, resulting in 
$R_{\mathrm{eff}}\approx 2G \Mp/u^{2}_{\mathrm{rel}}$. 
By using \cieq{eq:urel}, this approximation yields an efficiency
\begin{equation}
  \zeta\approx \frac{2}{\pi}\!\left(\frac{a_{p}}{H_{g}}\right)^{4}%
                       \frac{\left[1+(\tau_{s}\Omega_{\mathrm{K}})^{2}\right]^{2}}%
                              {1/4+(\tau_{s}\Omega_{\mathrm{K}})^{2}}%
                              \left(\frac{\Mp}{\Ms}\right).
 \label{eq:zeta_lj}
\end{equation}
In this case, a Moon-sized embryo orbiting at $5\,\AU$ would have 
$\zeta\approx 10^{-2}$ 
at maximum values of $dM_{s}/dt$ ($\tau_{s}\Omega_{\mathrm{K}}\approx 1$),
hence $d\Mp/dt \lesssim 10^{-5}\,\Mearth\,\mathrm{yr^{-1}}$,
much larger than the previous assessment.
However, both approximations for the effective radius neglect stellar gravity,
implying that in neither case can $R_{\mathrm{eff}}$ exceed $\approx \Rhill$,
hence $\zeta\lesssim (\Rhill/a_{p})/\pi$.
The accretion efficiency of a Moon-mass body would therefore be limited to
$\approx 10^{-3}$.
The remainder of the difference between the estimates provided by Equations~(\ref{eq:zeta})
and (\ref{eq:zeta_lj}) can be resolved by applying a more appropriate evaluation 
of $u_{\mathrm{rel}}$ close to the planet, as explained below.

If $R_{\mathrm{eff}}\lesssim H_{s}$, the two-dimensional approximation 
for the accretion of solids breaks down and the right-hand side of 
\cieq{eq:dotMp}, hence the efficiency $\zeta$ (see \cieq{eq:zetadef}), 
must be multiplied by $(\pi/4)(R_{\mathrm{eff}}/H_{s})$ to account for 
the vertical distribution of the solids.
In the small-particle regime, the thickness $H_{s}$ can be affected by 
turbulent stirring. 
If the kinematic viscosity $\nu_{g}$ is written in terms of $\alpha_{g}$,
then \citep[][]{dubrulle1995}
\begin{equation}
 \frac{H_{s}}{H_{g}}\approx\sqrt{\frac{\alpha_{g}}{\alpha_{g}+2\tau_{s}\Omega_{\mathrm{K}}}},
 \label{eq:Hs}
\end{equation}
where the numerical factor depends on the nature of the turbulence in the gas 
\citep[and may be somewhat different from $2$, see the discussion in][]{dubrulle1995}.
Typical values of the turbulence parameter $\alpha_{g}\sim 10^{-4}$--$10^{-2}$
would result in $H_{s}/H_{g}\sim 0.01$--$0.1$ for
$\tau_{s}\Omega_{\mathrm{K}}\approx 1$ 
\citep[see also][]{lambrechts2012}. Since the gas temperatures in planet-forming 
regions correspond to $H_{g}/a_{p}\sim 0.01$--$0.1$ \citep[e.g.,][]{dalessio1998}, 
the relative thickness of the solid layer would be 
$H_{s}/a_{p}\sim 10^{-4}$--$10^{-2}$. 
Thus, in the two-dimensional accretion regime (i.e., $R_{\mathrm{eff}}> H_{s}$),
$\zeta > (1/\pi) H_{s}/a_{p}$, and thus
$\gtrsim 10^{-5}$--$10^{-3}$, for $\tau_{s}\Omega_{\mathrm{K}}\approx 1$.

\begin{figure}
\centering%
\resizebox{\linewidth}{!}{\includegraphics[clip]{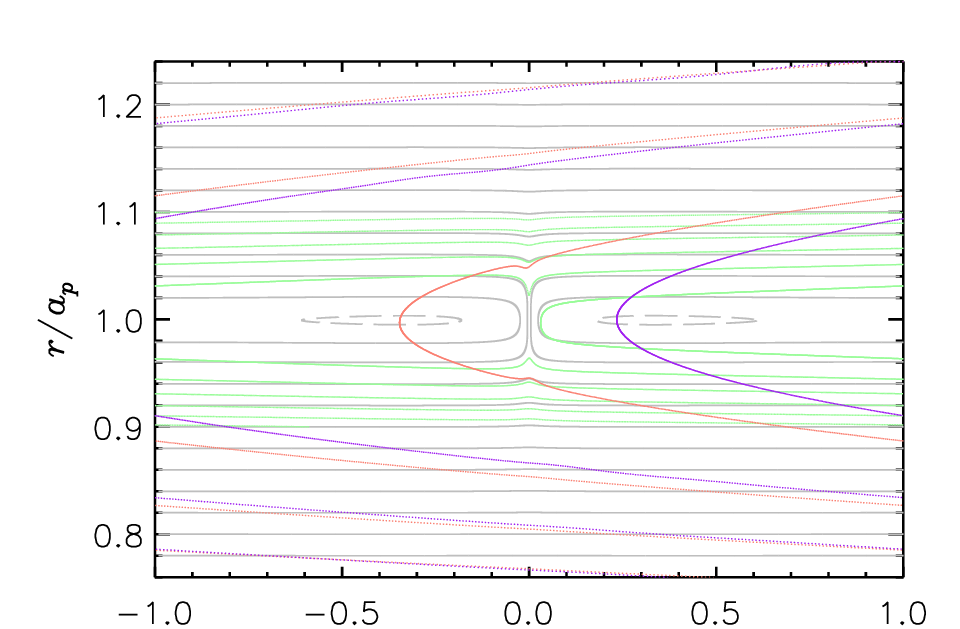}}
\resizebox{\linewidth}{!}{\includegraphics[clip]{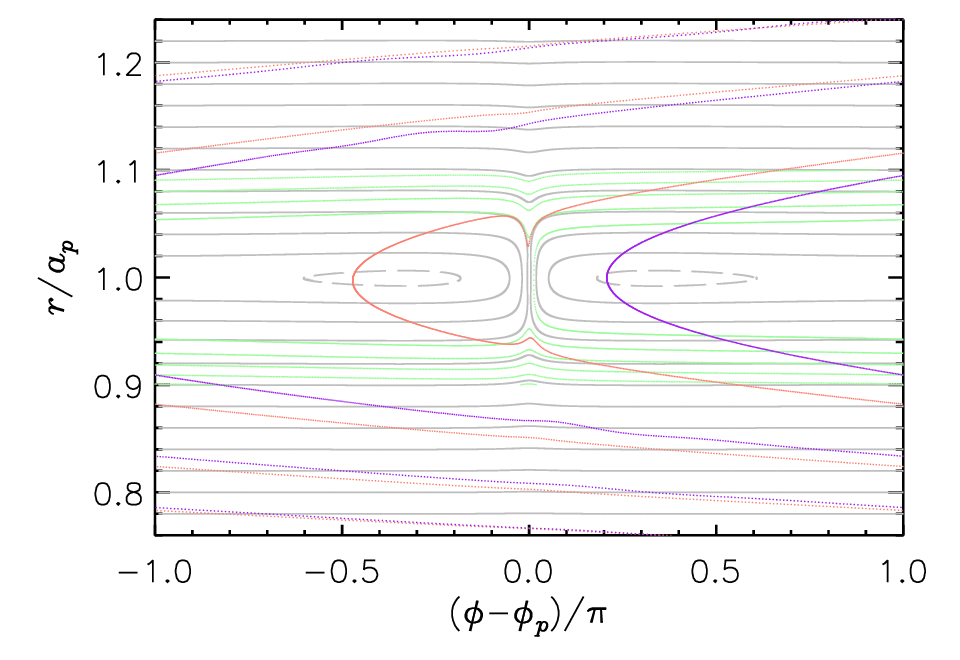}}
\caption{%
             Trajectories in the rotating frame of the planet of 
             $R_{s}=1$ (green), $10$ (orange), and $100\,\mathrm{cm}$ 
             (purple) particles, orbiting in a disk perturbed 
             by a $5$ (top) and a $15\,\Mearth$ (bottom) 
             planet on circular orbits. 
             The trajectory of the smallest particle is initiated 
             as the others, but a smaller portion is plotted.
             The gray lines represent streamlines of the perturbed gas. 
             The unperturbed disk properties are as in \cifig{fig:taud}
             at $5\,\AU$ (see also \citab{table:dat}).
             }
\label{fig:ff}
\end{figure}
Equations~(\ref{eq:zeta}) and (\ref{eq:zeta_lj}) neglect the gravitational
perturbations exerted by the planetary body, which alter the dynamics
of both solids and gas, i.e., 
$u_{\mathrm{rel}}$ is different from the unperturbed expression derived from 
Equations~(\ref{eq:ur_s}) and (\ref{eq:uphi_s}). 
An approximation to the perturbed rotation rate of the gas is 
still given by \cieq{eq:OPHI}, in which $\Phi$ must comprise the contributions 
due to the planet, including non-inertial terms (if applicable), i.e.,
 \begin{equation}
 \Phi=- \frac{G \Ms}{r}- \frac{G \Mp}{|\gvec{r}-\gvec{r}_{p}|}%
         + \frac{G \Mp}{r_{p}^{3}}%
          \,\gvec{r}\!\boldsymbol{\cdot}\!\gvec{r}_{p},
 \label{eq:phi}
\end{equation}
where $\gvec{r}$ and $\gvec{r}_{p}$ are the generic and planet position 
vectors. Neglecting contributions from gas pressure variations along
the azimuthal direction around the star, the perturbed radial component 
of the gas velocity is \citep{ogilvie2006}
\begin{equation}
 u^{r}_{g}\approx -\left(\frac{2}{r\Omega_{\mathrm{K}}}\right)%
                             \frac{\partial\Phi}{\partial\phi},
 \label{eq:ugr_ol}
\end{equation}
in which $\phi$ is the azimuthal angle.
The radial velocity arising from global transport, \cieq{eq:ugr},
should be added to the right-hand side of \cieq{eq:ugr_ol}, if relevant.
Given the importance of gas drag, corrections to the gas flow 
introduced by Equations~(\ref{eq:OPHI}), (\ref{eq:phi}) and
(\ref{eq:ugr_ol}) can have non-trivial effects on the dynamics 
of small solids.

The perturbed gas velocity can be used to integrate the trajectories 
of solid particles through the disk 
\citepalias[see, e.g.,][]{gennaro2015}. 
Some examples are shown in \cifig{fig:ff}, in a frame co-rotating 
with the planet, for $R_{s}=1$ (green line), $10$ (orange line),
and $100\,\mathrm{cm}$ (purple line) rocky particle drifting
through the gas perturbed by $5$ (top) and $15\,\Mearth$ (bottom)
planets on circular orbits. 
The gray lines represent perturbed gas streamlines. The particle 
motion is altered by the planet's gravity and by drag forces due 
to the perturbed gas velocity field, invalidating the approximations 
applied in deriving Equations~(\ref{eq:ur_s}) and (\ref{eq:uphi_s}).
Those equations remain valid only sufficiently far from the orbit 
of the planet.
Note that this approach does not account for planet-induced
tidal perturbations on the gas density, and hence pressure, 
which can also affect particle dynamics, as discussed later,
and are expected to become increasingly important as $\Mp$ grows. 

\begin{figure}
\centering%
\resizebox{\linewidth}{!}{\includegraphics[clip]{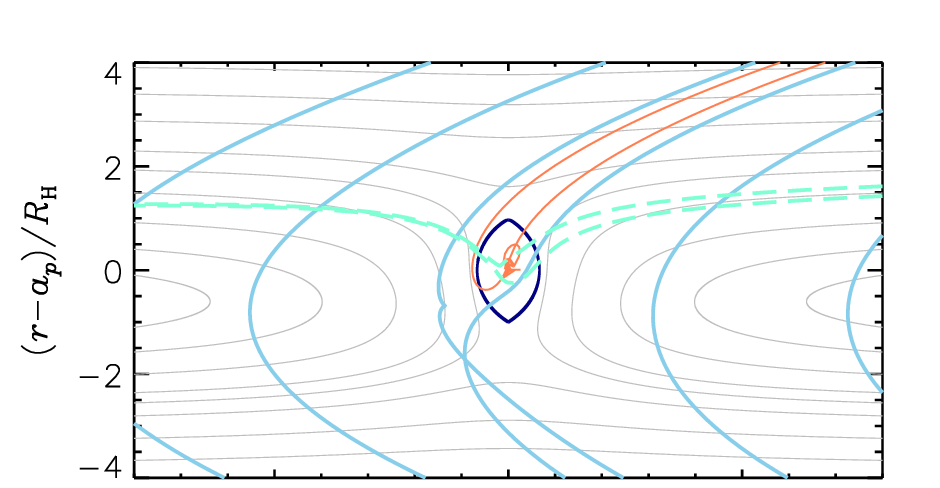}}
\resizebox{\linewidth}{!}{\includegraphics[clip]{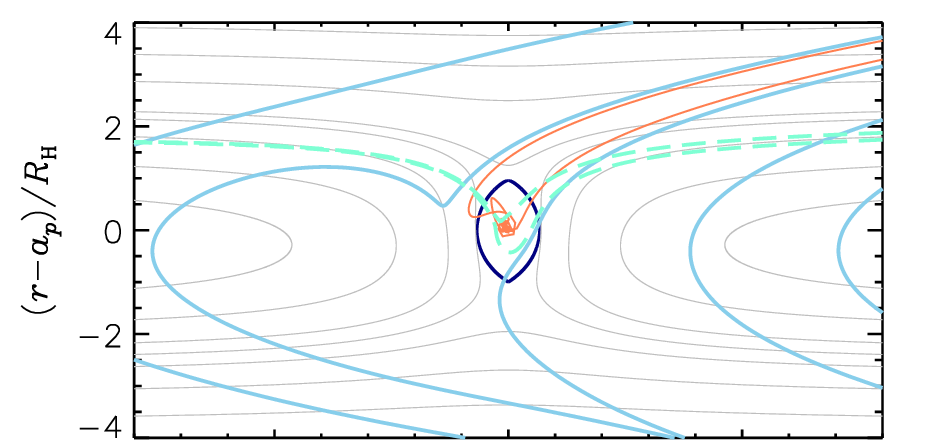}}
\resizebox{\linewidth}{!}{\includegraphics[clip]{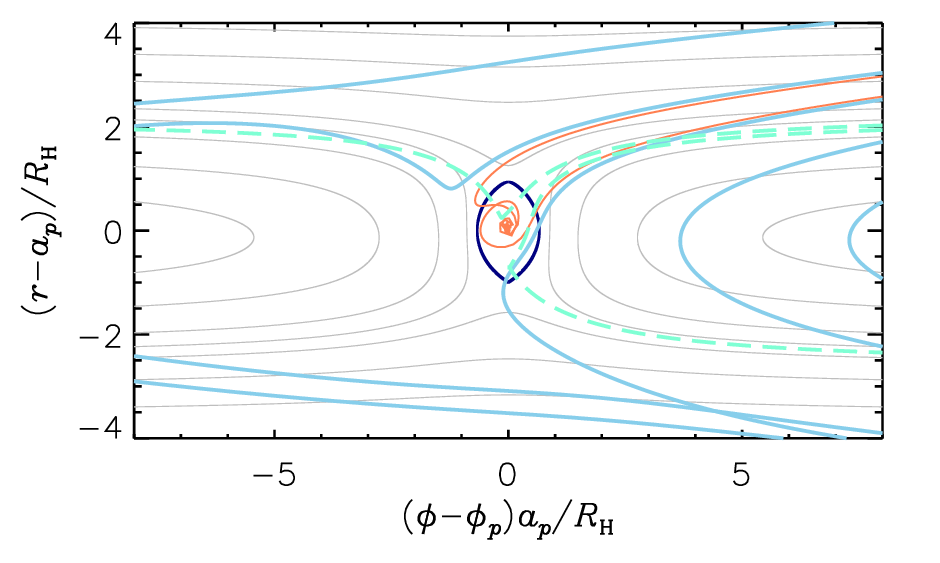}}
\caption{%
             Trajectories of $R_{s}=1$ (dashed lines) and
             $30\,\mathrm{cm}$ (solid lines) rocky particles in 
             the proximity of $0.01$ (top), $0.1$ (middle), and $1\,\Mearth$
             (bottom) planets, orbiting at $a_{p}=5\,\AU$ on 
             circular orbits. The trajectories are represented in 
             a frame co-rotating with the planet. 
             The gray lines represent gas streamlines.
             The planet's Roche lobe is also shown.
             The unperturbed disk properties are as in \cifig{fig:taud}
             at $5\,\AU$ ($\tau_{s}\Omega_{\mathrm{K}}\approx 1$
             for $R_{s}=30\,\mathrm{cm}$ and $\approx 10^{-2}$ for
             $R_{s}=1\,\mathrm{cm}$).
             }
\label{fig:close}
\end{figure}
\cifig{fig:close} illustrates results from similar experiments, 
at lower masses: $\Mp=0.01$ (top), $0.1$ (middle), and $1\,\Mearth$
(bottom) at $a_{p}=5\,\AU$.
The particle radius is $R_{s}=1$ (dashed lines) and $30\,\mathrm{cm}$ 
(solid lines), corresponding respectively to
$\tau_{s}\Omega_{\mathrm{K}}\approx 0.01$ 
and $\approx 1$ for the disk conditions at $5\,\AU$ in \cifig{fig:taud}. 
Gas streamlines are also plotted, along with the Roche lobe contours 
in the disk midplane. 
The calculation of the Roche lobe contours neglects modifications
induced by gas drag \citep[][]{murray1994}. 
The numerical results indicate that the estimate of $R_{\mathrm{eff}}$ 
used in \cieq{eq:zeta} may be a reasonable approximation in these cases. 
Although the accretion stream entering the Roche lobe is asymmetric 
with respect to the planet (see the orange trajectories), an average size 
of $R_{\mathrm{eff}}$ can be estimated as a few tenths of \Rhill.  
An outcome similar to that of the $R_{s}=30\,\mathrm{cm}$ particles 
is obtained from the trajectory of the $R_{s}=100\,\mathrm{cm}$ particles 
(not shown). 
In these experiments, the particle trajectories are integrated until 
they impact the planet, assumed to be a condensed object. 
As argued below,
an atmosphere could extend at most over some fraction of the Bondi 
radius, $\lesssim 0.2\,\Rhill$ in the $\Mp=1\,\Mearth$ case (and much 
less in the other cases), a length $\lesssim R_{\mathrm{eff}}$.

It must be pointed out that in the presence of a perturbing body,
the \emph{unperturbed} relative velocity provided by \cieq{eq:urel},
which is independent of the distance $\widetilde{r}$ from the 
perturbing object and of its mass $\Mp$, is appropriate for use only 
far away from the perturber.
However, both Equations~(\ref{eq:zeta}) and (\ref{eq:zeta_lj}) neglect three-body 
effects, and therefore the relative velocity should be sampled 
where the gravity field of the planetary body dominates, 
that is, where gravitational perturbations on $u_{\mathrm{rel}}$ are 
non-negligible.
If $\tau_{s}\Omega_{\mathrm{K}}\ll 1$, the \emph{perturbed} relative 
velocity can be estimated from Equations~(\ref{eq:OPHI}), (\ref{eq:phi}) 
and (\ref{eq:ugr_ol}), and it depends on $\Mp$, $\widetilde{r}$, and 
the direction of approach.
At a distance $\widetilde{r}\approx \Rhill$, its magnitude is roughly 
$\propto\Rhill$ and, for Moon-mass bodies, it is several times 
as large as the unperturbed velocity predicted by \cieq{eq:urel}.
For finite (non-zero) values of the particle stopping time, analytical derivations 
of the perturbed velocity are more involved, but it can be easily 
evaluated through numerical integration of the trajectories 
(as in \cifig{fig:close}).
For $\tau_{s}\Omega_{\mathrm{K}}\approx 1$, such experiments 
indicate that particles approaching a Moon-mass body can achieve 
relative velocities many times (up to a factor of ten) as large as 
the right-hand side of
\cieq{eq:urel} at distances $\widetilde{r}\lesssim \Rhill$.
Applying this correction, the efficiencies predicted by Equations~(\ref{eq:zeta})
and (\ref{eq:zeta_lj}) become similar.

\subsection{Simple Estimates of Accretion Efficiencies}
\label{sec:SEA}

\begin{figure}
\centering%
\resizebox{\linewidth}{!}{\includegraphics[clip]{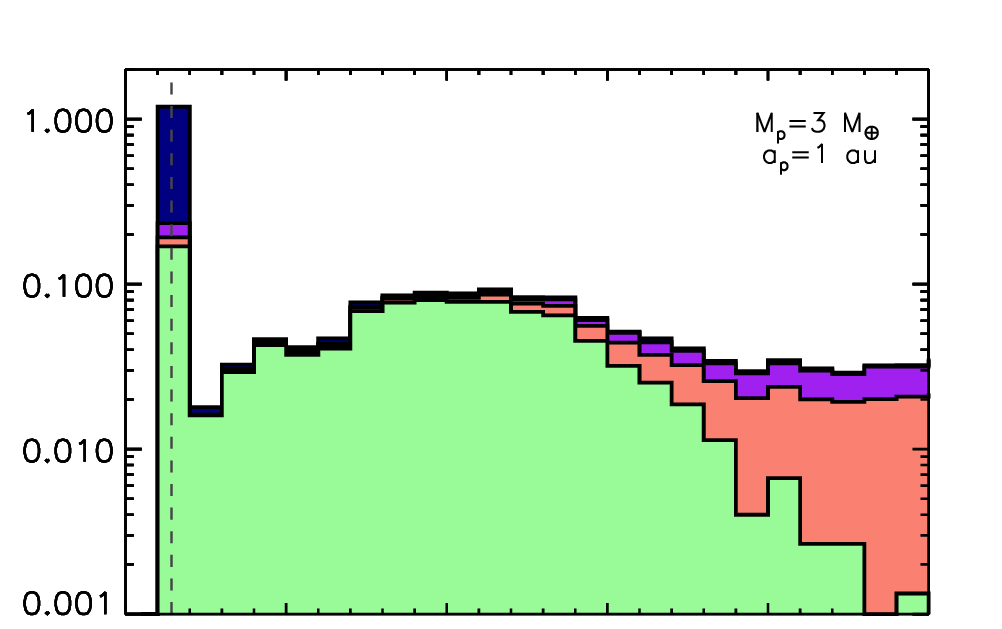}}
\resizebox{\linewidth}{!}{\includegraphics[clip]{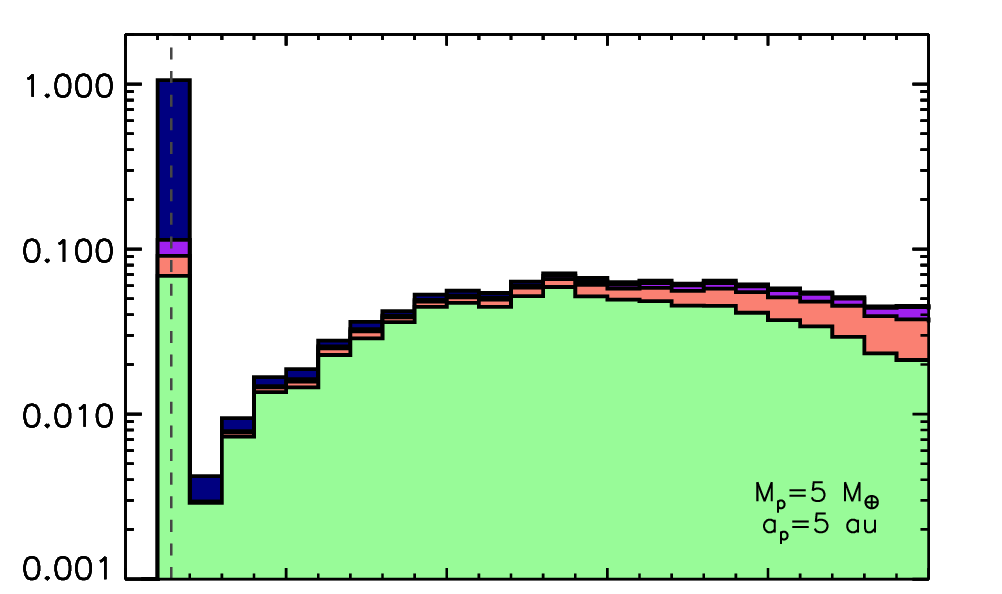}}
\resizebox{\linewidth}{!}{\includegraphics[clip]{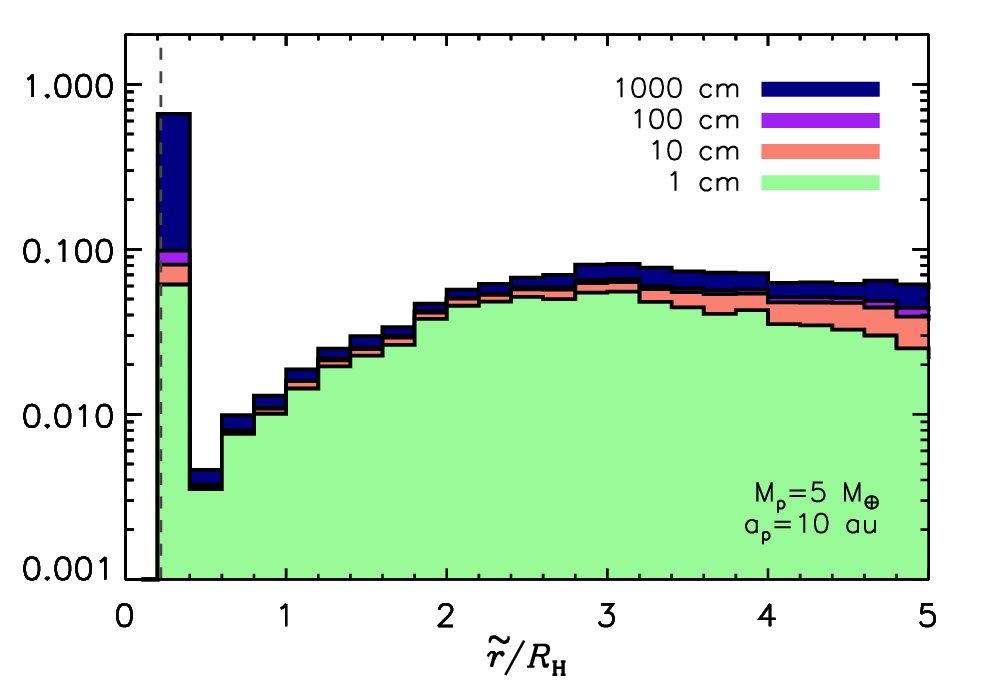}}
\caption{%
             Histograms of the closest approach, in units of the
             planet's Hill radius, along the trajectories of 
             $R_{s}=1$, $10$, $100$, and $1000\,\mathrm{cm}$ particles 
             (as indicated). Each bin is normalized by the total mass used 
             in the experiment for the given particle size. Bins are stacked 
             in order of increasing $R_{s}$, rendered by different colors.
             The dashed line indicates the planet's Bondi radius. 
             The unperturbed disk properties correspond to those of
             \cifig{fig:taud}, according to $a_{p}$.
             The case at $1\,\AU$ (top panel) uses
             the highest gas density in \cifig{fig:taud}.
             }
\label{fig:mdh}
\end{figure}
The approach used for the trajectory integration in Figures~\ref{fig:ff} 
and \ref{fig:close} is applied to conduct experiments on the accretion 
of particles, with radii of $1$, $10$, $100$, and $1000\,\mathrm{cm}$,
on $\Mp=3$ and $5\,\Mearth$ planets orbiting at $a_{p}=1$, $5$,
and $10\,\AU$.
These experiments provide a statistical estimate of the accretion 
efficiency $\zeta$ for the simplified gas flow given 
by Equations~(\ref{eq:OPHI}), (\ref{eq:phi}), and (\ref{eq:ugr_ol}).
The gas density is axisymmetric around the star and proportional 
to a power of $r$.
A summary of the results is displayed in \cifig{fig:mdh}. Histograms 
show the distributions of the closest approach distance along the particles' trajectories 
for the $\Mp=3\,\Mearth$ planet at $a_{p}=1\,\AU$ (top panel) and
for the $\Mp=5$ planets at $a_{p}=5$ and $10\,\AU$ 
(middle and bottom panels, respectively; see also the figure's caption). 
The gas densities are those quoted in \cifig{fig:taud} (see also
\citab{table:dat}), according to the planet's orbital radius 
(at $a_{p}=1\,\AU$, $\Sigma_{g}\approx 1100\,\sigu$). 
All trajectories begin as random circles at radial distances 
beyond the 2:3 mean-motion resonance with the planet, out to 
a distance somewhat beyond the 1:2 mean-motion resonance (in the orbital
plane of the planet).
These experiments assume equal masses of icy and rocky particles,
in each size bin, at $a_{p}=5$ and $10\,\AU$, and only rocks 
at $a_{p}=1\,\AU$. 
Accretion is occurs if a particle enters the planetary envelope,
$\widetilde{r}<R_{p}$ and $R_{p}$ is taken as the Bondi radius, 
\cieq{eq:RB}.
For a solid to be permanently captured, its relative 
velocity in the envelope must not exceed the escape velocity,
$u_{\mathrm{esc}}$. 
These experiments do not account for such requirement, which 
is generally irrelevant for the small particles considered 
here (as confirmed by the calculations discussed in \cisec{sec:RHD}).

The closest approach distance from the planet in \cifig{fig:mdh} 
is in units of  the Hill radius, $\Rhill$. For a given particle radius, the histogram 
bins are normalized to indicate the fraction of the initial mass 
of the solids with radius $R_{s}$ considered in the experiment. 
The bins are then overlaid in order of increasing solid size. 
The vertical dashed line indicates the distance
$\widetilde{r}=\Rbondi$, the assumed envelope radius.
Consistent with the results of \cifig{fig:close}, the histograms 
show that particles of all sizes can enter the planet's Hill sphere 
without necessarily being captured. The pile-up in the bin interior 
to or overlapping $R_{p}$ is caused by particles flagged as accreted. 
Since each color represents a normalized mass, the innermost bin 
in each panel provides a measure of the efficiency $\zeta$, estimated 
here as the accreted mass divided by the mass of the solids approaching 
and interacting with the planet. 

Particles with an initial $\tau_{s}\Omega_{\mathrm{K}}\sim 1$
have $R_{s}\approx 10$--$100\,\mathrm{cm}$. For both planet masses, 
the efficiency ranges from a few to several percent. 
For $1\,\mathrm{cm}$ particles 
($\tau_{s}\Omega_{\mathrm{K}}\lesssim 10^{-2}$ initially), 
$\zeta$ ranges from a few to $\approx 15$\%.
The largest, $1000\,\mathrm{cm}$ particles 
($\tau_{s}\Omega_{\mathrm{K}} > 10^{2}$ initially) can accrete 
efficiently: $\zeta>50$\% in these experiments.
For the two $\Mp=5\,\Mearth$ cases, $\zeta$ is comparable across 
the entire range of particle sizes. 

An estimate of the accretion rate of solids on the planet, $\dMp=\zeta \dot{M}_{s}$, 
can be obtained by combining the results in Figures~\ref{fig:dMsdt_rs} 
and \ref{fig:mdh}. 
The mass flux of $R_{s}=1\,\mathrm{cm}$ particles is sustained over 
the first
$\approx 0.5\,\Myr$ ($\dot{M}_{s}\sim 10^{-4}\,\mathrm{\Mearth\,yr^{-1}}$),
producing values of $\dMp$ in the range 
$10^{-6}$--$10^{-5}\,\mathrm{\Mearth\,yr^{-1}}$.
Particles with $R_{s}=10$--$100\,\mathrm{cm}$ have low $\zeta$
which, compounded with 
$\dot{M}_{s}\sim 10^{-5}\,\mathrm{\Mearth\,yr^{-1}}$ 
over the first $1\,\Myr$, would result in $\dMp$ of order
$10^{-7}$--$10^{-6}\,\mathrm{\Mearth\,yr^{-1}}$.
During the first $1\,\Myr$, $R_{s}=1000\,\mathrm{cm}$ solids also 
have $\dot{M}_{s} \sim 10^{-5}\,\mathrm{\Mearth\,yr^{-1}}$, 
which would generate values of $\dMp$ up to 
$\sim 10^{-5}\,\mathrm{\Mearth\,yr^{-1}}$.
It is important to notice that high values of $\zeta$ for 
a particular solid size imply an efficient removal of these solids, 
which then cannot significantly contribute to the growth of other 
planets orbiting closer to the star.

A number of effects are neglected in these experiments. Gas density
perturbations driven by the planet's tidal field can impact the drag 
forces exerted on the particles, possibly causing a size-dependent
segregation effect. 
Additionally, the complex three-dimensional flow circulation in 
the proximity of the planet can alter the balance of forces and 
change the outcome of the capture process. Moreover, for the icy 
solids, variations of the gas temperature close to the planet can 
promote ablation and vaporize some fraction of the local  mass
of solids. 
These effects, which require more realistic disk-planet interaction 
models and more sophisticated calculations of the solids' evolution 
in disks, are included in the simulations discussed below.

\section{Radiation-Hydrodynamics Calculations of Solids' Accretion}
\label{sec:RHD}

\subsection{Gas Thermodynamics}
\label{sec:GDT}

\begin{deluxetable*}{ccccccccc}
\tablecolumns{9}
\tablewidth{0pc}
\tablecaption{Properties of Planets and Disks in the RHD Models\label{table:dat}}
\tablehead{
\colhead{$a_{p}$ [\AU]}&\colhead{$\Mp/\Mearth$}&\colhead{$\Rbondi/a_{p}$}&\colhead{$R_{p}/\Rbondi$\tablenotemark{a}}&\colhead{$R_{p}/\Rhill$\tablenotemark{a}}&\colhead{$\langle T_{g}\rangle$\tablenotemark{b} [\K]}&\colhead{$\langle \rho_{g}\rangle$\tablenotemark{b} [$\rhou$]}
}
\startdata
$10$ & \phn$5$      & $0.0038$ & $0.69$ & $0.15$      & \phn$80$ & $5.0\times 10^{-12}$\\[-0.8ex]
$10$ & $10$           & $0.0075$ & $0.48$ & $0.17$      &                 &\\[-0.8ex]
$10$ & $15$           & $0.0113$ & $0.40$ & $0.18$      &                 &\\
\cline{1-7}\\[-3.5ex]
\phn$5$ & \phn$5$ & $0.0050$ & $0.92$ & $0.27$      & $124$      & $1.3\times 10^{-11}$\\[-0.8ex]
\phn$5$ & $10$      & $0.0097$ & $0.67$ & $0.30$      &                 &\\[-0.8ex]
\phn$5$ & $15$      & $0.0150$ & $0.64$ & $0.39$      &                 &\\
\cline{1-7}\\[-3.5ex]
\phn$1$ & \phn$3$ & $0.0091$ & $1.00$ & $0.61$      & $200$      & $2.0\times 10^{-11}$\\[-0.8ex]
\phn$1$ & \phn$3$ & $0.0055$ & $1.00$ & $0.37$      & $330$      & $1.5\times 10^{-10}$\\[-0.8ex]
\phn$1$ & \phn$3$ & $0.0039$ & $1.00$ & $0.26$      & $470$      & $6.7\times 10^{-10}$\\[-0.8ex]
\phn$1$ &      $20$ & $0.0602$ & $0.32$ & $0.70$      & $200$      & $2.0\times 10^{-11}$\\[-0.8ex]
\phn$1$ &      $20$ & $0.0365$ & $0.52$ & $0.70$      & $330$      & $1.5\times 10^{-10}$\\[-0.8ex]
\phn$1$ &      $20$ & $0.0256$ & $0.74$ & $0.70$      & $470$      & $6.7\times 10^{-10}$\\[-0mm]
\enddata
\tablenotetext{a}{Ratio measured for models at $5$ and $10\,\AU$, assumed at $1\,\AU$.}
\tablenotetext{b}{Midplane gas temperature and density at $r=a_{p}$,
averaged around the star.}
\end{deluxetable*}\vspace*{-5pt}

During the relatively rapid motion of small particles through the disk, 
the gas component of the disk can be assumed to be in a quasi-steady state, so that its 
thermodynamic fields, e.g., velocity $\gvec{u}_{g}$, temperature $T_{g}$, 
and density $\rho_{g}$, remain nearly constant in time when described in 
a reference frame co-rotating with an embedded planet on a circular orbit. 
This assumption is based on the fact that global disk evolution is mainly 
driven by viscous transport during the stages considered here. 
This study uses the thermodynamic steady-state fields 
$(\gvec{u}_{g},\rho_{g},T_{g})$ from the models of 
\citet[hereafter, \citetalias{gennaro2013}]{gennaro2013}, who performed 
global three-dimensional (3D) radiation-hydrodynamics (RHD) calculations 
of planets embedded in disks. They modeled planets of mass $\Mp=5$, 
$10$, and $15\,\Mearth$ on circular orbits at $a_{p}=5$ and $10\,\AU$ 
from a solar-mass star. In all cases, the envelope mass was a small 
fraction of \Mp. 
\citab{table:dat} lists some properties of these models.

Briefly, \citetalias{gennaro2013} adopted a spherical polar 
discretization of the disk, with coordinates $\{r,\theta,\phi\}$, solving
the Navier-Stokes equations for a compressible and viscous fluid in 
a reference frame rotating about the star at the same angular velocity 
as the planet's orbital frequency.
They applied a numerical method that is effectively second-order 
accurate in both space and time \citep[see, e.g.,][]{boss1992}. 
The disks extended in radius from $0.5\,a_{p}$ to $2\,a_{p}$, 
about $15^{\circ}$ in the meridional direction and $2\pi$ radians
in azimuth around the star.
The energy equation accounted for the advection of gas and radiation
energy, for the work done by gas and radiation pressure, and for viscous dissipation,
radiation transport, and energy released by accreted solids.
Radiative transfer was approximated via flux-limited diffusion 
\citep{levermore1981,castor2007} and solved by means of an implicit
algorithm second-order accurate in both space and time
\citep[an approach recently validated by][]{bailey2023}.
The gas equation of state was one of a mixture of hydrogen and helium 
in solar proportions that accounted for the ionization of atomic species, 
the dissociation of molecular hydrogen, and for the translational, 
rotational, and vibrational energy states of H$_{2}$. 

Once a thermodynamic quasi-steady state has been achieved,
as in the models 
of \citetalias{gennaro2013}, the disk evolution proceeds 
on the viscous timescale, $\sim r^{2}/\nu_{g}$. 
The calculations adopted a turbulent viscosity parameter 
$\alpha_{g}\approx 10^{-3}$, corresponding to a viscous timescale 
of several $10^{4}$ orbital periods at $r=a_{p}$. 
This level of turbulence is in accordance with observational
estimates \citep[e.g.,][]{flaherty2018}.
Toward the end of a disk's lifetime, when the gas density is very low,
evolution is mainly driven by stellar photo-evaporation and the gas
is cleared inside-out \citep[e.g.,][]{gorti2009a,ercolano2017}. 

For the purposes of this study, an important aspect of
\citetalias{gennaro2013}'s models is that they are global 
(disks extend many astronomical units in radius)
and resolve the actual envelopes of the planetary cores.
They did so by applying hierarchies of nested grids with 
finest resolutions comparable to the condensed core radii. 
In fact, a detailed comparison was performed between 3D 
envelopes and one-dimensional, spherically symmetric
envelopes obtained from planet structure and evolution 
calculations.
The high resolution of the models allows the trajectories of
the solids to be integrated until they enter the actual
planetary envelopes.

\begin{figure}
\centering%
\resizebox{\linewidth}{!}{\includegraphics[clip]{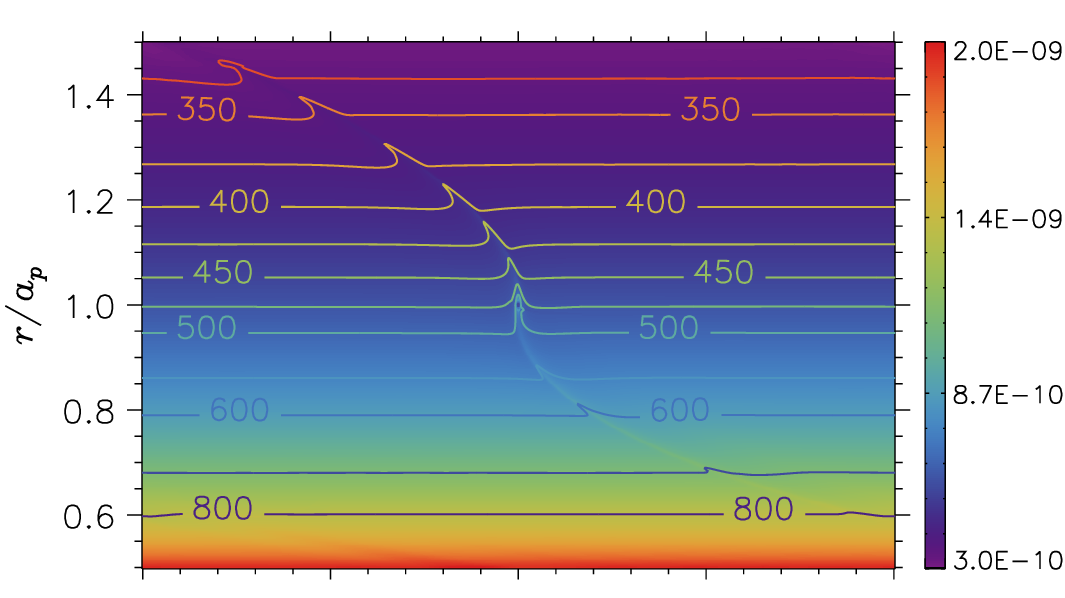}}
\resizebox{\linewidth}{!}{\includegraphics[clip]{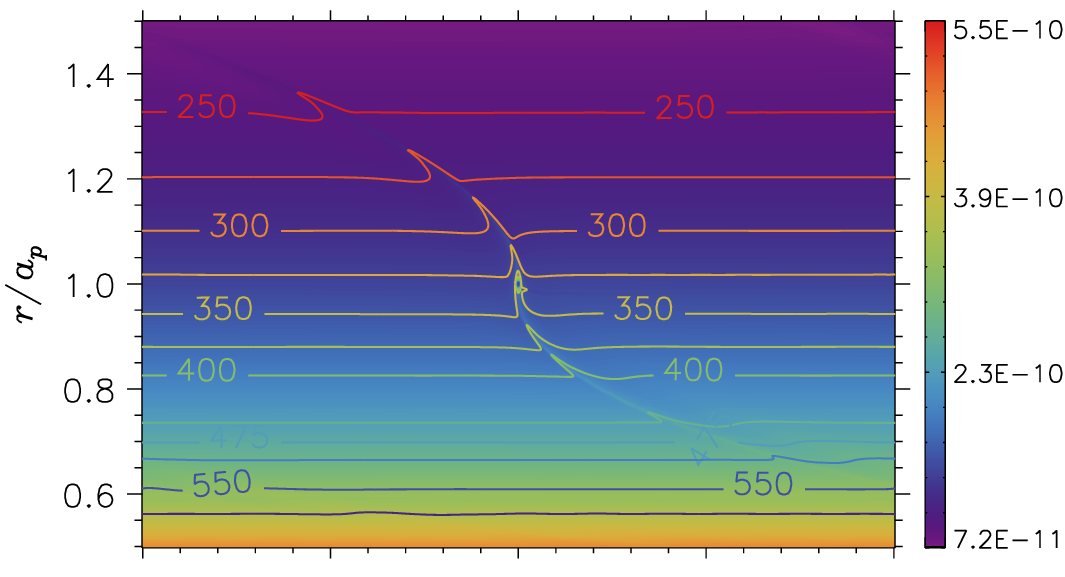}}
\resizebox{\linewidth}{!}{\includegraphics[clip]{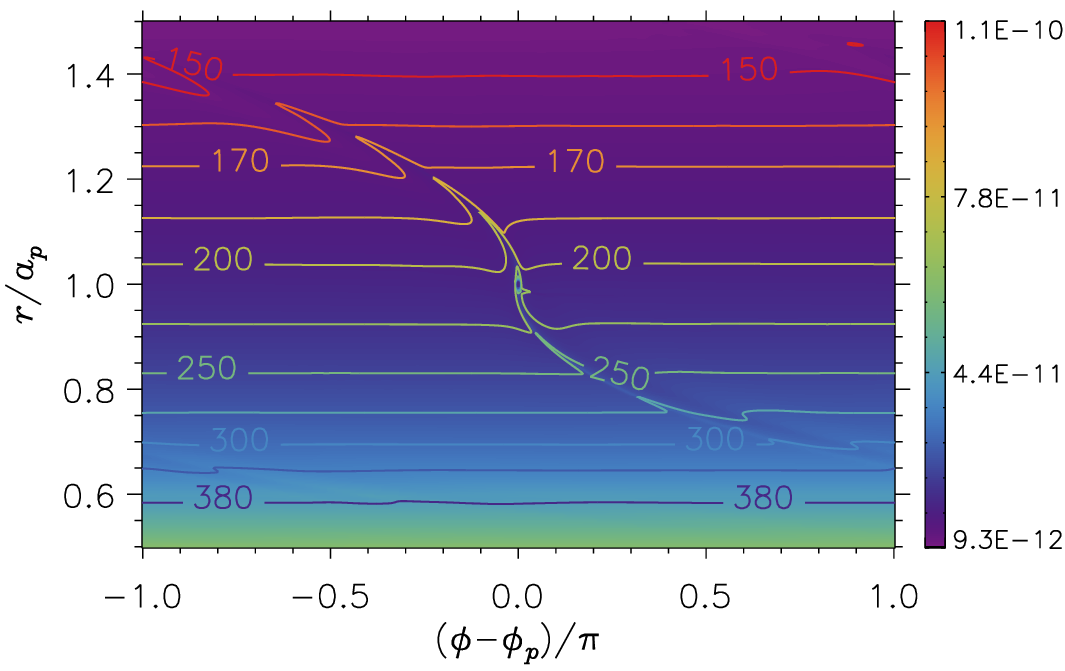}}
\caption{%
             Steady-state gas density (color scale) and temperature 
             (contour lines) 
             at the disk's midplane, perturbed by a (circular orbit) 
             $3\,\Mearth$ planet 
             located at $r/a_{p}=1$ and $\phi=\phi_{p}$.
             The density is in units of grams per cubic centimeter and the temperature is in 
             kelvin degrees. The azimuthally averaged surface density at $r=a_{p}$ is 
             $\Sigma_{g}\approx 1100$ (top), $\approx 220$ (middle), and 
             $\approx 20\,\sigu$ (bottom). See also \citab{table:dat}.
             }
\label{fig:mig}
\end{figure}
\begin{figure}
\centering%
\resizebox{\linewidth}{!}{\includegraphics[clip]{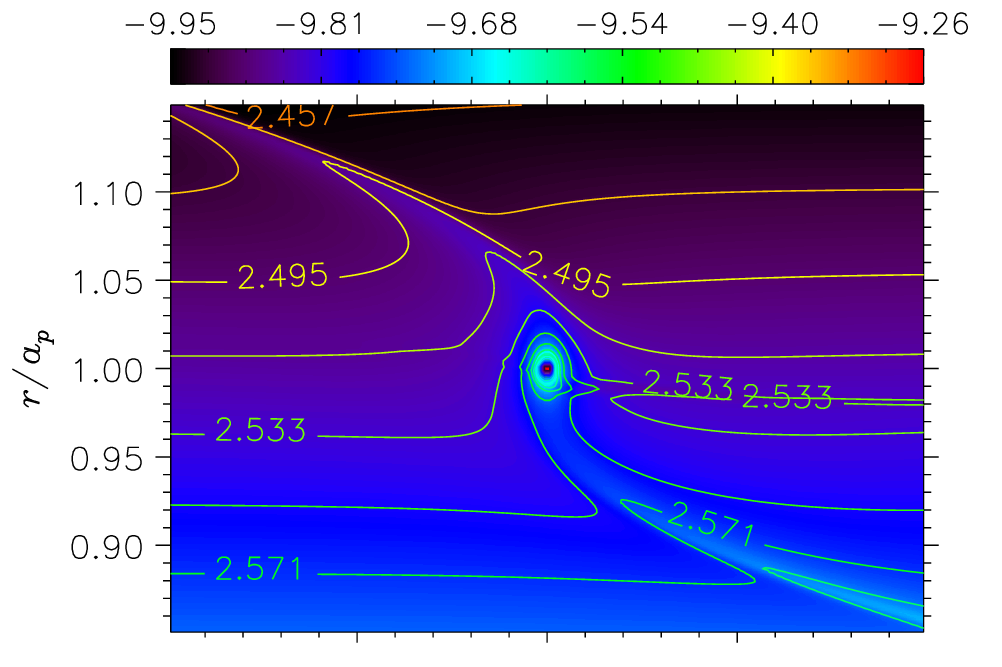}}
\resizebox{\linewidth}{!}{\includegraphics[clip]{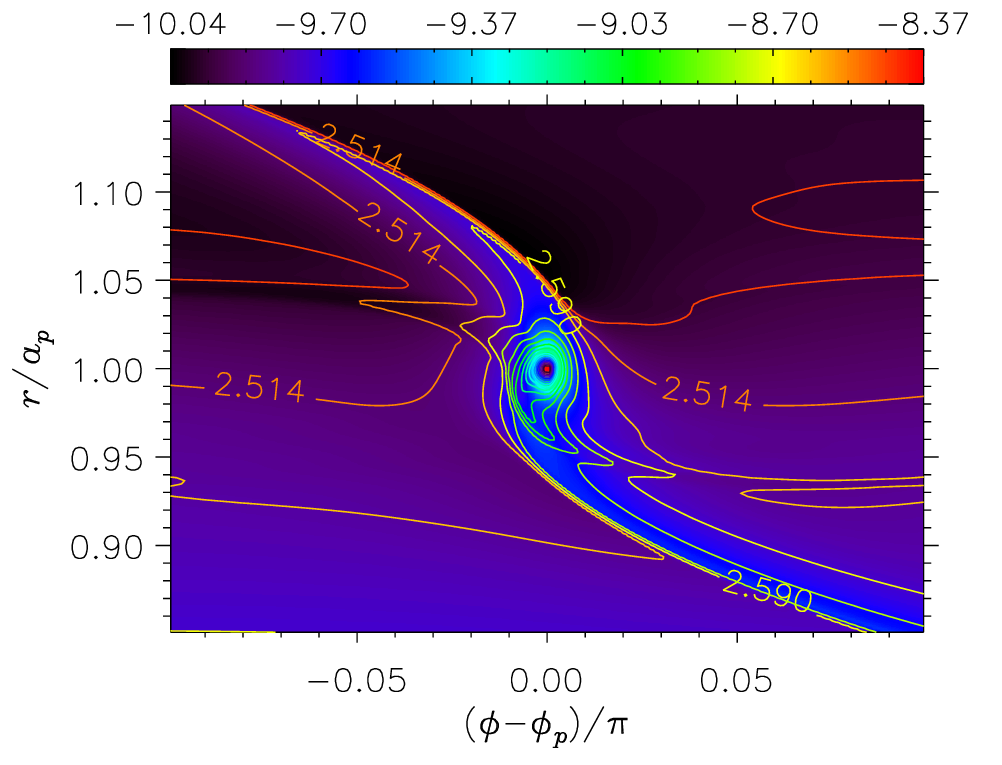}}
\caption{%
             Close-up of $\log{\rho_{g}}$ (color scale) and $\log{T_{g}}$ 
             (contour lines), 
             at the disk's midplane, for the $3$ (top) and $20\,\Mearth$ (bottom) 
             planet located at $a_{p}=1\,\AU$.
             The gas density and temperature are in units of grams per cubic centimeter
              and kelvin degrees, respectively.
             The unperturbed surface density at $r=a_{p}$ is 
             $\Sigma_{g}\approx 220\,\sigu$. 
             }
\label{fig:mig20}
\end{figure}
Models with planets at $a_{p}=5$ and $10\,\AU$ are complemented
by additional RHD calculations of $\Mp=3$ and $20\,\Mearth$ 
planets orbiting at $a_{p}=1\,\AU$ from a $1\,\Msun$ star. 
In these cases, however, the gas density is varied to mimic 
different epochs of the disk's evolution.
Values of $\Sigma_{g}$ at $r=a_{p}$ vary from $\approx 20$ to 
$\approx 1100\,\sigu$ (see \citab{table:dat}). 
The steady-state gas density and temperature in the disk midplanes
of $a_{p}=1\,\AU$ models are illustrated in \cifig{fig:mig}.
Contrary to the calculations of \citetalias{gennaro2013}, these models
do not resolve in detail possible gaseous envelopes bound to the cores. 
Therefore, the planetary cores are assumed to bear an envelope whose
radius is the smaller of \Rbondi\ and \Rhill.
Under this assumption, $R_{p}$ in these six models ranges from two 
to ten times the linear resolution of the grid.
A close-up around the planets is illustrated in \cifig{fig:mig20}. As a
reference, $\Rhill/a_{p}\approx 0.015$ and $0.027$ in the top and bottom
panels, respectively.

When the envelope volume is determined by an energy balance rather
than by gravity, $R_{p}$ is comparable to the Bondi radius
(see \citab{table:dat}).
This distance is found by equating the thermal velocity of the gas
around the planetary core and its escape velocity from the core,
\begin{equation}
 \Rbondi=\left(\frac{\pi}{4}\right)\frac{G \Mp \mu_{g} m_{\mathrm{H}}}{k_{\mathrm{B}}\langle T_{g}\rangle},
 \label{eq:RB}
\end{equation}
where $G$, $m_{\mathrm{H}}$ and $k_{\mathrm{B}}$ are standard physical 
constants, $\mu_{g}$ is the mean molecular weight of the gas mixture, 
and $\langle T_{g}\rangle$ is the average gas temperature at the disk
midplane at the radial distance $r=a_{p}$. 
\cieq{eq:RB} is valid when the energy of the gas is dominated by
thermal energy.
However, in the $20\,\Mearth$ planet models, $\Rbondi$ 
is similar (warmest disk model) or exceeds $\Rhill$. 
Therefore, in those three cases, it is assumed that $R_{p}=0.7\Rhill$,
the mean volume radius of the Roche lobe \citep{eggleton1983}.
The planetary radii in the models, compared to \Rbondi\ and \Rhill,
are reported in \citab{table:dat} 
\citep[see also][for a comparison]{kuwahara2024}.

The approach of simulating the evolution of the solids in the gas steady-state 
fields is dictated by the computational cost of these RHD calculations,
which can hardly be executed 
along with the thermodynamic evolution of solids for the required timescales.
Nonetheless, the validity of this approach is tested in Appendix~\ref{sec:test}, 
where results from two RHD calculations are compared. In the first calculation, 
a population of small solids evolves in the 3D steady-state fields 
$(\gvec{u}_{g},\rho_{g},T_{g})$ of a disk, as described in this section. In 
the second, the same population evolves together with the disk's gas, 
starting from the steady-state fields used in the first calculation. 
As discussed in Appendix~\ref{sec:test}, the outcomes of the two approaches are 
statistically consistent, with relative differences in accretion rates 
$\lesssim 10$\% and typically hovering around a few percent.

\subsection{Thermodynamics of Solids}
\label{sec:ST}

The physical model for the thermodynamic evolution of the solids
is described in \citetalias{gennaro2015}. Briefly, 
the equations of motion of a solid particle are written in terms 
of its linear and absolute angular momenta per unit mass. The particle 
is subjected to gravitational forces by the star and the planet, 
to non-inertial forces, 
to aerodynamic drag, and to the effect of mass loss due to ablation. 
The particle temperature, $T_{s}$, is determined through an energy 
equation that takes into account the work done by gas drag, 
the energy absorbed from the radiation field of the ambient gas, 
and black body emission from the particle's surface. 
The energy equation also accounts for the energy removed/supplied 
during phase transitions of the particle's outer layers. 
Details on the numerical methods and tests can be found in 
\citetalias{gennaro2015}.

Two types of material are considered here: ice (H$_{2}$O) 
and rock (SiO$_{2}$). 
Particles can lose mass by ablation. In this study, the saturated 
vapor pressure, $P_{v}$, of SiO$_{2}$ is upgraded to that reported 
by \citet{melosh2007}:
\begin{equation}
 \ln\left(\frac{P_{v}}{P_{\mathrm{cr}}}\right)=%
 A\left(1-\frac{T_{\mathrm{cr}}}{T_{s}}\right),
 \label{eq:PvSiO2}
\end{equation}
where the critical pressure is
$P_{\mathrm{cr}}=1.89\times10^{9}\,\mathrm{dyne\,cm^{-2}}$,
the critical temperature is $T_{\mathrm{cr}}=5398\,\K$ and 
$A=10.675270434$. The temperature of the solid $T_{s}$ is defined 
in \citetalias{gennaro2015}.
Ablation in the ambient disk's gas is important for ice (or an admixture of
ice and rock), but much less for rock owing to low gas temperatures. 
Ablation of rock becomes significant only inside the planetary 
envelopes (see Appendix~\ref{sec:dep}), when particles are already accreted.


\subsection{Distributions and Segregation of Solids}
\label{sec:SS}

\begin{figure*}
\centering%
\resizebox{\linewidth}{!}{\includegraphics[clip]{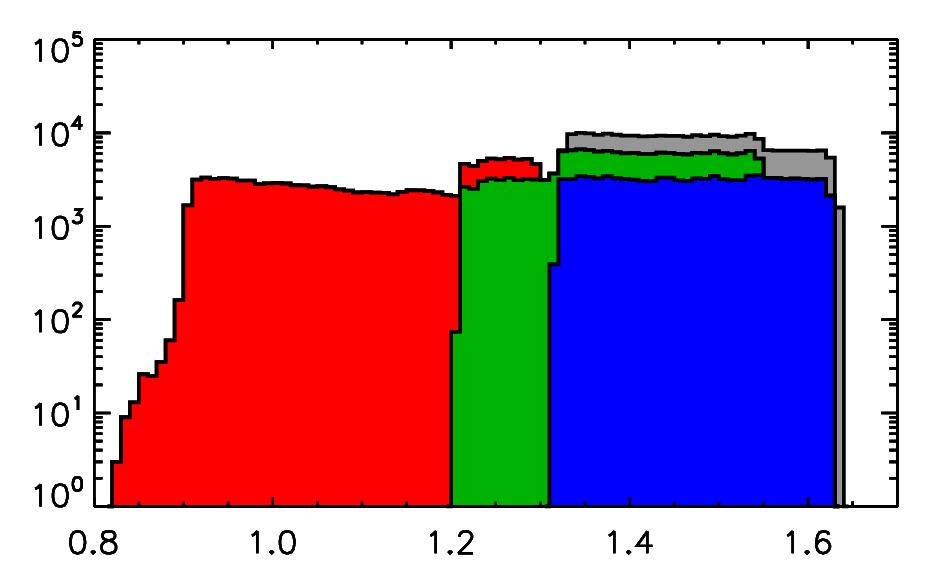}%
                                      \includegraphics[clip]{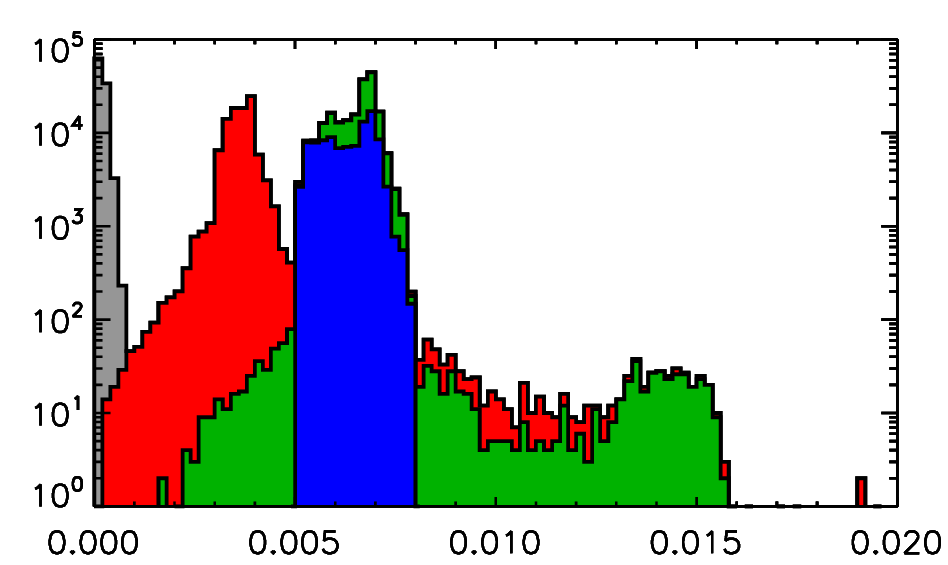}%
                                      \includegraphics[clip]{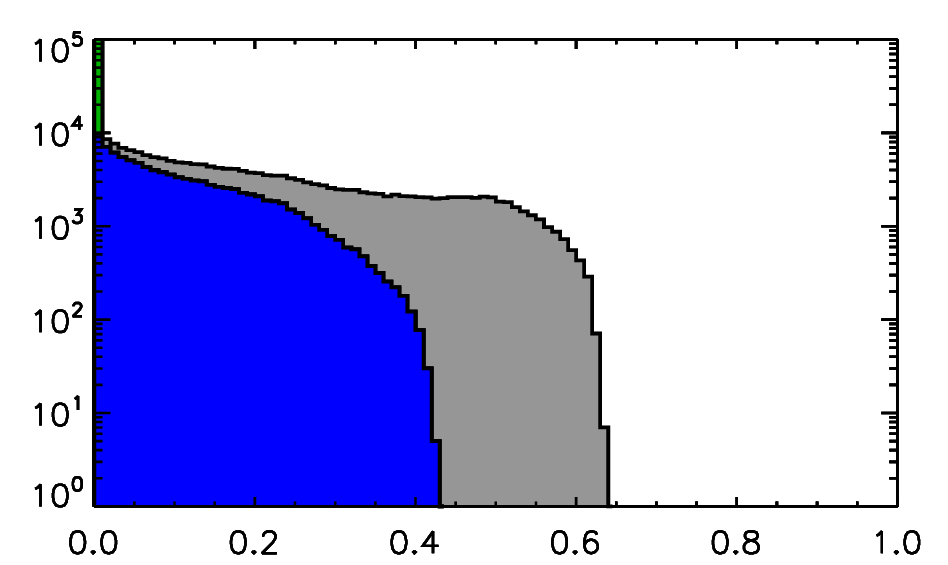}}
\resizebox{\linewidth}{!}{\includegraphics[clip]{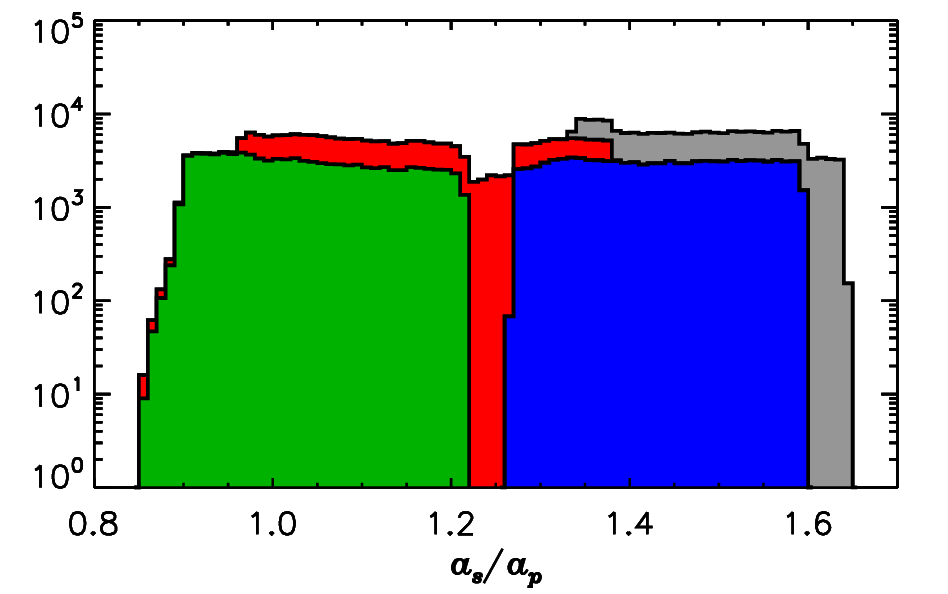}%
                                      \includegraphics[clip]{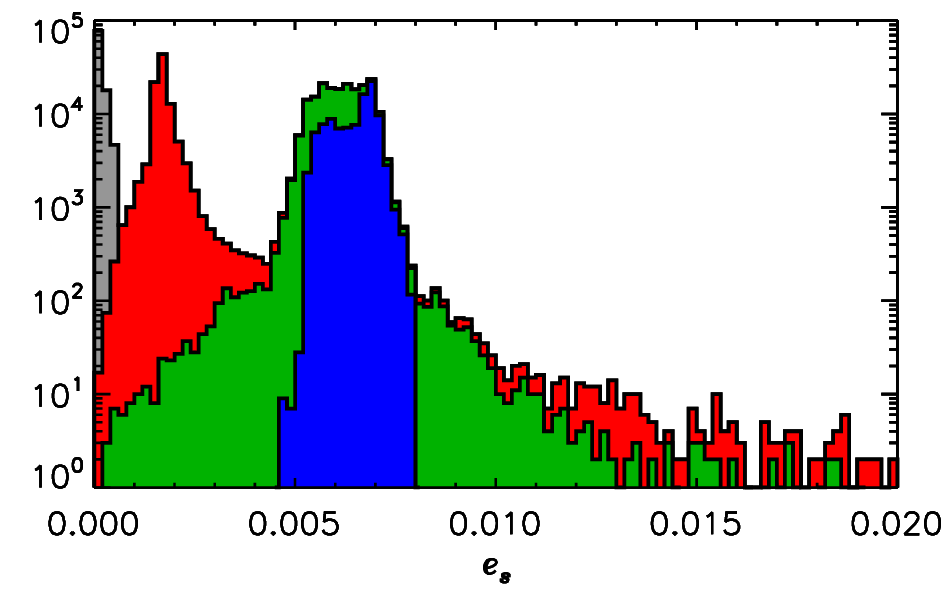}%
                                      \includegraphics[clip]{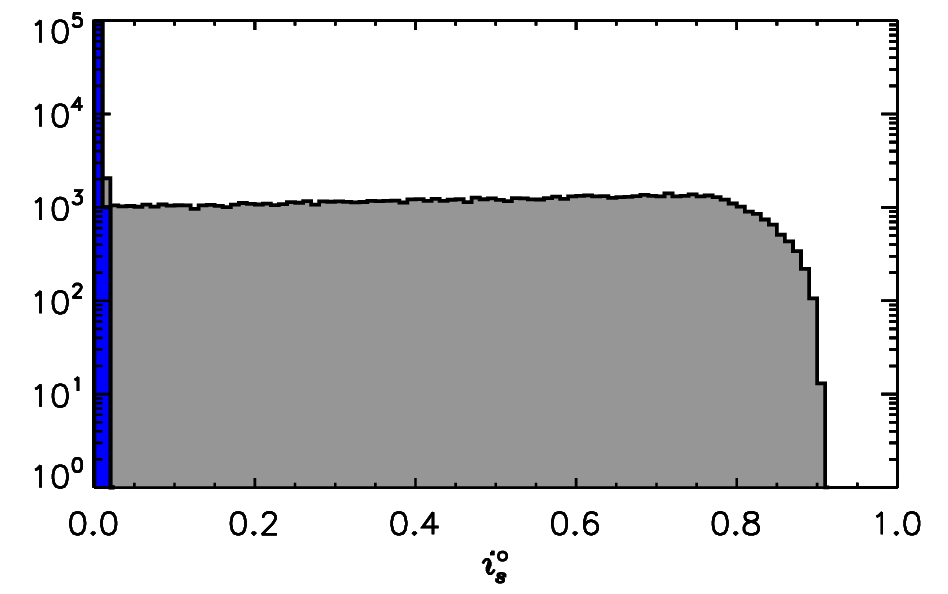}}
\caption{%
             Histograms of the particles' semi-major axes (left),
             eccentricities (center), and inclinations (right) as they
             approach and cross the planet's orbit ($a_{p}=5\,\AU$ and
             $\Mp=5\,\Mearth$). The top and bottom rows refer to icy and 
             rocky solids, respectively.
             The gray histograms include all particles 
             ($R_{s}\le 10\,\mathrm{m}$);
             the red histograms include particles with radii 
             $R_{s}\le 100\,\mathrm{cm}$;
             the green histograms include particles with radii 
             $R_{s}\le 10\,\mathrm{cm}$;
             and the blue histograms include particles with radii 
             $R_{s}\le 1\,\mathrm{cm}$.
            }
\label{fig:h55}
\end{figure*}
Four size bins are initially populated with $10^{5}$ particles each, 
with initial radii $R_{s}=1$, $10$, $100$, and $1000\,\mathrm{cm}$. 
The initial (circular) orbits of the solids are randomly distributed 
between $a_{s}=1.35$ and $1.65\, a_{p}$,
with orbital inclinations varying between $i_{s}=0$ and $1.7$ degrees
and random longitudes of the ascending node. Separate calculations 
are carried out for icy and rocky particles.

The normalized stopping time at deployment differs little between 
cases at $a_{p}=5$ and $10\,\AU$, ranging from 
$\tau_{s}\Omega_{\mathrm{K}}\sim 10^{-2}$ ($R_{s}\approx 1\,\mathrm{cm}$)
to $\sim 10^{2}$ ($R_{s}\approx 1000\,\mathrm{cm}$). 
The range is wider at $a_{p}=1\,\AU$ (see \cifig{fig:taud}).
\cifig{fig:h55} shows the distributions of the orbital properties of the solids
approaching and crossing the orbit of an $\Mp=5\,\Mearth$ planet at $5\,\AU$.
The histograms of different colors include particles grouped by a range 
of sizes, as specified in the caption.
Particles of $10$ and $100\,\mathrm{cm}$ in radius drift inward 
the fastest (see the left panels). As discussed later, in either case,
only a small fraction of these solids is actually accreted. 
The orbital eccentricity, $e_{s}$, remains small during the evolution,
efficiently damped by gas drag, and so does the orbital inclination 
(center and right panels, respectively). 
The inclination of the largest particles and of the smallest icy 
particles is also damped to small values, $i_{s}\lesssim 0.1^{\circ}$,
by the end of the simulations.
Similarly small values for $e_{s}$ and $i_{s}$, to those in 
\cifig{fig:h55}, are also found in the other models, with somewhat 
longer tails in the eccentricity distributions of higher-mass planet models. 

As particles radially drift inward, they cross mean-motion resonances 
with the planet. At those locations, resonant perturbations tend 
to raise a particle's semi-major axis, whereas gas drag tends 
to lower it, which may lead to capture in a stable orbit.
\citet{weidenschilling1985} found that capture is possible for 
a range of the parameter
\begin{equation}
  K=
  \frac{3}{8}C_{D}
  \left(\frac{\rho_{g}}{\rho_{s}}\right)
  \left(\frac{a_{p}}{R_{s}}\right).
  \label{eq:Kappa}
\end{equation}
It can be shown \citep[see][]{peale1993} that, in an unperturbed disk, 
the rate of change of $a_{s}$ of a solid on a near-circular orbit is 
\begin{equation}
  \frac{da_{s}}{dt}=-\left(\frac{K}{2}\right)\left(\frac{H_{g}}{a_{s}}\right)^{4}a_{s}\Omega_{\mathrm{K}}.
  \label{eq:apeale}
\end{equation}
If $K$ exceeds some critical value, the particle can break through 
the mean-motion resonance and inward migration continues. 
\citet{weidenschilling1985} quantified this critical value in 
the limit of small deviations from Keplerian orbits, i.e., for
$\tau_{s}\Omega_{\mathrm{K}}\gtrsim 2\pi$. 
(Note that, in \cieq{eq:Kappa}, their original definition is 
multiplied by $a_{p}$ so as to render the parameter non-dimensional.) 
\citet{kary1993} performed numerical experiments of particles' 
capture in exterior resonances, with both planets and smaller planetary 
embryos, and determined the values of $K$ for which capture is likely. 
For example, they found that an $\Mp\approx 0.3\,\Mearth$ planet 
orbiting at $a_{p}=5\,\AU$ in a minimum-mass solar nebula 
($\rho_{g}\sim 10^{-10}\,\rhou$) can trap 
$R_{s}\sim 100\,\mathrm{m}$ rocky bodies into mean-motion resonances. 
Since resonant perturbations increase with \Mp, smaller particles 
(i.e., larger $K$) can in principle be trapped by larger planets
\citep[see][]{kary1993}. In fact, for quasi-Keplerian orbits, 
the critical value of $K$ for resonance capture scales as 
$(\Mp/\Ms)(a_{p}/H_{g})^{2}$ \citep{weidenschilling1985}.
 
\begin{figure*}
\centering%
\resizebox{\linewidth}{!}{\includegraphics[clip]{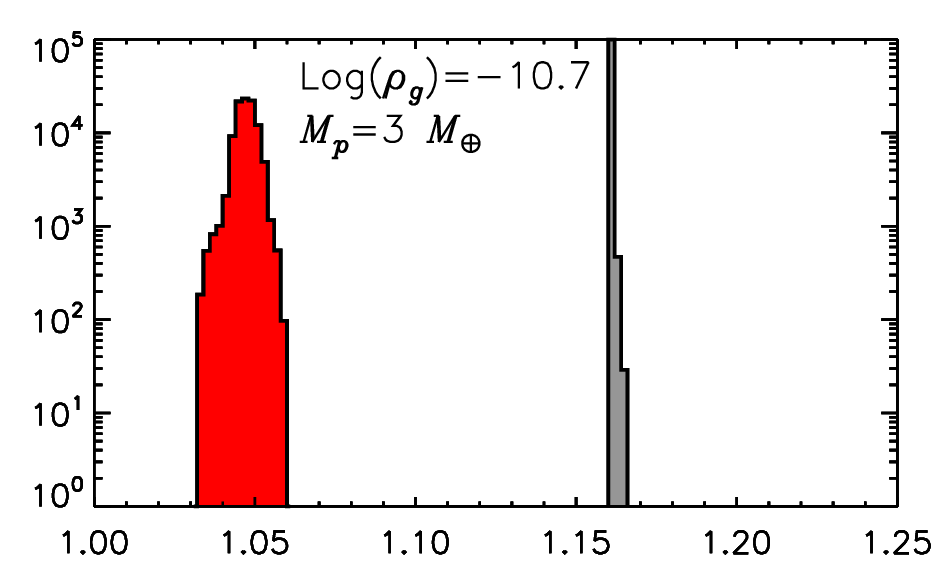}%
                                      \includegraphics[clip]{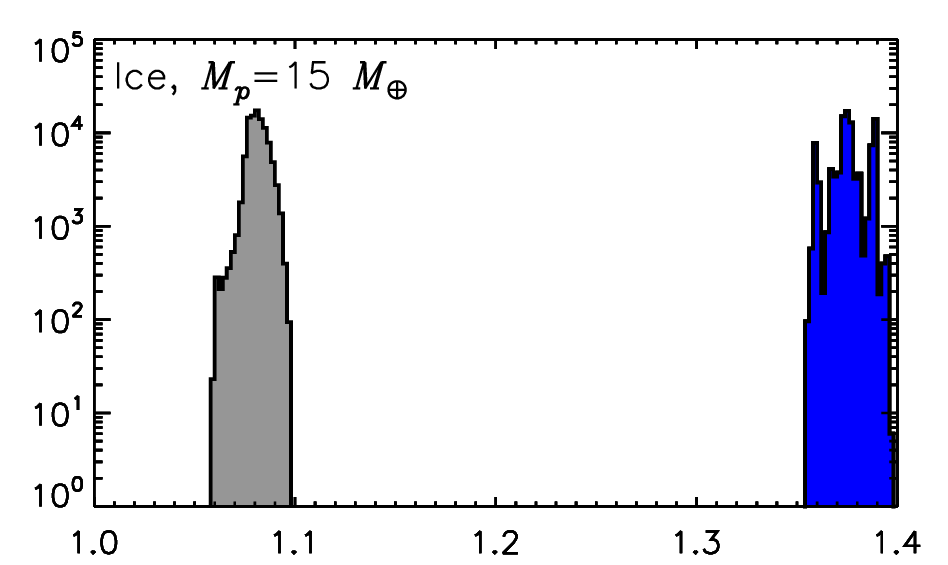}%
                                      \includegraphics[clip]{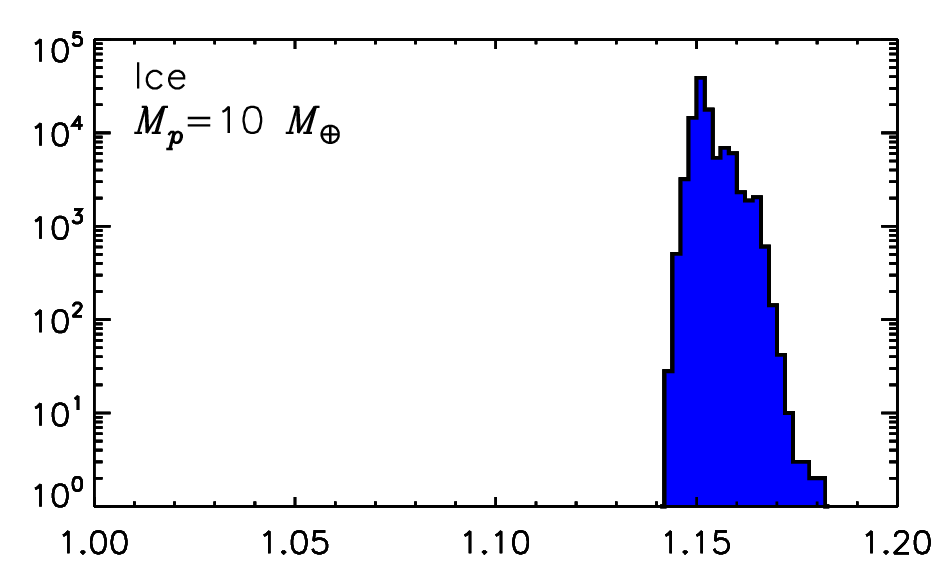}}
\resizebox{\linewidth}{!}{\includegraphics[clip]{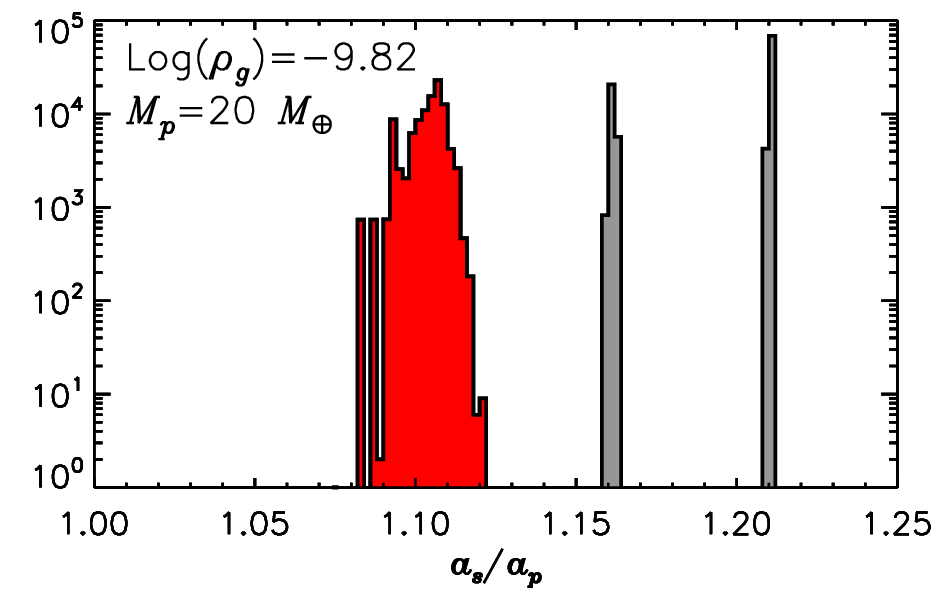}%
                                       \includegraphics[clip]{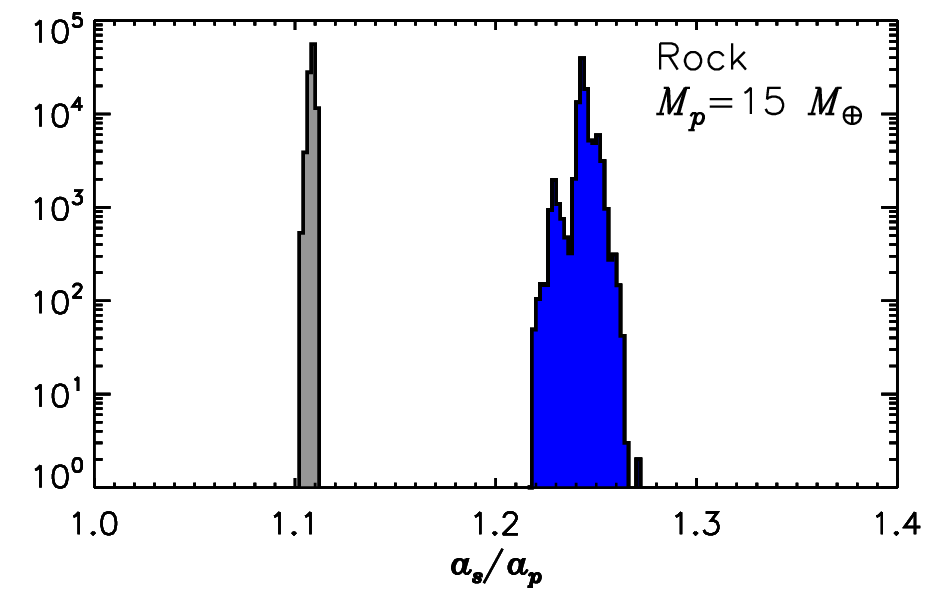}%
                                       \includegraphics[clip]{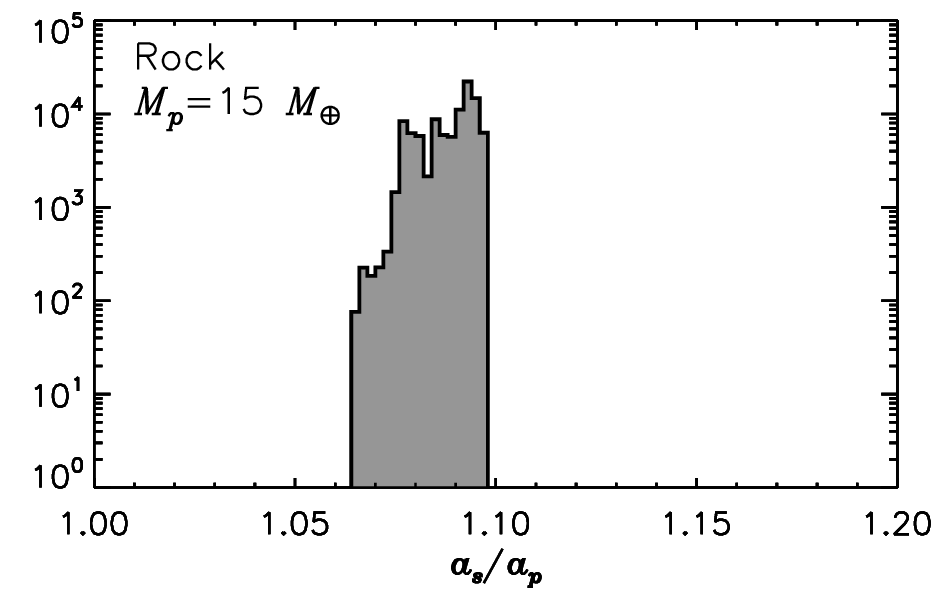}}
\caption{%
             Semi-major axis distributions of segregated particles 
             for models with
             $a_{p}=1$ (left), $5$ (center), and $10\,\AU$ (right). In the left panels,
             $\rho_{g}$ represents the unperturbed gas density at $r=a_{p}$, in units
             of grams per cubic centimeter. 
             The blue, red, and gray histograms refer to $R_{s}=1$, $100$, and
             $1000\,\mathrm{cm}$ particles, respectively.
             Segregation also occurs in other models not shown.
            }
\label{fig:as_seg}
\end{figure*}
In the two cases shown in \cifig{fig:h55} (icy and rocky particles), 
$K$ is large enough for gas 
drag to overcome the resonant forcing so that the simulated particles 
can drift inward, inside the planet's orbit. Other models, however, 
do result in particles segregated in exterior orbits (over the course 
of the calculation). Some examples are illustrated in \cifig{fig:as_seg}.
The trapped particles in the figure have orbits basically co-planar with 
the planet, and their eccentricities are typically small, 
$e_{s}\lesssim H_{g}/a_{s}$.
The semi-major axis distribution of the $10\,\mathrm{m}$ solids 
(gray histograms) can be very narrow, indicating segregation at locations 
of mean-motion resonances with the planet. Such are the cases shown 
in the left panels ($4:5$ and $3:4$ mean-motion resonances) and in 
the bottom center panel ($6:7$ mean-motion resonance), but other instances occur.
The equilibrium eccentricities of solids trapped in mean-motion resonances 
are $\lesssim H_{g}/(2 a_{s})$, in accordance with the estimate of
\citet{weidenschilling1985}.

The minimum distance $\Delta r=r-a_{p}$ at which particles can be trapped 
at resonant locations depends on resonance overlap, which may lead 
to chaotic orbits \citep{wisdom1980}. 
\citet{duncan1989} found that a limit for resonance overlap is given by 
$|a_{s}-a_{p}|/a_{p}\approx 1.5(\Mp/\Ms)^{2/7}$, ranging from $\approx 0.06$ 
to $\approx 0.09$ for the cases illustrated in \cifig{fig:as_seg}.
\citet{kary1993} found an additional limit related to orbital energy 
dissipation by gas drag and estimated that orbital stability is possible 
as long as 
$|a_{s}-a_{p}|/a_{p}\gtrsim 1.6[(H_{g}/a_{p})(\Mp/\Ms)]^{2/9}$, 
whose value is comparable to the limit of \citet{duncan1989} 
for the conditions found in the RHD calculations.
The semi-major axis distributions of $10\,\mathrm{m}$ particles 
trapped in mean-motion resonances satisfy these theoretical limits.

The distributions of smaller particles in \cifig{fig:as_seg} tend to be much 
broader, indicating that their segregation radii are associated to orbital 
locations where $\langle\Omega_{g}\rangle$ exceeds its unperturbed value. 
In the proximity but interior to these radii, drag torques tend to push 
solids outward. 
A planetary-mass body exerts a positive tidal torque on the exterior disk,
transferring angular momentum and hence augmenting the gas rotation rate
relative to that of an unperturbed disk. 
The torque density per unit disk mass is 
$\propto (\Mp/\Ms)^2 (a_{p}/\Delta r)^4$ 
\citep{lin1986b} and peaks for 
$\Delta r\approx H_{g}$ when $\Rhill \le H_{g}$ 
\citep{gennaro2010}.
When integrated over the disk mass exterior to the planet, one finds 
the well-known result that the one-sided torque exerted by 
the planet is $\propto (\Mp/\Ms)^2 (a_{p}/H_{g})^3$ \citep[see, e.g.,][]{lubow2011}.
Since the torque density function has a peak of finite width, 
particles well coupled to the gas can attain minimum trapping distances 
$a_{s}-a_{p}\sim H_{g}$ (if $\Rhill \le H_{g}$).
Angular momentum transfer beyond $|\Delta r|\approx 3 H_{g}$ 
becomes more inefficient \citep{gennaro2010}, even though tidal 
perturbations in the gas can extend farther (see \cifig{fig:mig}).
Gas density perturbations become increasingly prominent as $\Mp$ 
grows, eventually leading to gap formation, which begins when 
tidal torques exceed viscous torques, i.e., when
$(\Mp/\Ms)^{2}\gtrsim 3\pi\alpha_{g}(H_{g}/a_{p})^{5}$.
In the models presented herein, only the $\Mp=20\,\Mearth$ cases at 
$1\,\AU$ (marginally) satisfy this condition in the two coldest disks 
(see \citab{table:dat}).

The $10\,\mathrm{m}$ particles too can stop at equilibrium locations 
dictated by disk-planet tidal perturbations, as suggested by the broad 
distributions of $a_{s}$ in, e.g., the top center and bottom right
panels of \cifig{fig:as_seg}. 
In fact, the values of $K$ are comparable in these two cases, as is 
the width $\Delta r$ of the confinement region. 

Segregation would prevent further accretion, isolating the planet from 
the solids. 
\citet{lambrechts2014} argued that isolation can be achieved when
$\Mp/\Mearth\approx 20\,[H_{g}/(0.05 a_{p})]^{3}$. 
\citet{dipierro2017} found a comparable limit for
$\tau_{s}\Omega_{\mathrm{K}}\sim 1$, while \citet{bitsch2018}
estimated a mass larger by $\approx 25$\% for the conditions realized 
in the RHD calculations.
In the models with $a_{p}=5$ and $10\,\AU$, these ``isolation''
masses ($\approx 30\,\Mearth$ and $\approx 50\,\Mearth$, respectively)
would be much larger than the planet masses modeled herein.
Indeed, in none of the models are $10$--$100\,\mathrm{cm}$ particles
trapped in exterior orbits. However, at both $5$ and $10\,\AU$, 
$R_{s}=1\,\mathrm{cm}$ and $10\,\mathrm{m}$ particles can be
segregated by smaller, $10\,\Mearth$ planets (see \cifig{fig:as_seg}).

At $a_{p}=1\,\AU$, the proposed isolation limit would range 
from $\approx 8\,\Mearth$ (coldest, least dense disk) 
to $\approx 22\,\Mearth$ (warmest, most dense disk). 
At least two of the $20\,\Mearth$ models should thus result in 
the segregation of solids. 
At the lowest density ($\rho_{g}=2.0\times 10^{-11}\,\rhou$), 
$R_{s}=1\,\mathrm{cm}$ solids are efficiently accreted, while
$R_{s}\ge 10\,\mathrm{cm}$ solids are segregated.
At the intermediate density ($\rho_{g}\approx 10^{-10}\,\rhou$), 
$R_{s}\le10\,\mathrm{cm}$ particles are efficiently accreted, 
whereas larger ones are segregated. At the largest density,
$1\,\mathrm{cm}$ and $10\,\mathrm{m}$ solids are segregated. 
Nonetheless, even $3\,\Mearth$ planets can segregate solids
under appropriate conditions, as shown in the top left panel 
of \cifig{fig:as_seg}.

The continued accumulation of solids at segregation sites may enhance 
collisional comminution, possibly releasing some of the mass in 
the form of smaller particles, which may drift inward. 
Coagulation into larger bodies, less coupled to the gas, may also 
reactivate the accretion of heavy elements on the planet.

\subsection{Accretion Efficiencies}
\label{sec:AE}

\begin{figure}
\centering%
\resizebox{\linewidth}{!}{\includegraphics[clip]{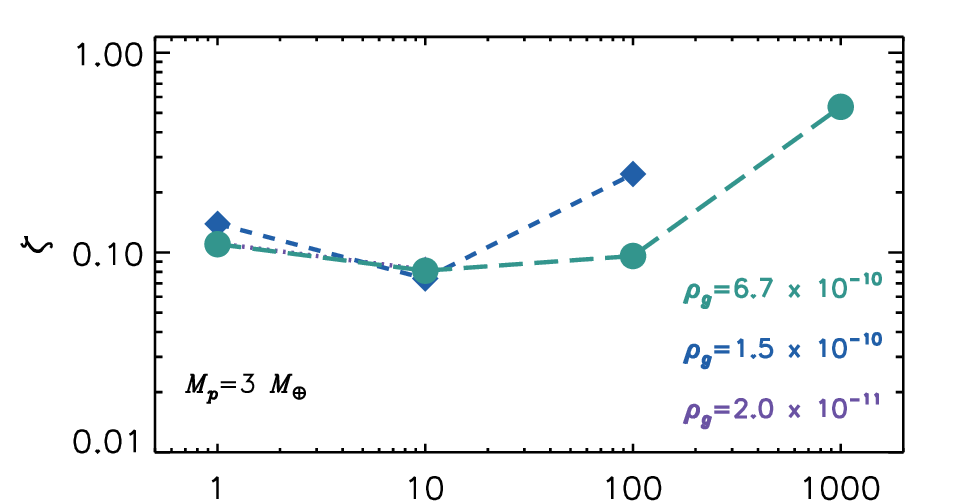}}
\resizebox{\linewidth}{!}{\includegraphics[clip]{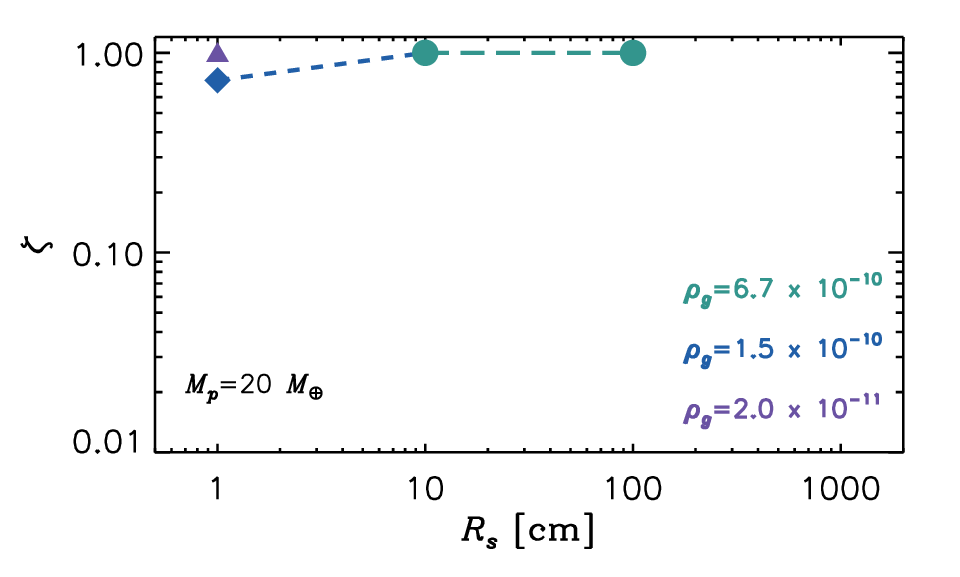}}
\caption{%
             Accretion efficiencies versus particle radius, obtained
             from the RHD calculations at $a_{p}=1\,\AU$, for an $\Mp=3$ 
             (top) and $20\,\Mearth$ (bottom) planet.
             The different colors/symbols refer to different values 
             of the reference gas density, as indicated (see also \citab{table:dat}).
             Non-accreted solids either are segregated in exterior
             orbits or drift toward the star.
             Missing symbols indicate $\zeta=0$.
             In the top panel, efficiencies are comparable, at all three
             values of the gas density, for $R_{s}=1$ and $10\,\mathrm{cm}$
             particles.
             }
\label{fig:zeta_rhd1}
\end{figure}
\begin{figure*}
\centering%
\resizebox{\linewidth}{!}{\includegraphics[clip]{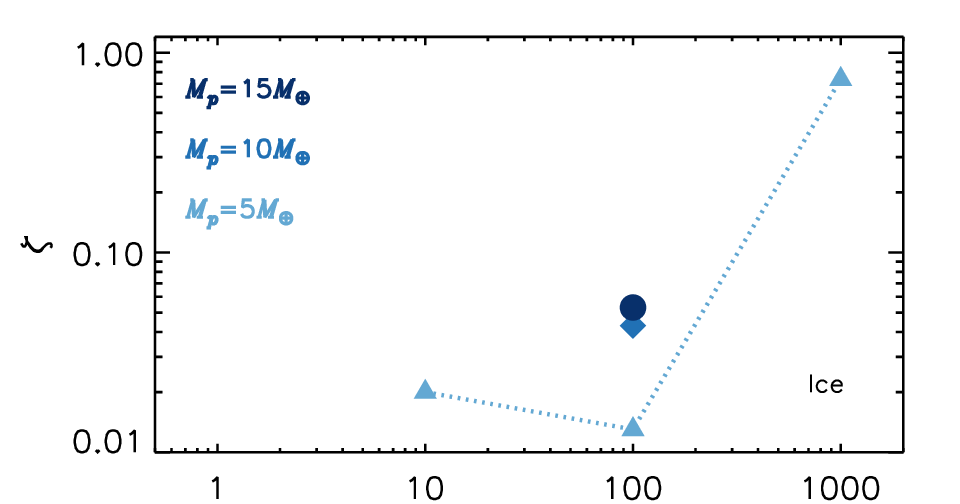}%
                          \includegraphics[clip]{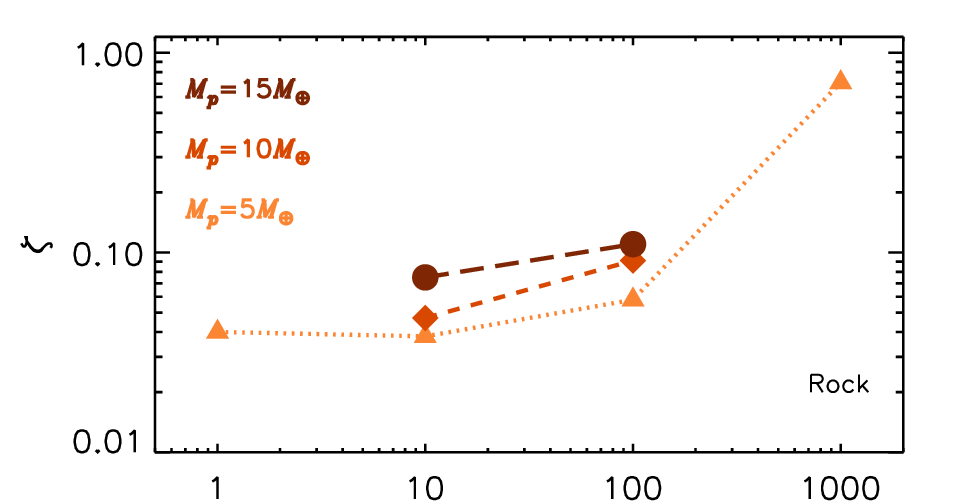}}
\resizebox{\linewidth}{!}{\includegraphics[clip]{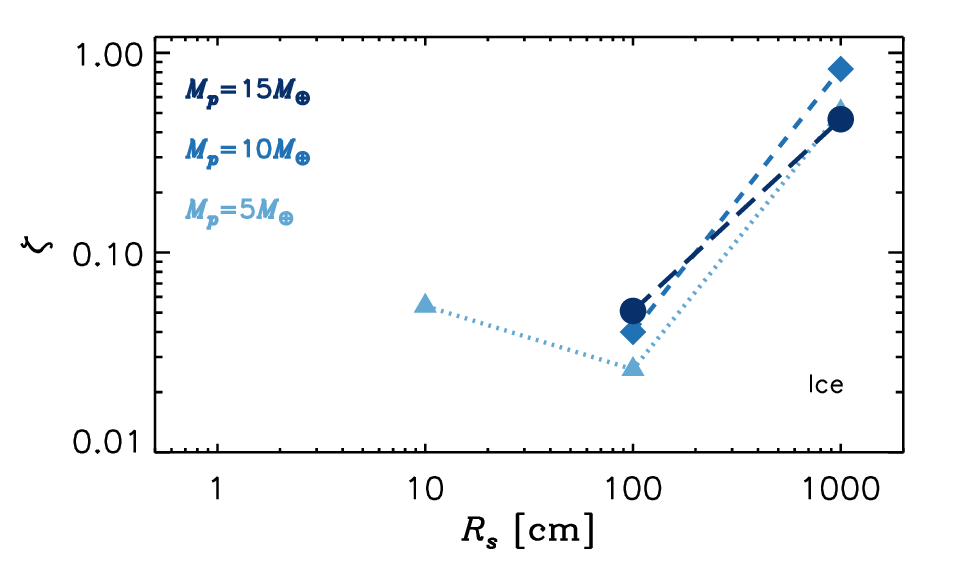}%
                          \includegraphics[clip]{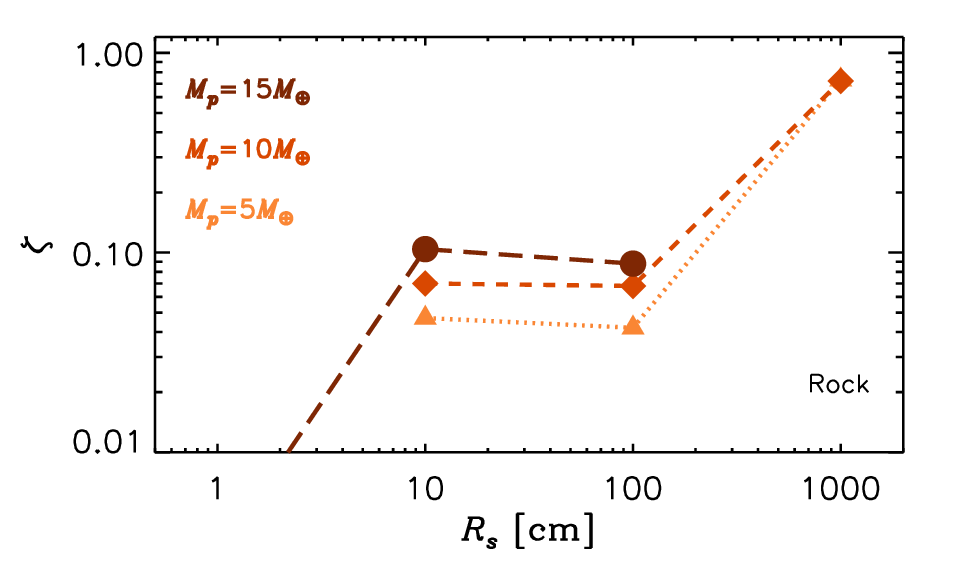}}
\caption{%
             Accretion efficiencies obtained from the RHD
             calculations at $a_{p}=5$ (top) and $10\,\AU$ (bottom).
             Icy and rocky solids are represented on the left and right,
             respectively. The different colors/symbols refer to different
             planet masses, as indicated.
             The different line styles connect symbols of the same 
             type and color.  Missing symbols indicate $\zeta=0$.
             Non-accreted solids either are segregated on
             exterior orbits or drift toward the inner disk.
             Icy particles can also evaporate.
             }
\label{fig:zeta_rhd510}
\end{figure*}

The efficiency or probability of accretion, $\zeta$, in these models
is computed as the ratio of the solids' mass accreted by the planet 
to the solids' mass interacting with the planet, as they drift inward.
Interacting particles can be accreted, transferred to inner orbits, 
or segregated in the outer disk. 
Scattering out of the computational domain is possible but rare.

\vspace*{2pt}
Icy particles can undergo ablation. In the models with $a_{p}=5\,\AU$, 
all particles drifting inward of $r\approx 4\,\AU$ 
($T_{g}\approx 150\,\K$) are consumed to some extent. 
For a fixed particle temperature $T_{s}$, the ablation timescale 
is $\propto R_{s}$ \citepalias{gennaro2015}. 
At $T_{s}\approx 150\,\K$, this time is in excess of $10^{4}$ years 
for $10\,\mathrm{m}$ solids, but much shorter for centimeter-sized
particles.
Since the gas temperature in the proximity of a planet is higher 
than the average disk temperature at $r=a_{p}$ \citepalias{gennaro2013}, 
particles can shed mass before being accreted. This process does 
indeed occur, typically reducing the accreted mass by 
$\lesssim 15$\% at $a_{p}=5\,\AU$ and by $\lesssim 7$\% at 
$a_{p}=10\,\AU$. The ablated mass is expected to increase 
in warmer disks and as $\Mp$ increases.

\vspace*{2pt}
The accretion efficiencies obtained from the RHD calculations are 
plotted in Figures~\ref{fig:zeta_rhd1} and \ref{fig:zeta_rhd510}. 
At $a_{p}=1\,\AU$, the $\Mp=3\,\Mearth$ planet typically accretes
rocks with efficiencies $\lesssim 10$\%, with few exceptions
(\cifig{fig:zeta_rhd1}, top panel).
The largest, $10\,\mathrm{m}$ particles can also be segregated, 
more often in the $\Mp=20\,\Mearth$ planet cases (see the bottom panel).
In these latter models, when particles accrete, they do so at high efficiency.

\vspace*{2pt}
At $a_{p}=5$ and $10\,\AU$, there are some similarities among 
the models in terms of accretion efficiency (see \cifig{fig:zeta_rhd510}). 
Particles of $1\,\mathrm{cm}$ in radius are accreted inefficiently
(a few percent or less). 
The missing symbols in the figure indicate zero efficiency, 
but not necessarily segregation in exterior orbits.
Solids of $10$ and $100\,\mathrm{cm}$ in radius are generally accreted 
at an efficiency $\zeta$ of a few to several percent, with peaks of $\approx 10$\%. 
The $10\,\mathrm{m}$ particles are accreted efficiently, when accreted, 
with $0.5\lesssim\zeta\lesssim 0.8$. 
However, in a number of cases ($\Mp\ge 10\,\Mearth$), $10\,\mathrm{m}$ 
icy and rocky solids are segregated in the exterior disk (see also
\cifig{fig:as_seg}).

\vspace*{2pt}
Overall, the results in Figures~\ref{fig:zeta_rhd1} and \ref{fig:zeta_rhd510}
are in general accordance with those from the simplified calculations shown in
\cifig{fig:mdh}. 
The values of $\zeta$ for $1$, $10$, and $100\,\mathrm{cm}$ solids 
are comparable. The typically large efficiency for the accretion 
of $10\,\mathrm{m}$ particles (when not segregated) is also in 
reasonable agreement.

\vspace*{2pt}
\citet{morbidelli2012} estimated small solids' accretion on $\Mp=1$ 
and $5\,\Mearth$ planets with $a_{p}=0.8\,\AU$. 
They found efficiencies ranging from a few to several percent 
for $R_{s}\approx 10$--$100\,\mathrm{cm}$ particles. 
They also found that $10\,\mathrm{m}$ solids have much larger
accretion efficiencies, in excess of $\approx 50$\%. 
These predictions are generally consistent with those shown in
Figures~\ref{fig:zeta_rhd1} and \ref{fig:zeta_rhd510}. 
Although they used an unperturbed surface density twice as large
as the largest density applied in the RHD calculations
(see \cifig{fig:mig}), the value of $\tau_{s}\Omega_{\mathrm{K}}$ 
for these particles was comparable, $\approx 0.1$--$100$.

\vspace*{2pt}
\citet{picogna2018} estimated particle accretion efficiencies
from hydrodynamics calculations of laminar and turbulent disks.
In models of $5$ and $10\,\Mearth$ planets,
they found $\zeta \approx 0.2$ at $\tau_{s}\Omega_{\mathrm{K}}\approx 0.01$
and $\approx 0.4$ at $\tau_{s}\Omega_{\mathrm{K}}\approx 100$, with a minimum
around a few percent at $\tau_{s}\Omega_{\mathrm{K}}\sim 1$.
Their results are similar to those presented in \cifig{fig:zeta_rhd510}
in the range $0.1\lesssim\tau_{s}\Omega_{\mathrm{K}}\lesssim 10$,
although their estimate of $\zeta$ is somewhat larger for $\approx 1\,\mathrm{cm}$ particles 
and somewhat smaller for $\approx 10\,\mathrm{m}$ particles than
reported herein.

\vspace*{2pt}
At an accretion rate of 
$\dot{M_{s}}\sim 10^{-5}\,\mathrm{\Mearth\,yr^{-1}}$
(see \cifig{fig:dMsdt_rs}) and $\zeta\approx 0.1$, a $3\,\Mearth$
planet at $1\,\AU$ would double its mass in $\sim 10^{6}\,\mathrm{yr}$, 
whereas a $20\,\Mearth$ planet could grow by up to $1\,\Mearth$ every 
$10^{5}\,\mathrm{yr}$ ($\zeta\approx 1$), if not isolated.
At $5$ and $10\,\AU$, by accreting either icy or rocky 
$R_{s}\le 100\,\mathrm{cm}$ solids, the accretion rate of 
a $5\,\Mearth$ planet would be 
$\dMp\lesssim 10^{-6}\,\mathrm{\Mearth\,yr^{-1}}$, 
but it could be many times as large if it accreted 
$10\,\mathrm{m}$ particles, resulting in a mass-doubling timescale 
of several times $10^{5}\,\mathrm{yr}$.
More massive, $10$--$15\,\Mearth$ planets would accrete at comparable 
rates, although the interaction with $10\,\mathrm{m}$ solids would 
typically lead to segregation rather than to accretion.

\section{Accretion Efficiency of Bare Cores}
\label{sec:SPC}

\citet{kary1993} found that the accretion efficiency is inversely 
proportional to the radial velocity of a particle, defined by
\cieq{eq:apeale}, in units of Hill radius per synodic period 
(at $\Delta r=2\sqrt{3}\Rhill$), i.e.,
\begin{equation}
  \zeta_{\mathrm{KLG}}\propto \frac{1}{K}\left(\frac{a_{p}}{H_{g}}\right)^{4}%
                                                 \left(\frac{\Rhill}{a_{p}}\right)^{2},
  \label{eq:zeta_kary}
\end{equation}
hence $\zeta_{\mathrm{KLG}}\propto 1/K$ (see \cieq{eq:Kappa}), 
for given planet and disk conditions, and proportional to 
$R_{s}/C_{D}$ for a given material. 
Their calculations and those presented in Sections~\ref{sec:DPK} and \ref{sec:SEA}
are most appropriate in the case of small tidal perturbations by the planet,
i.e., when
\begin{equation}
  \left(\frac{\Mp}{\Ms}\right)^{2}<3\pi\alpha_{g}\left(\frac{H_{g}}{a_{p}}\right)^{5}.
  \label{eq:notides}
\end{equation}
If $\alpha_{g}\approx 10^{-3}$ and $H_{g}/a_{p}=0.022(a_{p}/\AU)^{2/7}$, 
this limit would require $\Mp/\Ms< 2\times 10^{-5}$, or $\approx 6\,\Mearth$
at $5\,\AU$ from a solar-mass star.

A set of calculations, similar to those discussed in \cisec{sec:DPK}, 
were performed to evaluate the accretion efficiency of $R_{s}=1$, $10$, $100$, 
and $1000\,\mathrm{cm}$ rocky/icy solids on small cores, 
$\Mp\le 1\,\Mearth$, orbiting at $a_{p}=1$, $5$, and $10\,\AU$ from 
a $1\,\Msun$ star. The nebula model is based on that of \citet{chiang2010}.
The cores are ``bare'' in the sense that they do not bear a bound 
atmosphere  (of significant mass), an appropriate approximation at such low masses
\citep[see, e.g.,][]{kuwahara2024}. Therefore, these planets are assumed 
to be composed entirely of condensible materials and solids accrete when 
they hit the condensed surface of the planet. Although there is no local
enhancement of the gas density close to the planet, gas drag operates 
all the way down to its surface, effectively mimicking a low-density atmosphere.

\begin{figure}
\centering%
\resizebox{\linewidth}{!}{\includegraphics[clip]{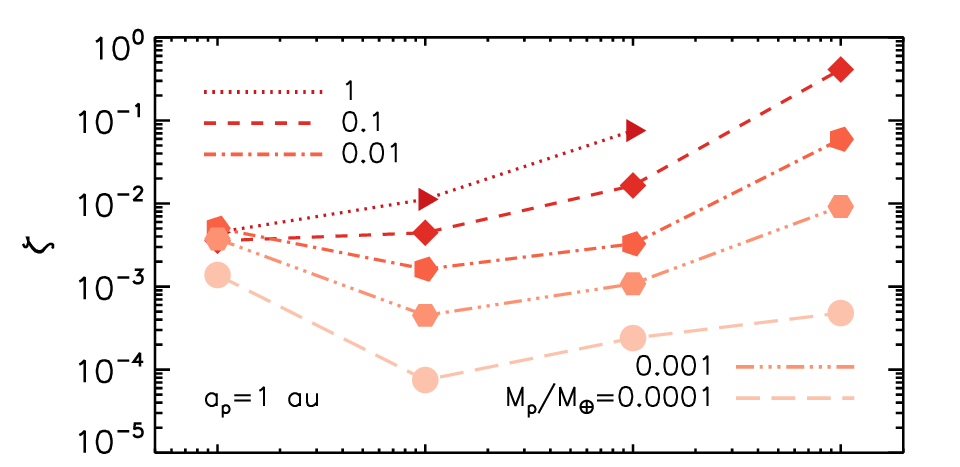}}
\resizebox{\linewidth}{!}{\includegraphics[clip]{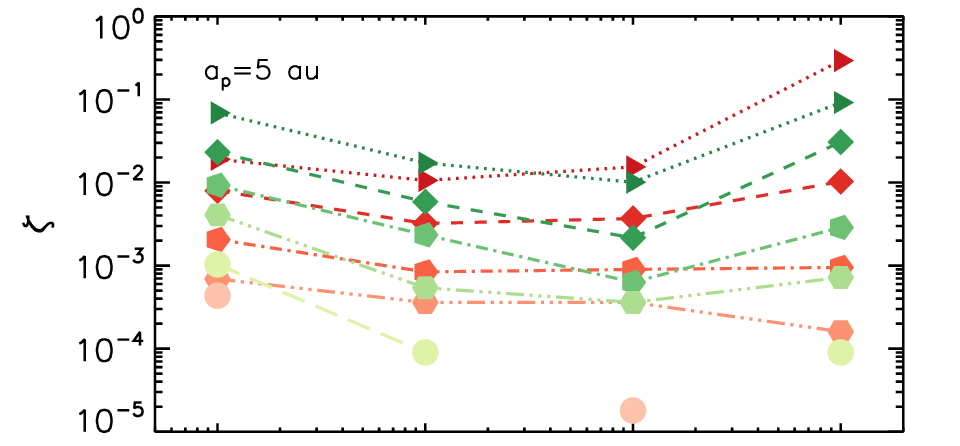}}
\resizebox{\linewidth}{!}{\includegraphics[clip]{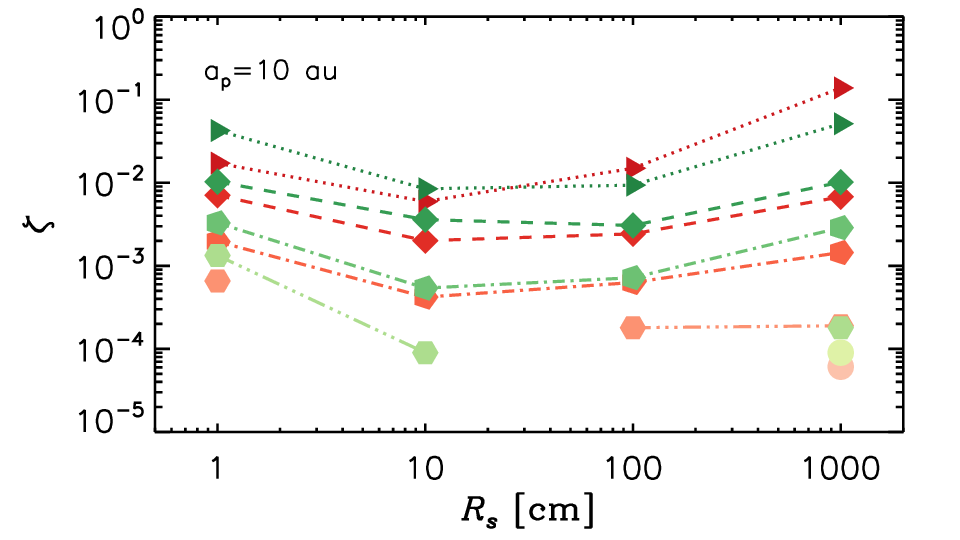}}
\caption{%
             Accretion efficiencies of bare cores (i.e., without an
             atmosphere), versus particle radius, for rocky (red colors)
             and icy particles (green colors).
             The planet's semi-major axis is $a_{p}=1$ (top), $5$ (middle), 
             and $10\,\AU$ (bottom), and the mass is indicated in the legend.
             The disk model is the same as that in \cifig{fig:sigs_rs}.
             Missing symbols indicate $\zeta=0$.
             }
\label{fig:peffi}
\end{figure}
Results from these calculations are illustrated in \cifig{fig:peffi}. 
The efficiency $\zeta$ is plotted as a function of $R_{s}$, for each 
core mass (see legend of the top panel). 
The missing symbols indicate that no collision was detected ($\zeta=0$).
The largest particles tend to accrete more efficiently, although 
they can become trapped in exterior mean-motion resonances 
(see the $\Mp=1\,\Mearth$ case in the top panel). 
The mass scaling in \cifig{fig:peffi} agrees reasonably well 
with the predictions of Equations~(\ref{eq:zeta}) and 
(\ref{eq:zeta_kary}), $\zeta\propto M_{p}^{2/3}$, at $a_{p}=5$
and $10\,\AU$ ($\zeta/\Mp^{2/3}$ varies by a factor $\lesssim 10$).
At $a_{p}=1\,\AU$, this scaling is restricted  to particles with
$\tau_{s}\Omega_{\mathrm{K}}\gtrsim 0.1$.

\section{Formation and Structure Models}
\label{sec:FSM}

The flow rate of $1$--$1000\,\mathrm{cm}$ solids through a disk, 
$\dot{M}_{s}=\dot{M}_{s}(t,R_{s})$, obtained from
\cieq{eq:sigs} (see, e.g., \cifig{fig:dMsdt_rs}) and the efficiencies 
illustrated in Figures~\ref{fig:zeta_rhd1}, \ref{fig:zeta_rhd510}, and \ref{fig:peffi}
(appropriately interpolated in planet mass), $\zeta=\zeta(\Mp,R_{s})$, 
are combined to provide a size-integrated accretion rate,
\begin{equation}
  \dot{M}_{Z}(\Mp,t)=\int\zeta(\Mp,R_{s})\,d\dot{M}_{s}(t,R_{s}),
  \label{eq:dotMZ}
\end{equation}
to formation calculations at $1$, $5$, and $10\,\AU$. 
Accretion of rocks ($a_{p}=1$, $5\,\AU$) and ice 
($a_{p}=5$, $10\,\AU$) is considered in separate models.
The total mass of the disk is $\approx 0.055\,\Msun$ at time $t=0$.
The initial gas-to-solids mass ratio is $100$ and $71.5$ 
\citep{pollack1994} in the rocky and icy disk, respectively.
Formation starts from Moon-mass embryos, $\Mp=10^{-2}\,\Mearth$
at $t=1000$ orbits (at each orbital distance).
Assuming mass equipartition among particle sizes, these embryos 
take $\approx 0.25$ to $\approx 0.55\,\Myr$ to attain 
$\approx 1\,\Mearth$, when structure calculations begin 
($\dot{M}_{p}=\dot{M}_{Z}$ at smaller masses). 

Planet evolution is modeled as in \citet{gennaro2014}, by solving 
the structure equations in spherical symmetry \citep{kippenhahn2013},
including mass and energy deposited by incoming solids and gas, 
particle break-up and ablation. Dust opacity in the envelope is
computed self-consistently with the sedimentation and coagulation 
of grains released by accreted gas and solids \citep{naor2010}.
Dissolution and mixing of heavy elements in H-He gas 
\citep[e.g.,][]{bodenheimer2018} is neglected and all heavy 
elements settle to the condensed core.
When the planet is in contact with the nebula, its radius \Rp\ is
evaluated as in \citet{lissauer2009}, and depends on both the Bondi 
and Hill radius. Mass loss is enabled when the envelope overflows
its boundary. Disk-limited accretion of gas is accounted for as in
\citet[][]{gennaro2016}. Disk evolution is modeled as outlined
in Section~\ref{sec:DUK} (see \cifig{fig:sigs_rs}).
Formation continues until gas dispersal, at $t\approx 3.5\,\Myr$. 

\begin{figure}
\centering%
\resizebox{\linewidth}{!}{\includegraphics[clip]{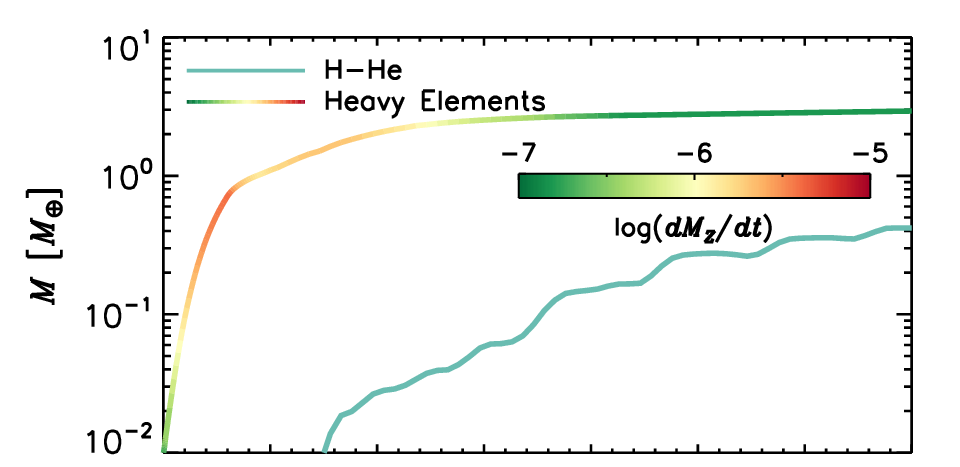}}
\resizebox{\linewidth}{!}{\includegraphics[clip]{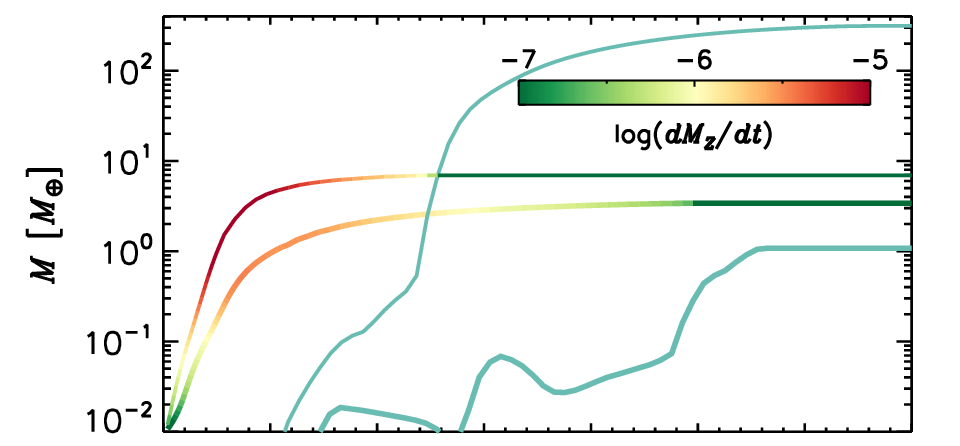}}
\resizebox{\linewidth}{!}{\includegraphics[clip]{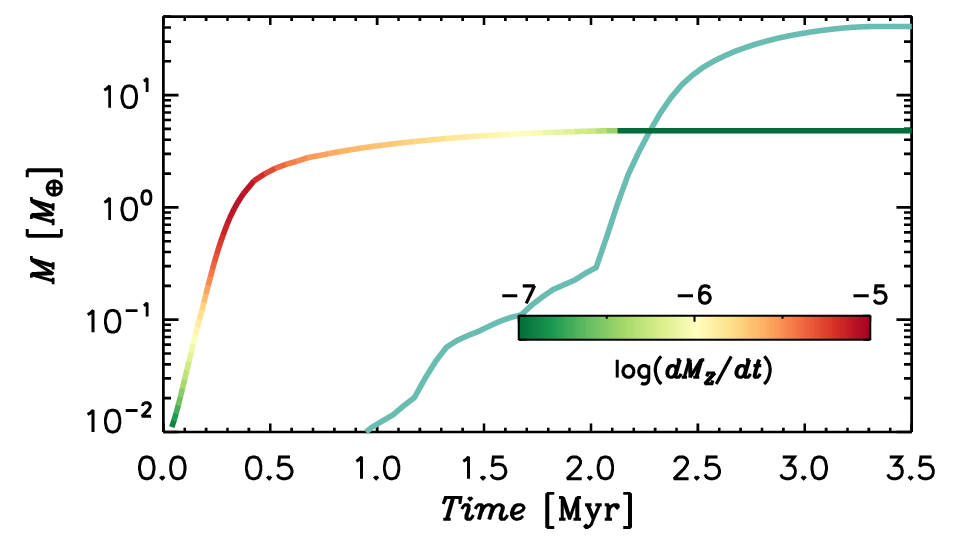}}
\caption{%
             Heavy-element and H-He masses obtained from formation
             models at $a_{p}=1$ (top), $5$ (middle), and $10\,\AU$ (bottom).
             The heavy-element mass is color-coded by the solids'
             accretion rate, in units of $\Mearth\,\mathrm{yr}^{-1}$.
             The thinner lines in the middle panel refer to the planet
             accreting icy particles.
             }
\label{fig:mp}
\end{figure}
The heavy-element and H-He masses of the planets are shown in
\cifig{fig:mp}.
Models at $1$ and $5\,\AU$ accreting silicates achieve heavy-element masses
$M_{Z}\approx 3$ and $3.4\,\Mearth$, and H-He inventories contributing
respectively $12$\% and $24$\% of the total masses. Instead, models
accreting ice become gas giants, $\Mp=321\,\Mearth$ 
($M_{Z}=6.9\,\Mearth$) at $5\,\AU$ and $\Mp=46\,\Mearth$
($M_{Z}=4.8\,\Mearth$) at $10\,\AU$.
For an initial gas-to-solids mass ratio equal to $100$, 
the heavy-element inventory at this latter distance would be 
limited to $\approx 1$ or $2.5\,\Mearth$ by accreting rocks 
or ice, respectively.

The different outcomes at $5\,\AU$ can be attributed to
the different amounts of solids transported across the planet's
orbit during formation: $\approx 150\,\Mearth$ in the case of ice
and $\approx 90\,\Mearth$ in the case of rocks. Had the initial 
gas-to-solids mass ratio in the former case been $100$ (as in the
rocky disk), this mass would have been $\approx 107\,\Mearth$. 
The cumulative accretion efficiency, i.e., $M_{Z}$ divided by the
amount of solids crossing the orbit of the planet, also contributes,
with that of icy particles ($\approx 0.046$) exceeding by about
$20$\% that of rocky particles.
In both models, most of the heavy-element mass ($\approx 70$\%)
is delivered by $10\,\mathrm{m}$ boulders/ice blocks. Only $10$\%--$15$\%
of $M_{Z}$ is supplied by $10$--$100\,\mathrm{cm}$ solids.

In the models at $1$ and $10\,\AU$, the total solid mass crossing 
the planet's orbit is $\approx 103\,\Mearth$ (of rocks) and $136\,\Mearth$
(of ice), respectively,
corresponding to cumulative accretion efficiencies of $0.03$ and $0.04$.
In these two calculations, about half of $M_{Z}$ is delivered by 
$100\,\mathrm{cm}$ boulders, at $1\,\AU$, and by $10\,\mathrm{m}$
ice blocks, at $10\,\AU$.
Note that a $50/50$ solid mixture (by mass) of ice and silicates
would have a material density closer to that of ice and, therefore,
is expected to behave somewhat closer to ice than to silicates
in terms of transport and accretion.

\begin{figure*}
\centering%
\resizebox{\linewidth}{!}{\includegraphics[clip]{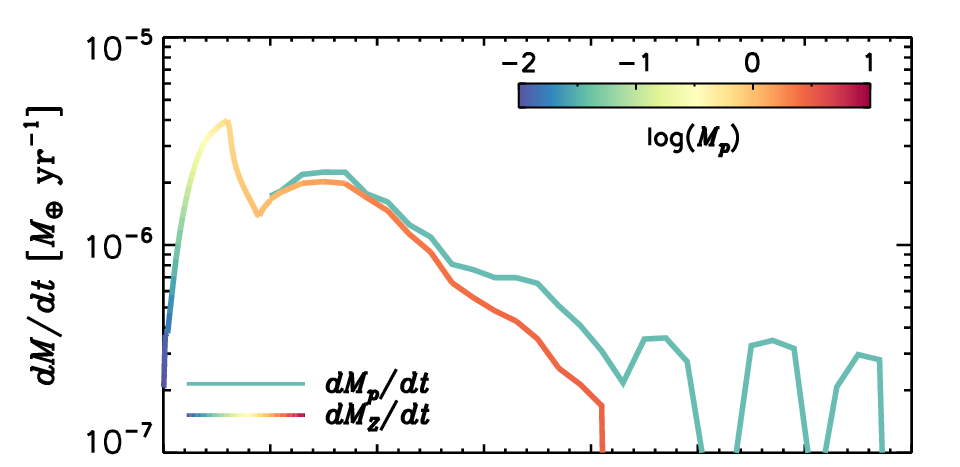}
                          \includegraphics[clip]{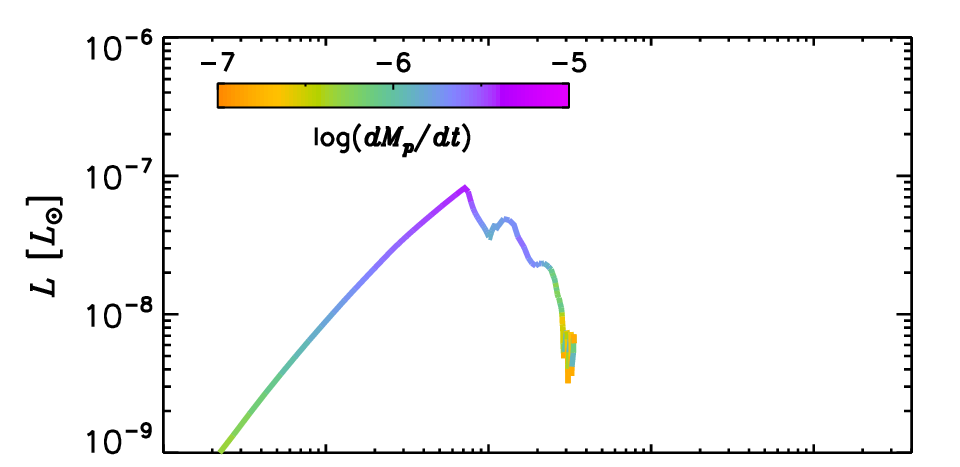}}
\resizebox{\linewidth}{!}{\includegraphics[clip]{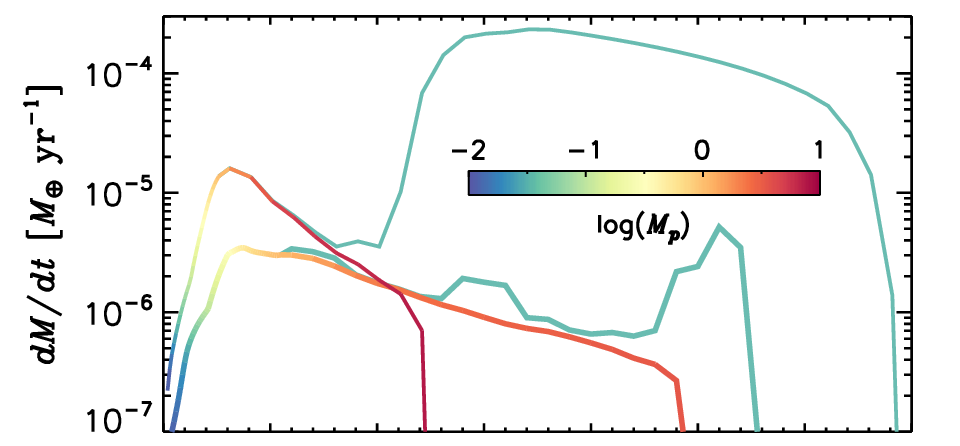}
                          \includegraphics[clip]{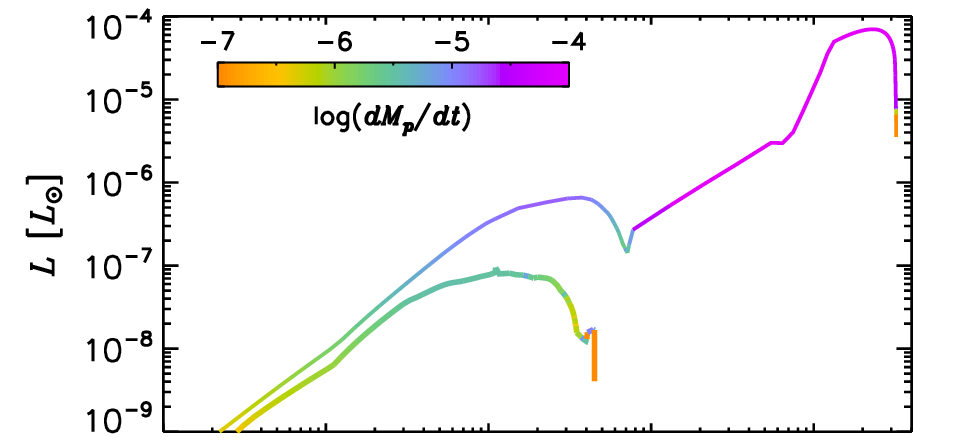}}
\resizebox{\linewidth}{!}{\includegraphics[clip]{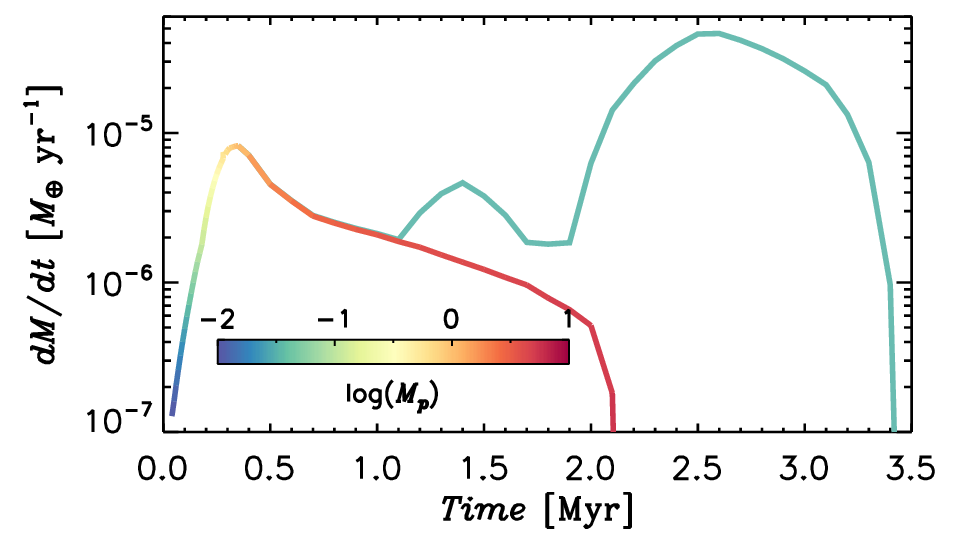}
                          \includegraphics[clip]{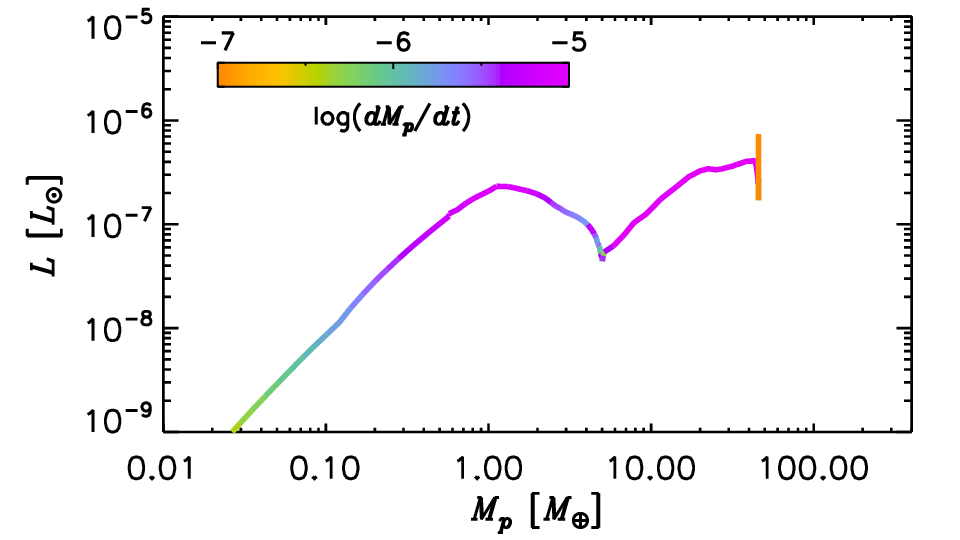}}
\caption{%
             Accretion rates versus time (left) and luminosity versus
             mass (right) from formation models at $a_{p}=1$ (top), $5$ (middle),
             and $10\,\AU$ (bottom).
             The accretion rate of solids is color-coded by the planet
             mass (in units of $\Mearth$) and the luminosity by the total
             accretion rate (in units of $\Mearth\,\mathrm{yr}^{-1}$).
             The thinner lines in the middle panels refer to the planet accreting
             icy particles.
             }
\label{fig:lp}
\end{figure*}
The H-He mass at the beginning of the structure calculations,
$\approx 10^{-5}\,\Mearth$, is negligible and remains so for 
$\sim 1\,\Myr$ (see \cifig{fig:mp}). In fact, the total accretion
rate $\dot{M}_{p}$ remains close to $\dot{M}_{Z}$ for a large part
of the formation epoch (see left panels of \cifig{fig:lp}).
The maximum of $\dot{M}_{Z}$ is achieved between $\Mp=0.5$ and $1\,\Mearth$
when heavy elements are supplied by silicates, and between $\Mp=1$ 
and $2\,\Mearth$ when they are supplied by ice.
At $1\,\AU$, the accretion rate of gas exceeds $50$\% that of solids
relatively late in the evolution, after $1.6\,\mathrm{Myr}$, 
when $\dot{M}_{Z}\lesssim 4\times 10^{-7}\,\mathrm{\Mearth\,yr^{-1}}$.
The planet accreting rocks at $5\,\AU$ first reaches this condition
around $1.5\,\mathrm{Myr}$ 
($\dot{M}_{Z}\approx 10^{-6}\,\mathrm{\Mearth\,yr^{-1}}$; see the middle panel),
and then again when
$\dot{M}_{Z}\lesssim 4\times 10^{-7}\,\mathrm{\Mearth\,yr^{-1}}$.
Mass loss by overflow is significant in this case during the
first $\approx 2\,\Myr$ of evolution.
In the models resulting in the formation of gas giants, the accretion rate of gas 
exceeds $\dot{M}_{Z}$ after $\approx 1\,\mathrm{Myr}$, when the heavy-element mass $M_{Z}$
has nearly attained its final value (see \cifig{fig:mp}).

Instead of mass equipartition, if all of the initial mass had been 
carried by particles of a single size, in general $M_{Z}$ would have been
largest by accreting $R_{s}=1$--$10\,\mathrm{m}$ boulders/ice blocks.
However, the final heavy-element inventory of a planet is not only
determined by the mass distribution of particles, but also by
the accretion of gas, since the accretion efficiency $\zeta$ is a function of \Mp.

The luminosity during formation is shown in the right panels
of \cifig{fig:lp}. Prior to beginning the structure calculations 
($\Mp\lesssim 1\,\Mearth$), the luminosity is computed as
$L=G M_{Z}\dot{M}_{Z}/\Rp$, where \Rp\ is the radius of the condensed
core. The curves are color-coded by the total accretion rate.
The first luminosity peak corresponds to the maximum of $\dot{M}_{Z}$.
For the gas giants, a second peak occurs around the maximum of the
gas accretion rate.
Although the planet accreting silicates at $5\,\AU$ exhibits a maximum
in gas accretion around $2.6\,\Myr$, the corresponding luminosity peak
is smaller than the first, because of the larger planet radius. In fact,
both lower-mass planets (i.e., those accreting silicates) remain in contact with the disk until
it completely disperses. Therefore, their envelopes are extended
throughout formation. The gas giants, instead, detach from the disk
when $t\approx 1.2$ and $2.2\,\Myr$ at $5$ and $10\,\AU$, respectively.
They undergo disk-limited accretion thereafter.

The dissolution of heavy elements in the interior of the planets may
produce a stratification in composition and possibly affect energy
transport and temperatures \citep[e.g.,][]{bodenheimer2018,stevenson2022}.
Although heavy-element deposition resulting from the accretion
of small solids can differ from that resulting from planetesimal
accretion, the radial composition of an interior may not provide
distinctive information on the original carriers after sustained
mixing. Moreover, the deposition of refractory material may not 
clearly differentiate between the accretion of small and large solids,
as deposition patterns in the planet's envelope can overlap (see Appendix~\ref{sec:dep}). 

\section{Discussion and Conclusions}
\label{sec:DC}

The accretion of small solids
($10^{-2}\lesssim \tau_{s}\Omega_{\mathrm{K}}\lesssim 10^{2}$) 
on a planet depends on the transport of solids through the disk,
$dM_{s}/dt$, and on the efficiency or probability of accretion, $\zeta$ 
(see \cieq{eq:dotMZ}). The former quantity is non-local, and 
determined by the global evolution of the disk's gas, the initial
distribution of solids, coagulation, and fragmentation. 
The latter quantity (i.e., the fraction of $dM_{s}/dt$ at $a_{p}$
that is accreted by the planet) is determined by multiple
processes, at orbital or smaller scales.
Small cores without an envelope of significant mass) allow for relatively straightforward calculations
of $\zeta$, including planet-induced velocity perturbations on 
the gas (see \cisec{sec:SPC}). However, as the planet grows larger,
the computation of $\zeta$ becomes increasingly complex. In fact,
tidal interactions between the planet and the disk have important
effects on the dynamics of the solids. Close-range, 3D gas thermodynamics is also
expected to significantly affect $\zeta$.
Tidal segregation (and capture in mean-motion resonances) can impede 
the flux of solids through the planet's orbit, a process that
depends on solid size and composition, planet mass, and on the
gas thermal state (and, therefore, on disk age and mass).
The epoch of accretion is also limited by the duration of the solids' 
transport across the planet-forming region.
Boulder-sized solids tend to accrete more efficiently than smaller solids,
unless they become segregated. 
A disk of initial mass $0.055\,\Msun$ in gas and $\approx 260\Mearth$ 
in solids would provide typical accretion rates of heavy elements
(in the $1$--$10\,\AU$ region) between 
$10^{-6}$ and $10^{-5}\,\mathrm{\Mearth\,yr^{-1}}$, for the first 
$1$--$1.5\,\Myr$ of evolution, and $\lesssim 10^{-6}\,\mathrm{\Mearth\,yr^{-1}}$
thereafter (see \cifig{fig:lp}).

The results presented in Figures~\ref{fig:sigs_rs} and \ref{fig:dMsdt_rs}
are specific to the applied (global) disk conditions, but appropriate
solutions can be found by integrating Equations~ \ref{eq:dotM_s} and 
\ref{eq:sigs} in relevant gaseous disks. The efficiencies of accretion
presented in Figures~\ref{fig:zeta_rhd1} and \ref{fig:zeta_rhd510}
are more general, since they only depend
on local quantities, and can be used as long as the gas densities and temperatures
are comparable to those found in the RHD calculations (see \citab{table:dat}).
The same considerations apply to the efficiencies of small cores with no significant envelope
reported in \cifig{fig:peffi}.

The formation calculations presented herein apply accretion rates of solids
based on simple, but consistent, models of gas and solid disk evolution
and first-principles determinations of accretion efficiencies.
The structure calculations include detailed computations of grain opacity
as the solids travel through the envelope and release material.
The outcomes encompass a broad range, from low-mass, heavy-element-dominated
planets to H-He gas giants. 
Heavy-element masses from $3$ to $7\,\Mearth$ require an amount of
solids from $\approx 90$ to $150\,\Mearth$ to cross the planet's orbit.
The maximum luminosity emitted during the epoch of solids' accretion
ranges from $\approx 10^{-7}$ to $\approx 10^{-6}\,L_{\odot}$.

Cumulative efficiencies of accretion in the formation calculations
are below $5$\%. Multiple planets may form if solids can drift inward,
past a planet. Heavy elements would be delivered
during the epoch when $d{M}_{s}/dt$ at a given distance is non-negligible.
As outer planets remove solids, lesser amounts would be available to
inner planets. The higher the efficiency of accretion in a size range, the
lower is the availability of solids to planets downstream.
In addition, because of segregation, a planet large enough (a few to several
times \Mearth; see \cifig{fig:as_seg}) could hinder or terminate 
the supply of solids (in some size ranges) to inner planets.
In summary, predictions of heavy-element delivery (to one or more planets)
require accounting for both local and remote processes, over the disk lifetime.
Considerations of the evolving size distribution of solids are also necessary,
including at segregation sites, where enhanced comminution and/or growth may
release solids and promote further accretion.

\section*{Acknowledgments}

We thank an anonymous reviewer whose comments helped clarify and 
improve this manuscript. 
Support from NASA's Research Opportunities in Space and Earth Science, 
through grants 80HQTR19T0086 and 80HQTR21T0027 (G.D.) and grant NNX14AG92G
(P.B.), is gratefully acknowledged.
Computational resources supporting this work were provided by the NASA 
High-End Computing Capability (HECC) Program through the NASA Advanced 
Supercomputing (NAS) Division at Ames Research Center.



\appendix

\section{Evolution of Solids in Quasi-Steady State Gas Fields}
\label{sec:test}

\begin{figure*}
\centering%
\resizebox{\linewidth}{!}{\includegraphics[clip]{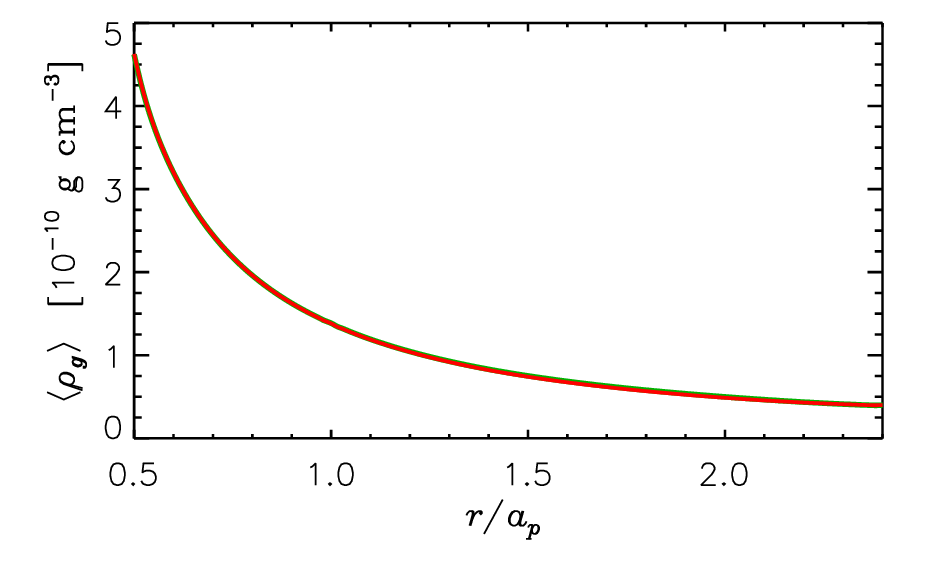}%
                          \includegraphics[clip]{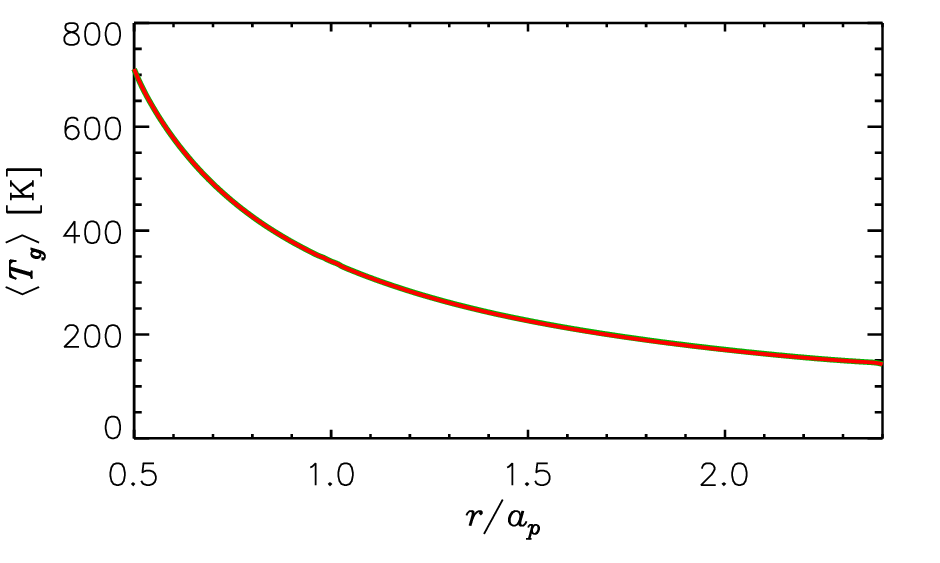}}
\caption{%
             Azimuthal averages (around the star) versus
             radial distance of the 
             midplane gas density (left) and midplane gas
             temperature (right) at the quasi-steady state
             $\mathcal{S}_{i}$ (green) and at the end 
             of the test calculation (red), several hundred 
             planet's orbits later. Curves at the two epochs
             overlap one another in both panels (for better
             visibility, the green curves are thicker than the red
             curves).
             }
\label{fig:sstst}
\end{figure*}

\begin{figure*}
\centering%
\resizebox{\linewidth}{!}{\includegraphics[clip]{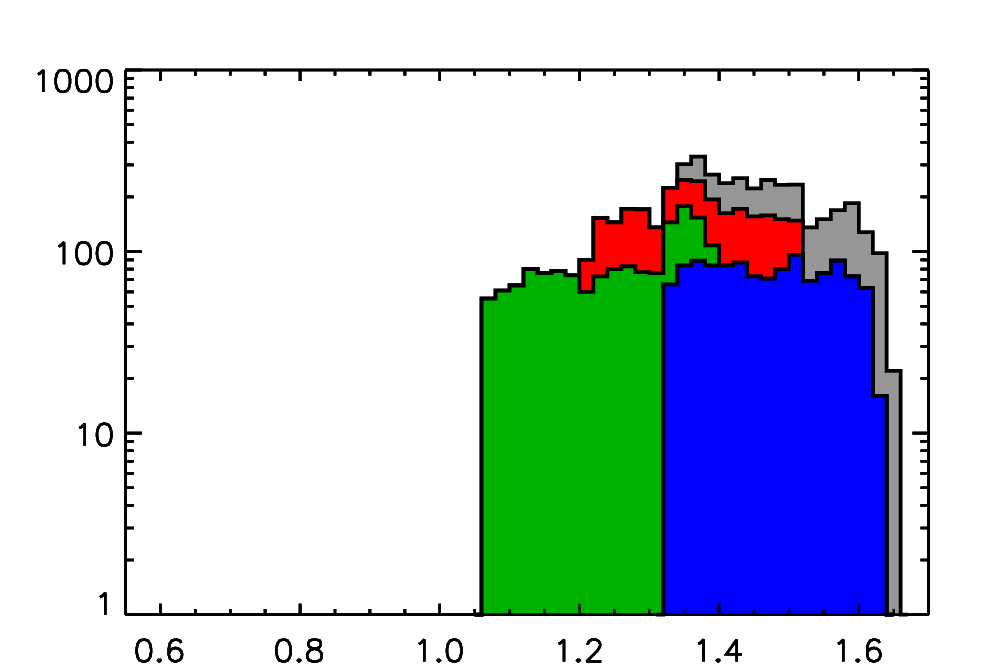}%
                          \includegraphics[clip]{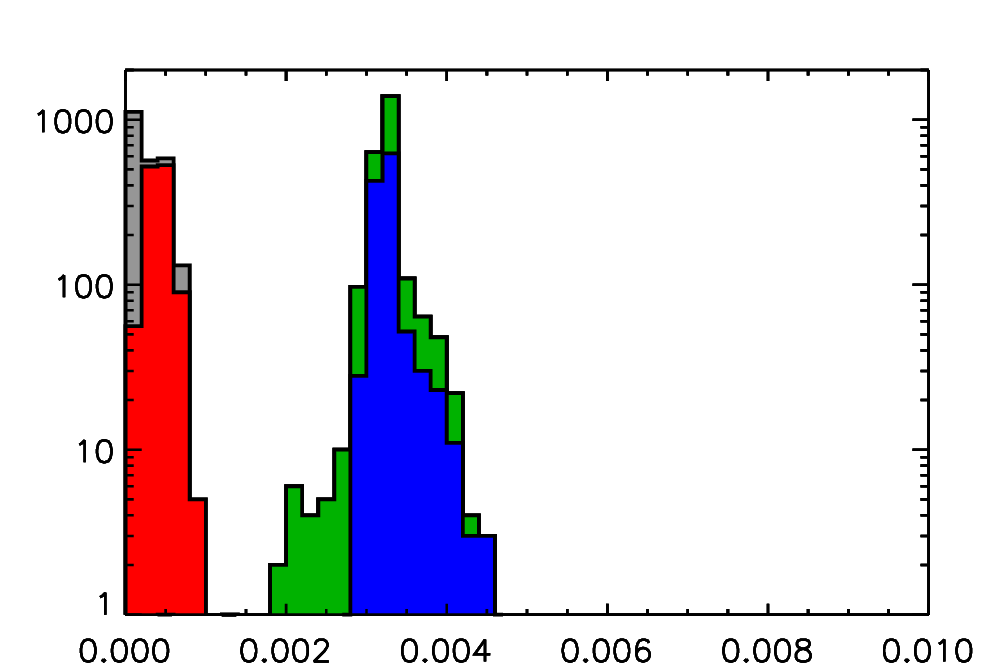}%
                          \includegraphics[clip]{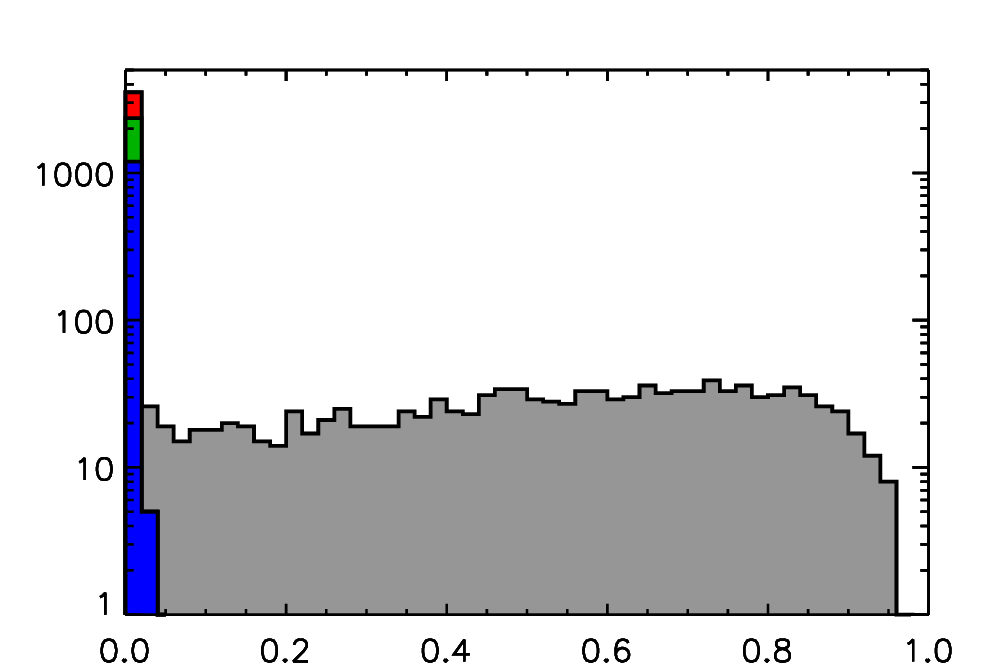}}
\resizebox{\linewidth}{!}{\includegraphics[clip]{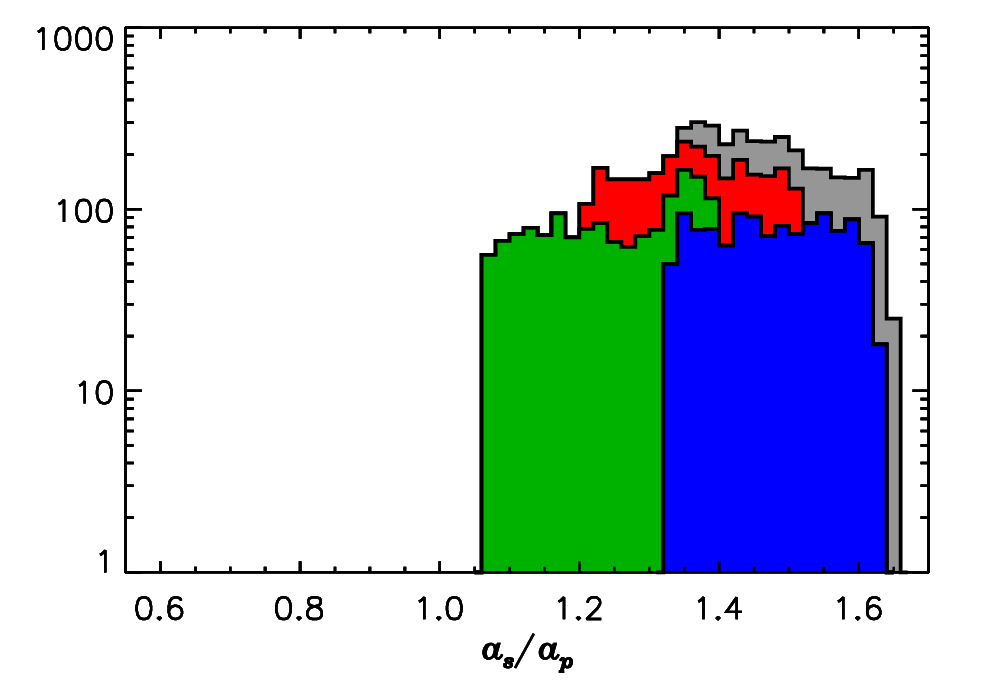}%
                          \includegraphics[clip]{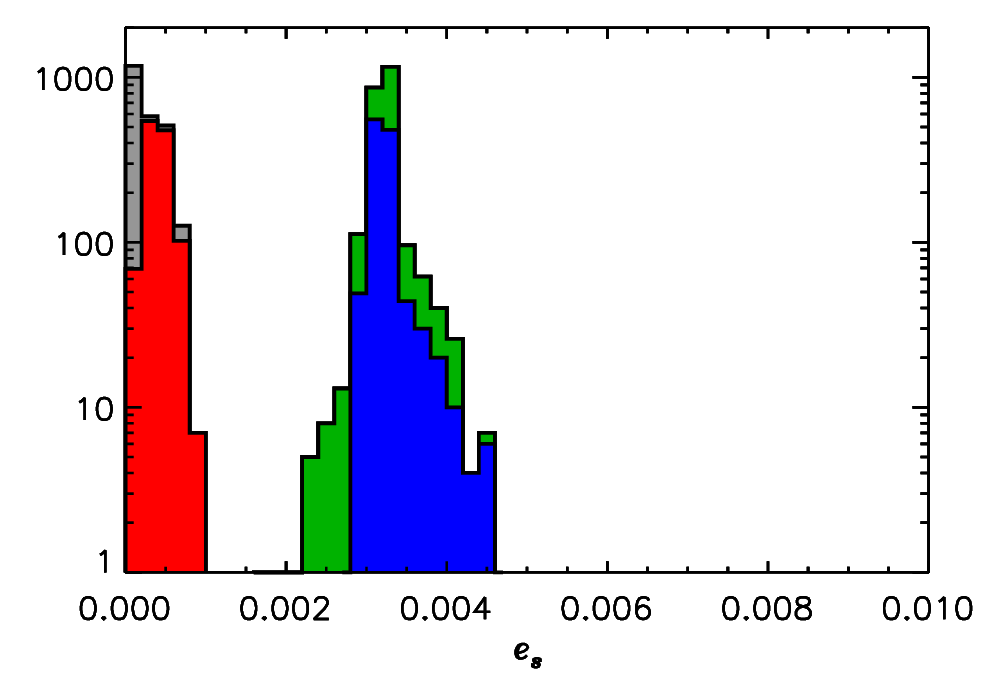}%
                          \includegraphics[clip]{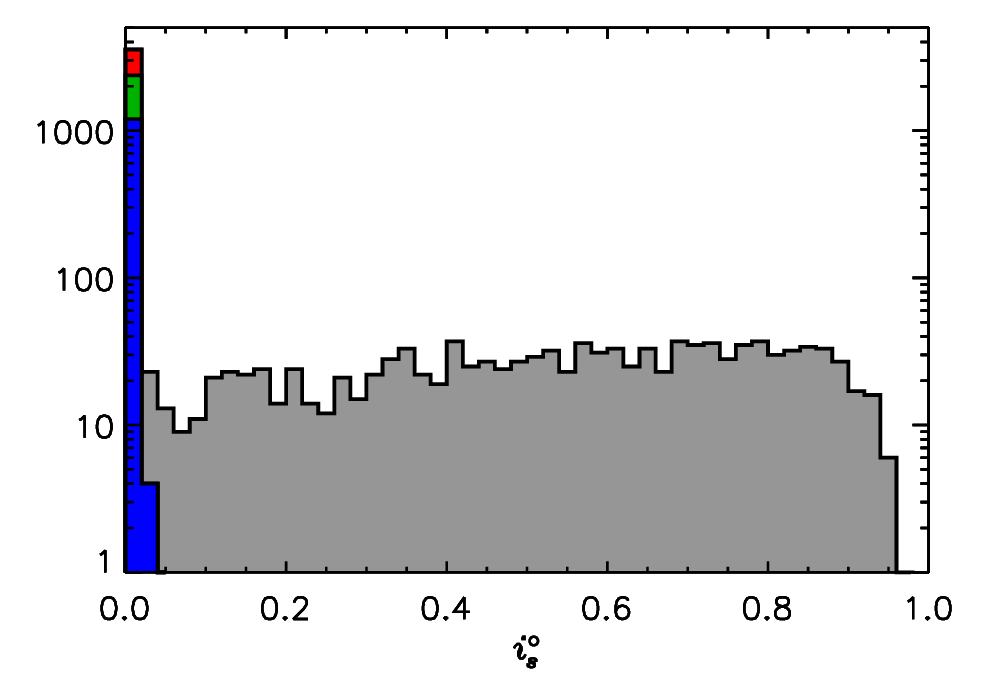}}
\caption{%
             Histograms of the particles' semi-major axes
             (left), eccentricities (center), and inclinations (right) 
             as they approach the planet's orbit. The top panels refer
             to the model in which the evolution of both the gas and 
             the solids is calculated. 
             In the bottom panels, the gas quasi-steady state
             $\mathcal{S}_{i}$ (defined in the text) is applied
             and only the evolution of the solids is calculated.
             The gray histograms include all particles;
             the red, green, and blue histograms include particles with
             radii $R_{s}\le 100\,\mathrm{cm}$,
             $R_{s}\le 10\,\mathrm{cm}$, and $R_{s}= 1\,\mathrm{cm}$,
             respectively.
             }
\label{fig:comp}
\end{figure*}

\begin{figure}
\centering%
\resizebox{\linewidth}{!}{\includegraphics[clip]{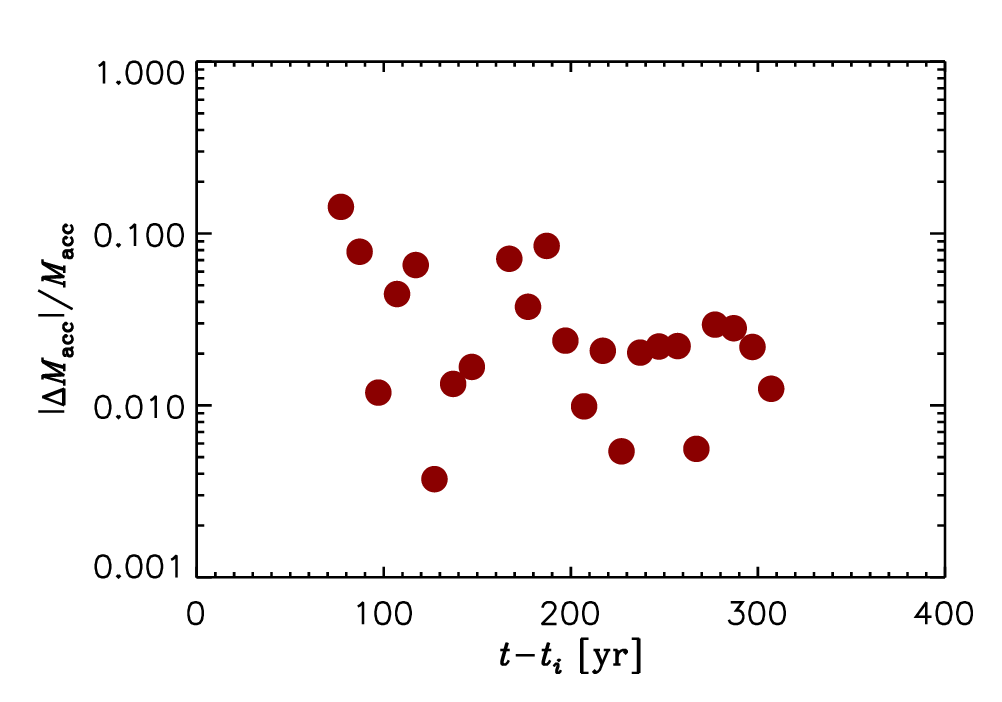}}
\caption{%
             Normalized relative difference in the accretion rate 
             of solids on the planet versus time, since the reference 
             time of state $\mathcal{S}_{i}$ (defined in the text), 
             between the two calculations described in the text. 
             Accretion first involves $R_{s}=10\,\mathrm{cm}$ particles and 
             then $R_{s}= 100\,\mathrm{cm}$ particles.
             }
\label{fig:acctst}
\end{figure}

The results of the RHD calculations presented in \cisec{sec:RHD} rely 
on the assumption that the evolution of the disk's gas can be neglected
while the solids evolve through the gas.
The assumption is justified by the fact that the gas fields
$(\gvec{u}_{g},\rho_{g},T_{g})$ 
are in a near-steady state in the frame co-rotating with the planet and
do not change significantly over the course of the calculations. 
The approach is intended to mitigate the computational requirements of
those RHD models and render the problem more tractable. Here a test is
presented on the validation of this approach.

An RHD calculation of a $3\,\Mearth$ planet, embedded in a disk
at $r=a_{p}=1\,\AU$ from a $1\,\Msun$ star, is performed and evolved
until the system attains a state of quasi-equilibrium, hereafter
referred to as state $\mathcal{S}_{i}$.
The setup of the model is equivalent to that of the models with
$a_{p}=1\,\AU$ discussed in \cisec{sec:RHD} 
(reference density $\langle \rho_{g}\rangle=1.5\times 10^{-10}\,\rhou$),
but at a somewhat lower grid resolution.
The planet radius is assumed to be approximately equal to \Rbondi.
Four size bins, 
$R_{s}=1$, $10$, $100$, and $1000\,\mathrm{cm}$, are populated with
equal numbers of rocky particles. They are deployed outside 
the planet's orbit by applying the same random process as described 
in \cisec{sec:SS}. The calculation is then continued, from state
$\mathcal{S}_{i}$, by evolving \emph{both} the gas and the solids for
several hundred orbits of the planet. During this time, the disk's 
gas density and temperature (red curves) do not deviate significantly
from those of state $\mathcal{S}_{i}$ (green curves), as illustrated
in \cifig{fig:sstst} (in which curves basically overlap and are difficult to distinguish).
In another calculation, the same distribution of solids evolves
in the gas fields provided by state $\mathcal{S}_{i}$.

The stopping times of the solids at deployment range from 
$\tau_{s}\Omega_{\mathrm{K}}\approx 10^{-2}$ ($R_{s}=1\,\mathrm{cm}$) 
to 
$\tau_{s}\Omega_{\mathrm{K}}\approx 10^{3}$ ($R_{s}=1000\,\mathrm{cm}$).
Intermediate size particles
($0.1\lesssim\tau_{s}\Omega_{\mathrm{K}}\lesssim 10$)
drift toward the planet relatively quickly.
For solids in the various size ranges,
\cifig{fig:comp} shows the histograms of the particles' 
semi-major axes (left),
orbital eccentricities (center), and inclinations (right) during 
the approaching phase (see the figure's caption for further details). 
The orbits remain nearly circular, although the smallest particles 
are somewhat more eccentric than the largest ones (see also
\cifig{fig:h55}). 
The inclination with respect to the midplane of the disk is basically zero, 
except for that of the largest particles, which is damped over 
a longer timescale 
\citep[$\propto R_{s}$,][]{adachi1976}.
In statistical terms, the orbital elements of the solids in the two 
calculations remain quantitatively similar. By the end of the 
simulations, the $R_{s}=10$ and $100\,\mathrm{cm}$ particles have 
either accreted on the planet or moved toward interior orbits,
whereas the smallest and largest
particles continue drifting toward the planet. 

In order to gauge variations in the accretion rates of solids on 
the planet,
\cifig{fig:acctst} shows the relative difference of the accreted mass 
versus time, since state $\mathcal{S}_{i}$, between the two calculations.
The $R_{s}=10\,\mathrm{cm}$ particles intercept the planet's orbit 
first, followed by the $R_{s}=100\,\mathrm{cm}$ particles. 
The differences in accreted mass are below $\approx 10$\%,
and typically at the level of a few percent on average. 
The accretion efficiencies $\zeta$ are also in close agreement, 
within $\approx 1$\%. The fact that the accretion rates of the
particles are
statistically consistent during close encounters with the planet 
implies that the particle dynamics is indeed similar in the two
simulated scenarios, even during the phases of the evolution
preceding accretion.

\section{Dissolution of Solids in Planetary Envelopes}
\label{sec:dep}
\begin{figure*}[]
\centering%
\resizebox{\linewidth}{!}{\includegraphics[clip]{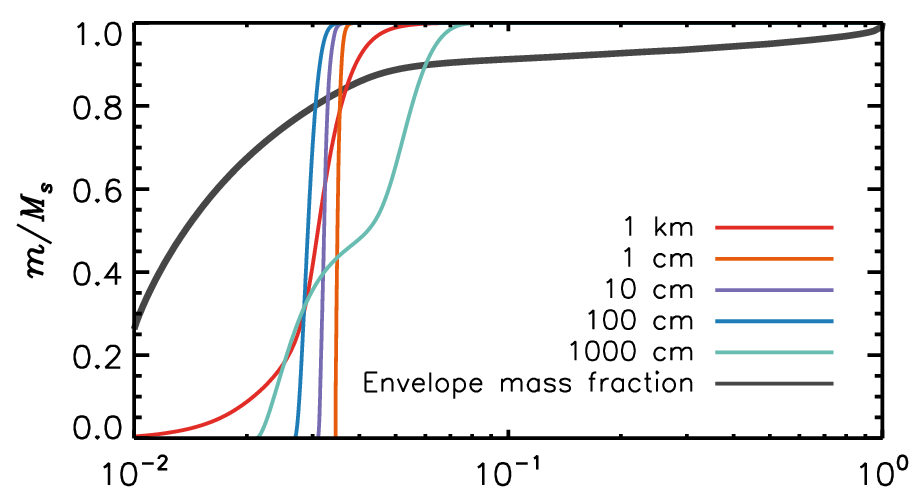}%
                          \includegraphics[clip]{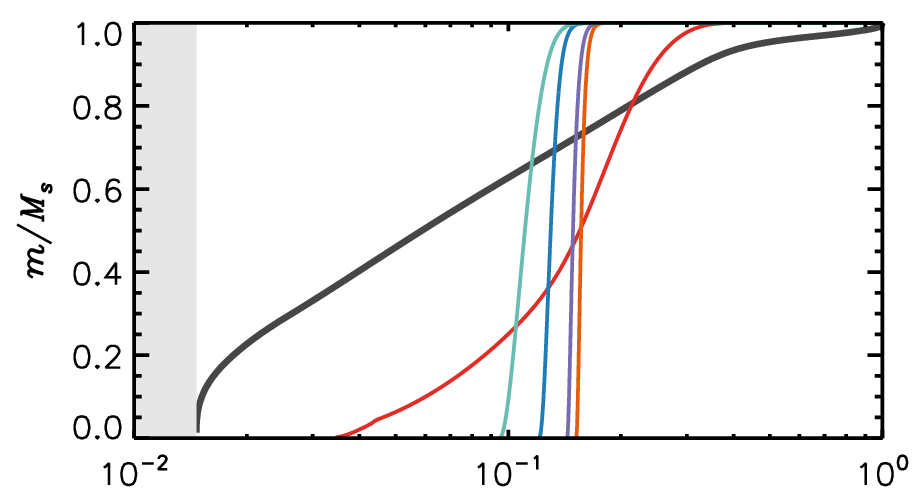}}
\resizebox{\linewidth}{!}{\includegraphics[clip]{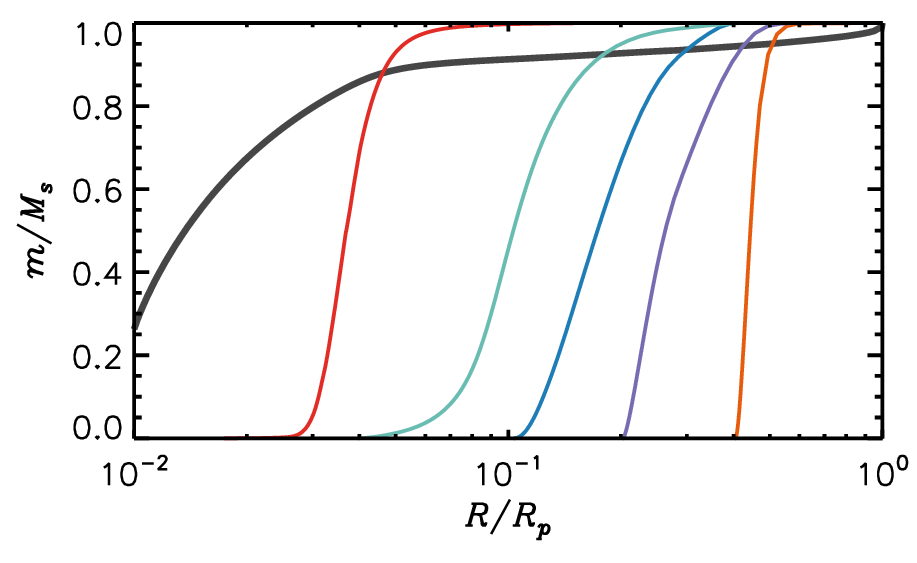}%
                          \includegraphics[clip]{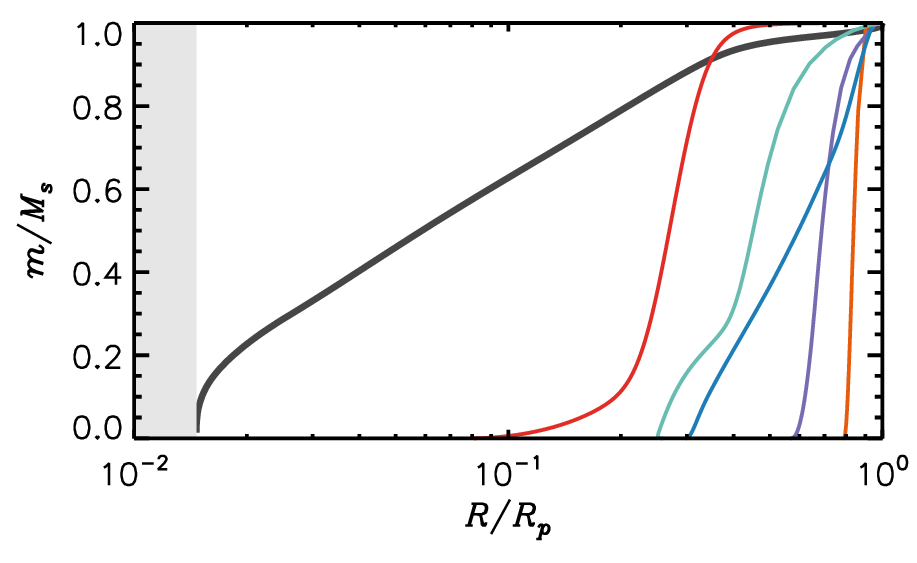}}
\caption{%
             Mass evolution of rocky (SiO$_{2}$; top) and icy solids (bottom),
             versus planet radius, as they ablate while descending
             in a low-mass ($0.15\,\Mearth$; left) and high-mass 
             ($6.9\,\Mearth$; right) envelope. The interior structures are taken
             from the giant planet model at $5\,\AU$, discussed in
             \cisec{sec:FSM}. The solids have different initial sizes,
             as indicated in the legend of the top left panel.
             The quantity $M_{s}$ refers to the initial mass entering
             the envelope.
             The thick gray lines indicate the envelope mass fraction.
             The shaded areas in the right panels represent the
             condensed core region. See the text for further details.
             }
\label{fig:dep}
\end{figure*}

Solid bodies moving through an atmosphere warm up and ablate,
disseminating heavy elements along their path. This process can
affect the composition of planetary atmospheres and interiors. We performed
experiments of solids' dissolution in planetary envelopes taken
from the giant planet formed at $5\,\AU$ and presented in \cisec{sec:FSM}.
The envelopes refer to times at which their masses are $0.15\,\Mearth$
($\Mp=6.8\,\Mearth$) and $6.9\,\Mearth$ ($\Mp=13.8\,\Mearth$).  
To circumvent segregation issues, solids were deployed around
the planet's $L_{2}$ point (or close enough to allow for accretion).
The spread of the results due to possible differences in entry
velocities was not assessed. Outside of the envelope, the disk
is modeled as in the planet structure calculation 
(see \cisec{sec:FSM}).

The energy budget of a particle includes heating by frictional work
and by the ambient thermal field of the envelope, and cooling
by black-body radiation at the surface temperature and by latent
heat release through phase change.
Above the critical temperature of the material, all energy input
drives mass loss \citepalias[see][ for details]{gennaro2015}.
A body can break-up when subjected to a dynamical pressure larger than 
the compressive strength of the material, although break-up did not
occur in these experiments.

The heavy-element deposition profiles of the rocky (SiO$_{2}$; top)
and icy solids (H$_{2}$O; bottom) are illustrated in \cifig{fig:dep}.
The high-mass
envelope planet is shown on the right (the shaded areas represent
the condensed core region). Upon entry in the envelope, and prior
to significant phase change, the surface temperature of
$R_{s}\lesssim 10\,\mathrm{cm}$ particles tends to attain values
similar to gas temperatures. In these cases, significant ablation
begins only after envelope temperatures rise above some value,
dictated by the vapor pressure curve of the material.
When frictional heating exceeds warming by the local thermal field,
deposition may begin at lower envelope temperatures. In the top
panels of \cifig{fig:dep}, because of the contribution of frictional
heating, $1\,\mathrm{km}$ planetesimals can disseminate heavy 
elements over a broad range of depths, including layers in which 
small solids would ablate. 
At all sizes considered in these experiments, about $50$\% of the input silicate mass would be released 
in the outer $15$\%--$20\%$ (top left) and outer $25$\%--$35$\% (top right)
of the envelope mass. 
Small icy solids (bottom panels) would be disseminated farther
up in the planet, compared to $1\,\mathrm{km}$ planetesimals.
Nonetheless, for all sizes shown in the figure, $50\%$ of 
the accreted ice mass would be dispersed in the outer $5$--$15$\% 
of the envelope mass.
Thus, in all cases shown in \cifig{fig:dep}, most of the accreted
solids would not settle to the core, as assumed in \cisec{sec:FSM},
but would result in an envelope enriched in heavy elements.




\end{document}